\documentclass[12pt]{iopart}

\newcommand{\be}{\begin{eqnarray}}
\newcommand{\ee}{\end{eqnarray}}
\newcommand{\bea}{\begin{eqnarray}}
\newcommand{\eea}{\end{eqnarray}}

\usepackage{graphicx}
\usepackage{bm}
\usepackage{color}
\usepackage{slashed}
\usepackage{cite}
\usepackage{graphicx}
\usepackage{multicol}
\usepackage{graphicx}
\usepackage{color}
\usepackage{slashed}
\usepackage{CJKutf8}
\begin{document}
\begin{CJK}{UTF8}{<font>}
\title{Unveiling the unconventional optical signatures of regular black holes within accretion disk}

\author{Sen Guo$^{*1}$, \ Yu-Xiang Huang$^{2}$, \ Yu-Hao Cui$^{3}$, \ Yan Han$^{*4}$, \ Qing-Quan Jiang$^{2}$, \ En-Wei Liang$^{*1}$, \ Kai Lin$^{*4}$}

\address{
$^1$Guangxi Key Laboratory for Relativistic Astrophysics, School of Physical Science and Technology, Guangxi University, Nanning 530004, People's Republic of China\\
$^2$School of Physics and Astronomy, China West Normal University, Nanchong 637000, People's Republic of China\\
$^3$Hubei Subsurface Multi-scale Imaging Key Laboratory, School of Geophysics and Geomatics, China University of Geosciences, Wuhan 430074, People's Republic of China\\
$^4$College of Mathematics and Information, China West Normal University, Nanchong 637000, People's Republic of China}

\ead{sguophys@126.com}
\vspace{10pt}
\begin{indented}
\item[]Oct. 2023
\end{indented}

\begin{abstract}
The optical characteristics of three types of black holes (BHs) surrounded by a thin accretion disk are discussed, namely the Schwarzschild BH, Bardeen BH, and Hayward BH. We calculate the deflection angle of light as it traverses the vicinity of each BH using numerical integration and semi-analytical methods, revealing that both approaches can effectively elucidate the deflection of light around the BH. We investigate the optical appearance of the accretion disk and its corresponding observational images at various viewing angles, discovering that the luminosity in the region near the BH on the inner side of the accretion disk is higher than that on the outer side owing to higher material density in closer proximity to the BH. We observe a significant accumulation of brightness on the left side of the accretion disk, attributed to the motion of matter and geometric effects. Our findings emphasize the significant influence of the observation inclination angle on the observed outcomes. An increase in the observation inclination angle results in the separation of higher-order images. With the improvement in EHT observation accuracy, we believe that the feature of a minimal distance between the innermost region of the direct image of the Hayward BH and the outermost region of the secondary image can be used as an indicator for identifying Hayward BHs.
\end{abstract}

\section{Introduction}
\label{intro}
\par
Black hole (BH), often regarded as the most mysterious object in the universe, are theoretically predicted to arise from the general relativity (GR). Despite the formidable challenge of directly observing BHs, a substantial body of astronomical evidence supports their presence in the cosmos. The Laser Interferometer Gravitational-Wave Observatory (LIGO) has successfully detected gravitational waves generated by the merger of BHs, not only corroborating Einstein's GR but also furnishing a novel tool for BH research \cite{1}. Subsequently, numerous other gravitational wave events have been detected, yielding a wealth of data to further substantiate the existence of BHs \cite{2,3,4,5}. The most compelling evidence for BH existence stems from direct observations of properties associated with the event horizon. Utilizing the capabilities of the Event Horizon Telescope (EHT), the image of the supermassive BH in Messier (M) 87$^{*}$ elliptical galaxy is obtained, which provided the first direct observational evidence for the existence of BHs \cite{6}. Recently, the EHT collaboration released an image of the BH at the core of our galaxy, Sagittarius (Sgr) A$^{*}$ \cite{7}. These achievements not only demonstrate the spacetime structure of the central BH but also encode dynamic information about the distribution of surrounding matter \cite{8}.

\par
An observable astrophysical BH is typically encircled by a luminous accretion disk, which forms a ring-shaped region where material accumulates around the BH and is considered one of the primary energy sources for the BH. Through numerical simulation, the magnetic field and jet characteristics on the BH accretion disk were investigated, revealing the influence of angular momentum on the accretion-ejection theory \cite{9}. To elucidate the formation of seed BHs in the early universe, a super-exponential accretion disk evolution model has been proposed. It has been demonstrated that a dynamic mechanism can trigger hyper-exponential accretion when a BH seed is trapped in a star cluster that is fed by ubiquitous dense, cold gas streams \cite{10}. By establishing analytic force-free solutions of rotating stars and BHs within the magnetic field, the influence of BH accretion disk and spin on the extraction of electromagnetic energy has been investigated \cite{11}. It is noteworthy that a substantial research team employs numerical methods to study the accretion flow of BHs. In particular, Porth $et~al.$ provided a list of some of the most advanced codes still in use for this purpose as of 2019 \cite{Porth}. Furthermore, Noble $et~al.$ obtained numerical results when the accumulated heat energy is fully radiated. Their findings revealed that the total luminosity of the accretion flow could be approximately $\simeq$ 20$\%$ greater than the value predicted by the Novikov-Thorne model \cite{12}.

\par
In addition to numerical simulations investigating the characteristics of BH accretion disks, there have been notable studies focusing on imaging these disks. For instance, Luminet employed a semi-analytical method to examine the image of a standard accretion disk within the framework of a Schwarzschild BH and obtained direct and secondary images of the disk \cite{13}. Bambi utilized ray-tracing code to fit X-ray data from current and future BH candidates, effectively constraining possible deviations from Kerr geometry in the spin-deformation parameter plane \cite{14}. Grenzebach $et~al.$ derived an analytical formula for the shadow in the Kerr-Newman-NUT-(anti-)de Sitter BH spacetime, allowing them to visualize the photon sphere regions and shadow radii under different parameter values \cite{15}. The EHT Collaboration has successfully conducted rigorous tests on BH image general relativistic radiative transfer codes, resulting in a minor 1$\%$ flux error, primarily attributed to software camera setup discrepancies. These findings provide assurance that numerical uncertainties are unlikely to hinder parameter estimation in current EHT observations \cite{Gold}. Prather $et~al.$ assessed various general relativistic radiative transfer codes used to predict polarized emissions from BH accretion systems and found them to be highly consistent. Whether employing an analytic accretion model or a magnetohydrodynamic simulation, these codes consistently produced images with minimal deviations, ensuring reliable results for interpreting polarized BH images \cite{Prather}. Other intriguing studies on imaging BH accretion disks can be found in references \cite{16,17,18,19,20,21,22}. Gralla $et~al.$ proposed a refined description of the BH shadow, lensing ring, and photon ring within the framework of optically thin emission from a geometrically thin or thick disk. They observed that the brightness of the photon rings exhibits a logarithmic divergence at the shadow radius and that the observable characteristics of the BH shadow are influenced by the details of the accretion disk \cite{23}. Subsequently, researchers have investigated image formation in BH accretion disks within various gravity spacetime backgrounds \cite{24,25,26,27,28,29,30,Wang,Zhang,Hou,Yan,Meng,Meng-1}.

\par
These studies predominantly center on optically thin emission from geometrically thin or thick accretion disk, despite the fact that most observed accretion disks in astrophysics exhibit optically thick. In one particular study, an optically thick and geometrically thin accretion disk surrounding a BH is derived under the influence of Einstein-Gauss-Bonnet gravity. The results reveal that the image of the Gauss-Bonnet BH bears similarity to that of the Schwarz-schild BH, while the naked singularity exhibits distinct characteristics \cite{31}. The accretion disk model employed in this context adheres to the Novikov-Thorne model, which is a standard model characterized by geometric thinness and optical thickness. By examining the observational characteristics of strongly naked static Janis-Newman-Winicour singularity in the Novikov-Thorne model context, Gyulchev $et~al.$ provided insights into properties of the resulting multi-ring structure and elucidated the reasons for the appearance of this ring structure by unveiling its underlying physical formation mechanism \cite{32}. Furthermore, Liu $et~al.$ discussed the physical properties and optical appearance of the electromagnetic radiation emitted from a thin accretion disk around a charged and slowly rotating BH within the context of the Einstein-{\ae}ther theory \cite{33}. Accretion disk images have also been explored in other background spacetimes, including $\gamma$-metric BH \cite{34}, wormholes \cite{35}, Schwarzschild BH penetrated by cosmic string \cite{36}, string cloud Schwarzschild BH \cite{aa}, Schwarzschild-MOG BH in scalar-tensor-vector gravity \cite{bb}.

\par
In addition to studying the characteristics of BH images in singular spacetimes, exploring BH images in regular spacetimes represents an intriguing area of research. Different from irregular BHs, which feature inherent singularities at the origin of spacetime, Bardeen introduced a BH solution unaffected by spacetime singularities \cite{37}. Building upon Bardeen's idea, Hayward proposed a regular Hayward BH solution by combining nonlinear electrodynamics with Einstein's field equation \cite{38}. In our prior research, we examined the observable attributes of the shadow and photon rings of the Hayward BH, uncovering the influence of accretion morphology and the BH's magnetic charge on its optical appearance \cite{27}. Another study focused on a thin-shell wormhole with a Hayward profile \cite{39}. Additionally, we examined the shadow images and observed luminosity of the Bardeen BH under various accretion scenarios \cite{40}. Regrettably, our work only consider regular BHs surrounded by an optically thin emission from geometrically thin accretion disk. Our findings indicated that the disappearance of singularities does not significantly impact the optical appearance of the regular BH, and the lensed ring and photon ring do not reflect spacetime singularity.

\par
In this analysis, our objective is to investigate the appearance of regular BH within the Novikov-Thorne accretion disk context. This investigation aims to provide a comprehensive understanding of the optical morphology of regular BHs, enabling us to distinguish them from singular BHs by investigating their observable characteristics. To achieve this, we employed the semi-analytical approach initially introduced by Luminet \cite{13}, complemented by the ray tracing technique as outlined by Falcke and collaborators \cite{Falcke}. Subsequently, we utilized these methodologies to develop our own \emph{Mathematica} code for the Novikov-Thorne accretion disk model. This enabled us to delve into the investigation of both the primary and secondary images originating from three distinct categories of BHs enveloped by an optically thick accretion disk. We analyze the optical morphology of these BHs, taking into account redshift and flux distribution, and provide a comparative analysis. The structure of this paper is as follows: In Section \ref{sec:2}, we briefly review three types of BHs and discuss their effective potential. In Section \ref{sec:3}, we derive the direct and secondary images of the target BHs using numerical integration and semi-analytical methods. In Section \ref{sec:4}, we present the optical appearance of the accretion disk as well as the corresponding observational images under different observation angles. Finally, we draw conclusions in Section \ref{sec:5}.

\section{Effective potential and ray-tracing}
\label{sec:2}
The static spherically symmetric BH metric can be described as follows:
\begin{equation}
\label{2-1}
{\rm d}s^{2}=-A(r){\rm d}t^{2}+B(r){\rm d}r^{2}+C(r)({\rm d}\theta^{2}+\sin^{2}\theta {\rm d}\phi^{2}),
\end{equation}
where $A(r)$ is BH metric potential, $B(r)$ is defined as the reciprocal of $A(r)$, and $C(r)$ is defined as $r^2$. In this article, we consider Schwarzschild BH, Bardeen BH, and Hayward BH, whose metric descriptions are \cite{37,38}\\
$\emph{Schwarzschild~BH}:$
\begin{equation}
\label{2-2}
A_{1}(r) = 1 - \frac{2M}{r},
\end{equation}
$\emph{Hayward~BH}:$
\begin{equation}
\label{2-3}
A_{2}(r) = 1-\frac{2 M r^{2}}{r^{3}+g^{3}},
\end{equation}
$\emph{Bardeen~BH}:$
\begin{equation}
\label{2-4}
A_{3}(r) = 1-\frac{2 M r^{2}}{(r^{2} + g^{2})^{\frac{3}{2}}},
\end{equation}
in which $M$ represents the mass of BH, while $g$ denotes the BH magnetic charge. As the magnetic charge approaches zero, both the Hayward and Bardeen BHs transition into a Schwarzschild BH. Consequently, the magnetic charge in the regular BH does not alter the spacetime causal structure or the Penrose diagram. The discussion of the causal structure of regular BHs has been elaborated upon in several sources \cite{Lemos,Frolov,Leon,Masa}. However, since the causal structure of BHs does not provide insights into their observable features, we will not delve into Penrose diagrams in this context. Instead, our attention is directed toward examining the optical characteristics of BHs.

\par
To study the motion trajectory of photons around a BH, it is necessary to analyze the evolution of particles in the BH spacetime. Our analysis begins with the Lagrangian of these particles
\begin{eqnarray}
\label{2-5}
\mathcal{L}=-\frac{1}{2}g_{\rm \mu \nu}\frac{{\rm d} x^{\rm \mu}}{{\rm d} \lambda}\frac{{\rm d} x^{\rm \nu}}{{\rm d} \lambda} =\frac{1}{2} \big(A(r)\dot{t}^{2}-B(r)\dot{r}^{2}-C(r)(\dot{\theta}^{2}+\sin^{2}\theta \dot{\phi}^{2})\big),
\end{eqnarray}
where $\lambda$ denotes an affine parameter and $\dot{x}^{\mu}$ represents the four-velocity of the particle. The Lagrangian $\mathcal{L}$ governing the motion of a particle depends on its mass. For massive particles, the Lagrangian is negative, whereas for massless particles, such as photons, it vanishes, i.e., $\mathcal{L}=0$. The generalized momentum $p_{\rm \mu}$ associated with the particle can be obtained from the Lagrangian by applying the Euler-Lagrange equations of motion,
\begin{equation}
\label{2-6}
p_{\rm \mu} = \frac{\partial \mathcal{L}}{\partial \dot{x}^{2}} = g_{\rm \mu \nu} \dot{x}^{\rm \nu},
\end{equation}
which leads to four equations of motion for a particle with energy $E$ and angular momentum $L$, we have
\begin{eqnarray}
\label{2-7}
&&p_{\rm t} = g_{\rm tt} \dot{t} + g_{\rm t \phi} \dot{\phi} = - E,\\
\label{2-8}
&&p_{\rm \phi} = g_{\rm \phi t} \dot{t} + g_{\rm \phi \phi} \dot{\phi} = L,\\
\label{2-9}
&&p_{\rm r} = g_{\rm rr} \dot{r},\\
\label{2-10}
&&p_{\rm \theta} = g_{\rm \theta \theta} \dot{\theta}.
\end{eqnarray}
We only consider the particle that move on the equatorial plane ($\theta_{0}=\pi/2$, $\dot{\theta_{0}}=0$ and $\ddot{\theta}=0$). In addition, we assume that the test particle is a photon with zero angular momentum ($\mathcal{L}=0$). Using the four equations mentioned above, we can obtain the following results:
\begin{eqnarray}
\label{2-11}
&&\dot{t} = \frac{E g_{\rm \phi \phi} + L g_{\rm t \phi}}{g^{2}_{\rm t \phi} - g_{\rm tt}g_{\rm \phi \phi}},\\
\label{2-12}
&&\dot{\phi} = - \frac{E g_{\rm t \phi}+L g_{\rm tt}}{g_{\rm t \phi}^{2} - g_{\rm tt} g_{\rm \phi \phi}},\\
\label{2-13}
&&g_{\rm rr}\dot{r}^{2}= \frac{E^{2}g_{\rm \phi \phi} + 2 E L g_{\rm t \phi} +L^{2} g_{\rm tt}}{g^{2}_{\rm t \phi} - g_{\rm tt} g_{\rm \phi \phi}} -1 = \mathcal{V}_{\rm eff},
\end{eqnarray}
where $\mathcal{V}_{\rm eff}$ is the effective potential. For the static spherically symmetric BH ($g_{\rm t \phi}=0$), one can get
\begin{equation}
\label{2-14}
\dot{r}^{2}=\frac{1}{b^{2}}-\mathcal{V}_{\rm eff},
\end{equation}
in which $b$ is the impact parameter, defining as $b \equiv L/E=\frac{r^{2}\dot{\phi}}{A(r)\dot{t}}$. The effective potential can be further expressed as
\begin{equation}
\label{2-15}
\mathcal{V}_{\rm eff} = L^{2} \frac{A(r)}{r^{2}}.
\end{equation}
By utilizing Eq. (\ref{2-15}), we can determine both the radius of the photon sphere and the critical impact parameter by setting $\mathcal{V}_{\rm eff}=\frac{1}{b_{\rm c}^2}$ and $\mathcal{V}_{\rm eff}'=0$. Fig. 1 illustrates the effective potential of a BH as a function of radius, showcasing three different BH solutions. It is noticeable that the peak of the effective potential for a Schwarzschild BH occurs at approximately $\simeq 3r_{\rm g}$, where $r_{\rm g}$ signifies the Schwarzschild radius. For the same set of parameters, the effective potential for a Bardeen BH is greater, and the radius corresponding to the peak of the effective potential is smaller in comparison to that of a Hayward BH.
\begin{center}
\includegraphics[width=8cm,height=6cm]{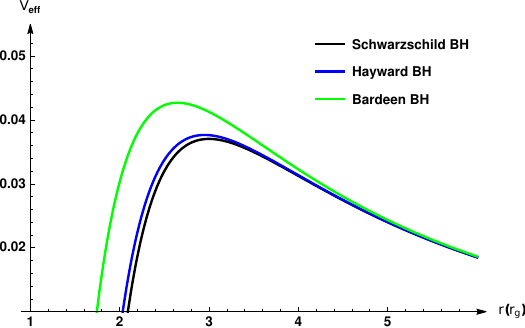}
\parbox[c]{15.0cm}{\footnotesize{\bf Fig~1.}  
Effective potential as a function of radius for three types of BHs. The black line corresponds to the Schwarzschild BH, the green line corresponds to the Bardeen BH, and the blue line represents the Hayward BH. The BH mass is set as $M = 1$ and magnetic charge is $g=0.5$.}
\label{fig1}
\end{center}

\par
To better illustrate the distinctions among the three types of BHs, we derived the deflection angles for various scenarios. Utilizing Eqs. (\ref{2-8}) and (\ref{2-9}), we obtained the following results:
\begin{equation}
\label{2-16}
\frac{{\rm d} r}{{\rm d} \phi}=r^{2}\sqrt{\frac{1}{b^{2}}-\frac{1}{r^2}A(r)}.
\end{equation}
By introducing a parameter $u \equiv 1/r$, the above equation is written as
\begin{equation}
\label{2-17}
\Omega(u) \equiv \frac{{\rm d} u}{{\rm d} \phi}=\sqrt{\frac{1}{b^{2}}-u^{2}A(u)}.
\end{equation}
For a distant observer, the deflection angle of light is
\begin{equation}
\label{2-18}
\psi = \int^{u_{\rm obs}}_{u_{\rm source}} \frac{{\rm d} u}{\Omega(u)}  =  \int^{u_{\rm obs}}_{u_{\rm source}} \frac{{\rm d} u}{\sqrt{\frac{1}{b^{2}}-u^{2}A(u)}},
\end{equation}
where the upper and lower limits of the integral represent the photon's emission point $u_{\rm source}$ and the observer's position $u_{\rm obs}$, respectively. It's essential to note that in some instances, rays with an impact parameter exceeding the critical value may possess a radial inflection point $u_{\rm o}$. In such scenarios, the motion of photons can be described by splitting the integral into two separate integrals.
\begin{equation}
\label{2-19}
\psi = \int^{u_{\rm source}}_{u_{\rm o}} \frac{{\rm d} u}{\sqrt{\frac{1}{b^{2}}-u^{2}A(u)}} + \int^{u_{\rm o}}_{u_{\rm obs}} \frac{{\rm d} u}{\sqrt{\frac{1}{b^{2}}-u^{2}A(u)}},
\end{equation}
in which $u_{\rm o}$ is photon turn point, satisfying
\begin{equation}
\label{2-20}
u_{\rm o}^{2} = \frac{1}{b^{2} A(u)}.
\end{equation}
Hence, Eq. (\ref{2-19}) effectively guarantees the tracking of all light near a BH. Apart from employing numerical integration algorithms to investigate light deflection angles, it's also valuable to delve into the semi-analytical algorithm introduced by Luminet \cite{13}. In this article, we also employ a semi-analytical approach to derive the deflection angle of light, with the goal of comparing it with numerical methods.

\par
Using Eq. (\ref{2-17}), one can get\\
$\emph{Schwarzschild~BH}:$
\begin{equation}
\label{2-21}
\Bigg(\frac{{\rm d} u_{\rm s}}{{\rm d} \phi}\Bigg)^{2} = 2 M u^{3} - u^{2} + \frac{1}{b^{2}},
\end{equation}
$\emph{Hayward~BH}:$
\begin{equation}
\label{2-22}
\Bigg(\frac{{\rm d} u_{\rm h}}{{\rm d} \phi}\Bigg)^{2} = \frac{2M}{g^3+\frac{1}{u^{3}}} - u^{2} + \frac{1}{b^{2}},
\end{equation}
$\emph{Bardeen~BH}:$
\begin{equation}
\label{2-23}
\Bigg(\frac{{\rm d} u_{\rm b}}{{\rm d} \phi}\Bigg)^{2} = \frac{2M}{\Big(g^2+\frac{1}{u^{2}}\Big)^{\frac{3}{2}}} - u^{2} + \frac{1}{b^{2}},
\end{equation}
where the lower indicators $u_{\rm s}$, $u_{\rm h}$, and $u_{\rm b}$ represent the Schwarzschild, Hayward, and Bardeen BHs, respectively. Using the Cardano formula, we can rewrite Eqs. (\ref{2-21})-(\ref{2-23}) as follows:\\
$\emph{Schwarzschild~BH}:$
\begin{equation}
\label{2-24}
2M G_{\rm S}({\rm u}) = 2M (u_{\rm s}-u_{\rm s1})(u_{\rm s}-u_{\rm s2})(u_{\rm s}-u_{\rm s3}),
\end{equation}
$\emph{Hayward~BH}:$
\begin{equation}
\label{2-25}
2M G_{\rm H}({\rm u}) = 2M (u_{\rm h}-u_{\rm h1})(u_{\rm h}-u_{\rm h2})(u_{\rm h}-u_{\rm h3}),
\end{equation}
$\emph{Bardeen~BH}:$
\begin{equation}
\label{2-26}
2M G_{\rm B}({\rm u}) = 2M (u_{\rm b}-u_{\rm b1})(u_{\rm b}-u_{\rm b2})(u_{\rm b}-u_{\rm b3}),
\end{equation}
where the cubic polynomial $G({\rm u})$ contains two positive roots and one negative root, denoted by $u_{1} \leq 0 < u_{2} < u_{3}$. The specific forms of these roots can be determined using the method proposed by Luminet \cite{13} and expressed in terms of the periastron distance $P$ as follows
\begin{equation}
\label{2-27}
u_{1}=\frac{P-2M-Q}{4MP},~~u_{2}=\frac{1}{P},~~u_{3}=\frac{P-2M+Q}{4MP},
\end{equation}
in which the $Q^{2} \equiv (P-2M)(P+6M)$. The impact parameter in three different BH scenarios can be written as\\
$\emph{Schwarzschild~BH}:$
\begin{equation}
\label{2-28}
b_{s}^{2}=\frac{P^{3}}{P-2M},
\end{equation}
$\emph{Hayward~BH}:$
\begin{equation}
\label{2-29}
b_{h}^{2}=\frac{1}{\frac{1}{P^{2}}+2M\Big(u^{3}-\frac{1}{P^{3}}+\frac{1}{g^{3}+\frac{1}{u^{3}}}\Big)},
\end{equation}
$\emph{Bardeen~BH}:$
\begin{equation}
\label{2-30}
b_{b}^{2}=\frac{1}{\frac{1}{P^{2}}+2M\Bigg(u^{3}-\frac{1}{P^{3}}+\frac{1}{\big(g^{2}+\frac{1}{u^{2}}\big)^{\frac{3}{2}}}\Bigg)}.
\end{equation}
Based on the equations provided above, it is possible for an observer located at infinity can capture the light that arrives here at a given periastron distance $P$. As a result, the bending angle of the light ray can be calculated, which is
\begin{equation}
\label{2-31}
\psi (u) = \sqrt{\frac{2}{M}} \int_{0}^{u_{2}} \frac{{\rm d}u}{\sqrt{(u-u_{1})(u-u_{2})(u-u_{3})}} - \pi.
\end{equation}
By introducing the elliptic integral, Eq. (\ref{2-31}) can be transformed into
\begin{equation}
\label{2-32}
\psi (u) = \sqrt{\frac{2}{M}} \Bigg(\frac{2 F(\Psi_{1},k)}{\sqrt{u_{3}-u_{1}}} - \frac{2 F(\Psi_{2},k)}{\sqrt{u_{3}-u_{1}}}\Bigg) - \pi,
\end{equation}
where $F(\Psi_{\rm i},k)$ represents an incomplete elliptic integral of the first kind, with amplitudes as parameter variables. The parameter variables are $\Psi_{1}=\frac{\pi}{2}$, $\Psi_{2}=\sin^{-1}\sqrt{\frac{-u_{1}}{u_{2}-u_{1}}}$, and length $k=\sqrt{\frac{u_{2}-u_{1}}{u_{3}-u_{1}}}$. Therefore, the total change in bending angle can be simplified as follows
\begin{equation}
\label{2-33}
\psi (u) = 2 \sqrt{\frac{P}{Q}} \Big(K(k)- F(\Psi_{2},k)\Big) - \pi,
\end{equation}
in which $K(k)$ is the complete elliptic integrals of the first kind.

\par
We computed the deflection angles of light in three types of BHs using two different methods, both of which, despite their distinct forms, all yield good results. Fig. 2 shows the trajectory of light near a BH using the ray-tracing code, where the yellow line represents $b<b_{\rm c}$, the red line represents $b>b_{\rm c}$, and the blue line represents $b=b_{\rm c}$. It's important to note that in this scenario, photons rotate multiple times around the BH. We can observe that, in comparison to the Schwarzschild BH, the black disk radius of the Hayward BH and Bardeen BH are slightly smaller, and the light density received by a distant observer decreases.
\begin{center}
\includegraphics[width=5cm,height=5cm]{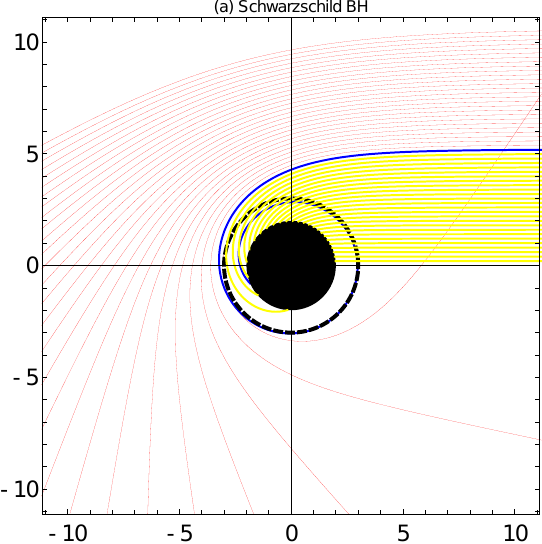}
\includegraphics[width=5cm,height=5cm]{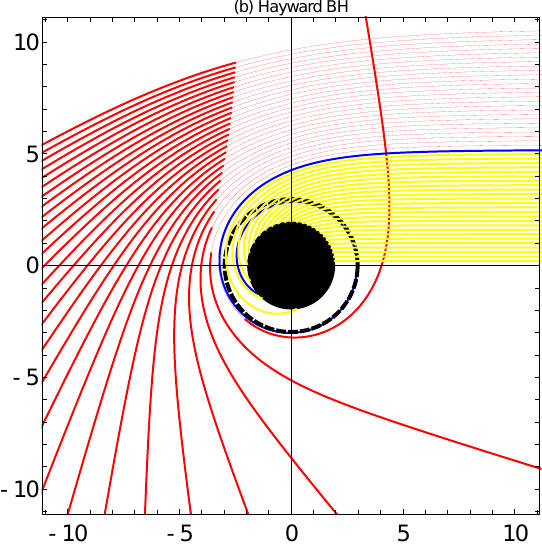}
\includegraphics[width=5cm,height=5cm]{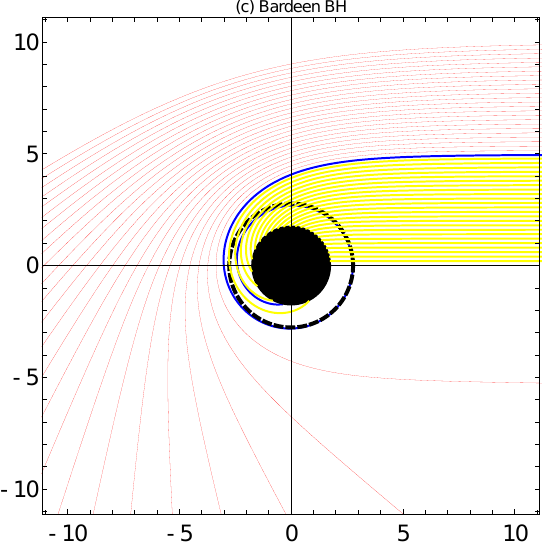}
\parbox[c]{15.0cm}{\footnotesize{\bf Fig~2.}  
The light trajectories of different BHs in the polar coordinate $(b,\phi)$. The BHs are shown as the black disks, and the dashed black lines represent the BHs photon ring orbits. {\em Left Panel}: Schwarzschild BH. {\em Middle Panel}: Hayward BH. {\em Right Panel}: Bardeen BH. The BH mass is taken as $M = 1$ and magnetic charge is $g=0.5$.}
\label{fig2}
\end{center}

\section{Direct and secondary images of BH}
\label{sec:3}
\par
We consider a BH enveloped by an optically thick and geometrically thin accretion disk, as described by the Novikov-Thorne thin disk model. This model originates from research conducted in the previous century \cite{Novikov,Page}. The details of this model are not discussed here (see Refs. \cite{13,Novikov,Page} for details). Our primary focus is on contrasting the images of three distinct types of BHs within the framework of this model.

\par
The radiation originates from a point with coordinates $(r,\phi)$ on the emitting plane $M$ and propagates to a point with coordinates $(b,\alpha)$ on the observation plane $m$. Two images of the accretion disk can be obtained on the observation plane: the direct image at coordinates $(b^{(d)},\alpha)$ and the secondary image at coordinates $(b^{(s)},\alpha+\pi)$ \cite{13}. In this section, we derive the direct and secondary images of BHs surrounded by an accretion disk using a numerical integration algorithm and semi-analytical methods, respectively.

\par
Following the references \cite{31,32,33,34}, we assume that there is a relationship between the azimuth angle $\phi$ of the photon emission point, the celestial angle $\eta$, and the observer inclination angle $\theta_{0}$, which is
\begin{eqnarray}
\label{3-1}
\cos \phi = - \frac{\sin \eta \tan \theta_{0}}{\sqrt{\sin^{2} \eta \tan^{2} \theta_{0}+1}},\\
\sin \phi = - \frac{1}{\sqrt{\sin^{2} \eta \tan^{2} \theta_{0}+1}},
\end{eqnarray}
where the celestial angle $\eta$ ranges from $0$ to $2\pi$. Using these two trigonometric functions and and Eq. (\ref{2-19}), we can determine the impact parameters of all photon trajectories that are emitted from a specific circular orbit with radial coordinates $r=r_{\rm source}$. It should be noted that these trajectories can reach the observer whose observation inclination angle is $\theta_{0}$ and position is $r=r_{\rm obs}$. Therefore, we obtain the following expression:
\begin{eqnarray}
\label{3-2}
\phi= \int^{u_{\rm obs}}_{u_{\rm source}}\frac{{\rm d} u}{\sqrt{\frac{1}{b^{2}}-u^{2}A(u)}} = - \arccos \frac{\sin \eta \tan \theta_{0}}{\sqrt{\sin^{2} \eta \tan^{2} \theta_{0}+1}}.
\end{eqnarray}
This formula only applies to a specific ray of light and cannot account for all possible deflections that may occur. It is possible that some of the light emitted from the disk could be deflected at a greater angle and could even orbit the BH multiple times before reaching a distant observer. To account for these more complex scenarios, the above equation must be extended to a more general situation
\begin{eqnarray}
\label{3-3}
\int^{u_{\rm obs}}_{u_{\rm source}}\frac{{\rm d} u}{\sqrt{\frac{1}{b^{2}}-u^{2}A(u)}} = k \pi - \arccos \frac{\sin \eta \tan \theta_{0}}{\sqrt{\sin^{2} \eta \tan^{2} \theta_{0}+1}},
\end{eqnarray}
where $k$ is a positive integer that can represent the image order. $k=0$ describes the direct image of the accretion disk, while $k=1,~2,~3...$ representing second order, third order, and higher order situations.

\par
The same effect can also be demonstrated using Luminet's semi-analytic method. In this approach, we assume that the deflection angle from the source to the observer is denoted by $\gamma$, and the observer's inclination angle is denoted by $\theta_{0}$. This derivation also involves the use of elliptic integrals. By applying trigonometric function relationships, we obtain
\begin{equation}
\label{3-4}
\cos \alpha = \cos \gamma \sqrt{\cos^{2} \alpha + \cot^{2} \theta_{0}}.
\end{equation}
The Eq. (\ref{2-21}) can be rewritten as
\begin{equation}
\label{3-5}
\gamma = \frac{1}{\sqrt{2M}} \int_{0}^{1/r} \frac{{\rm d} u}{\sqrt{G(u)}} = 2 \sqrt{\frac{P}{Q}} \Big(F(\zeta_{\rm r},k) - F(\zeta_{\rm \infty}, k)\Big),
\end{equation}
where $F(\zeta_{\rm r},k)$ and $F(\zeta_{\rm \infty})$ represent the elliptical integrals, the parameter variables $\sin^{2}\zeta_{\rm r}=\frac{Q-P+2M+4MP/r}{Q-P+6M}$, $\sin^{2}\zeta_{\rm \infty}=\frac{Q-P+2M}{Q-P+6M}$, and length $k^{2}=\frac{Q-P+6M}{2Q}$ \cite{13}. Meanwhile, the radius $r$ can be written as a function of $\alpha$ and $P$, which is
\begin{equation}
\label{3-6}
\frac{1}{r} = \frac{P-2M-Q}{4MP} + \frac{Q-P+6M}{4MP} sn^{2}\Bigg(\frac{\gamma}{2}\sqrt{\frac{Q}{P}} + F(\zeta_{\rm \infty}, k)\Bigg),
\end{equation}
where $sn$ is Jacobi elliptic function. Depends on this equation, one can present the iso-radial curves for a specific angle $\theta_0$, which can be used to obtain a direct image of the accretion disk. To obtain the $(n+1)th$ order image of the accretion disk, the Eq. (\ref{3-5}) satisfies
\begin{equation}
\label{3-7}
2n \pi - \gamma = 2 \sqrt{\frac{P}{Q}} \Big(2K(k)- F(\zeta_{\rm r},k) - F(\zeta_{\rm \infty}, k)\Big),
\end{equation}
in which $K(k)$ is the complete elliptic integral, $n$ is a positive integer, which can represent the image order.

\par
Figure 3 present the direct ($k=0$) and secondary ($k=1$) images of circular rings in orbit around BHs. The image displays three different types of BHs: Schwarzschild BH, Hayward BH, and Bardeen BH. The ray trajectories are positioned at $r=6M$, $r=10M$, $r=15M$, and $r=20M$ (from innermost to outermost orbits), with inclination angles of $\theta_{0}=17^{\circ}$, $53^{\circ}$, and $75^{\circ}$. As the observation angle decreases, the secondary image becomes embedded inside the direct image, forming a structure similar to a ``photon ring''. This can be considered a viable alternative for a photon ring. As the inclination angle of the observation increases, the degree of separation between the direct and secondary images increases. Upon comparing the results of the three types of BHs, it can be observed that the orbital size of the Hayward BH is smaller than that of the Bardeen BH under the same parameters. Notably, the difference between the Bardeen BH and Schwarzschild BH is not significant.
\begin{center}
  \includegraphics[width=5cm,height=5cm]{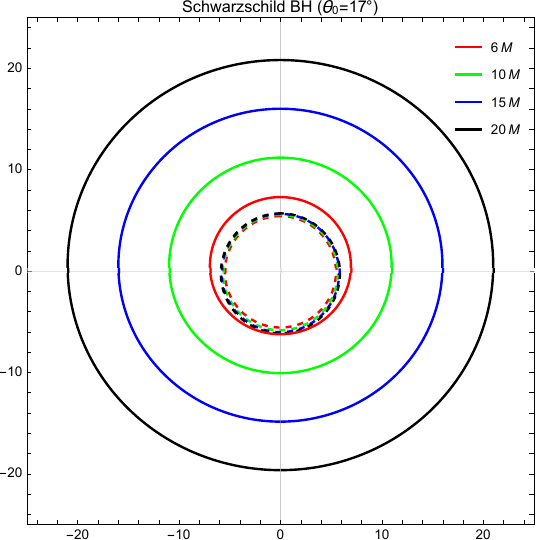}
  \includegraphics[width=5cm,height=5cm]{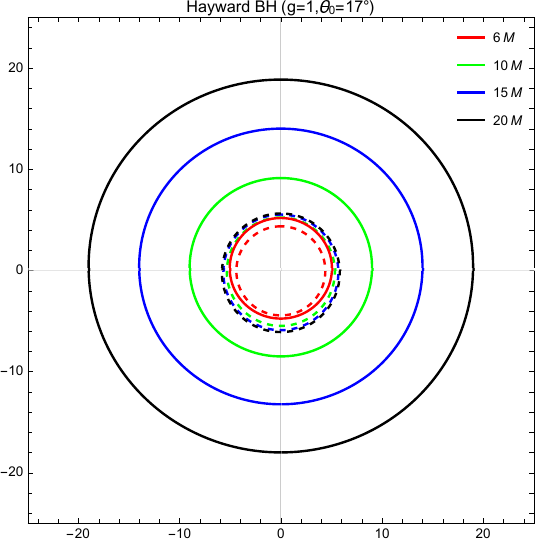}
  \includegraphics[width=5cm,height=5cm]{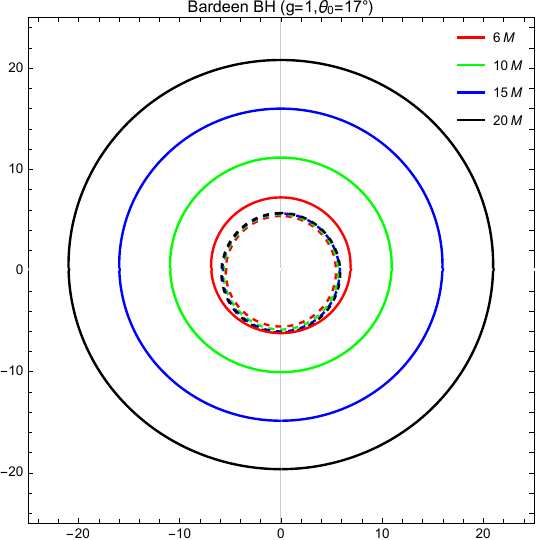}
  \includegraphics[width=5cm,height=5cm]{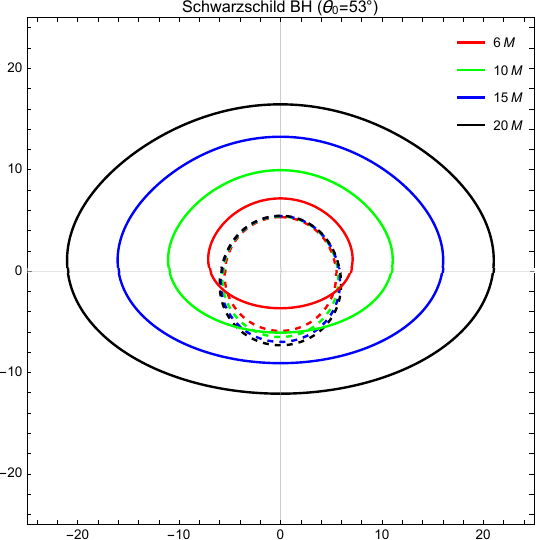}
  \includegraphics[width=5cm,height=5cm]{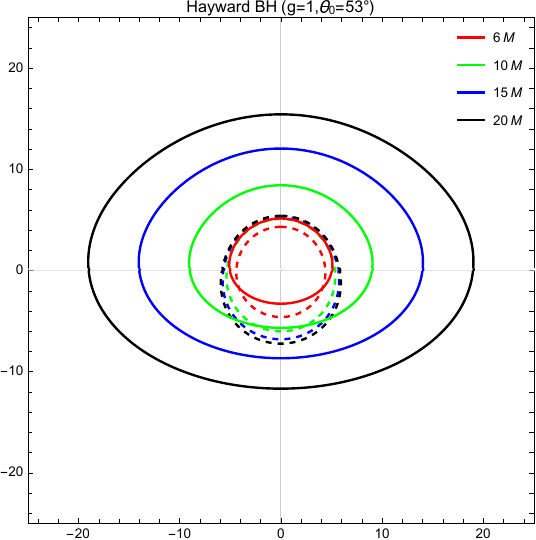}
  \includegraphics[width=5cm,height=5cm]{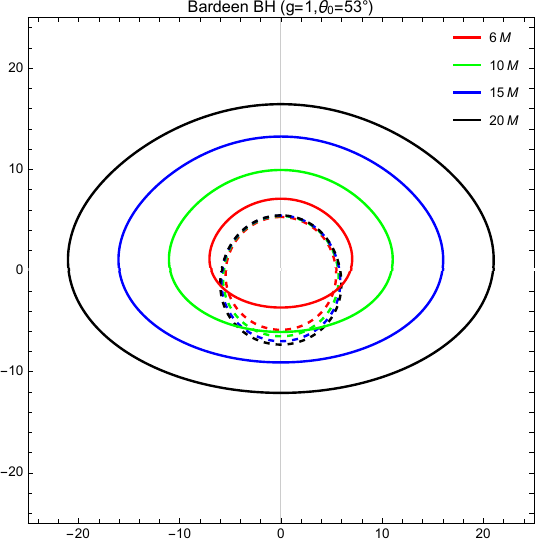}
  \includegraphics[width=5cm,height=5cm]{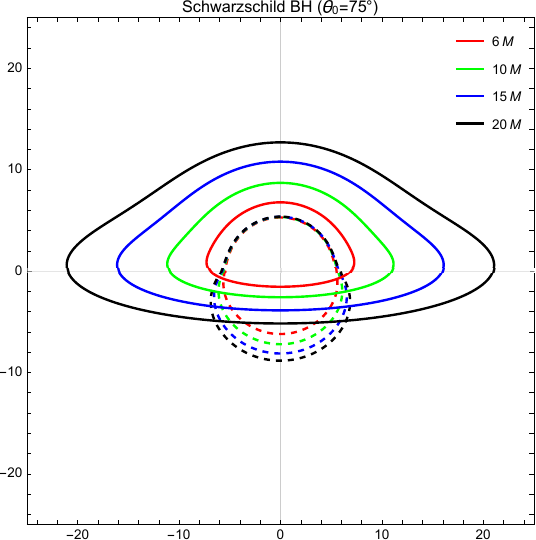}
  \includegraphics[width=5cm,height=5cm]{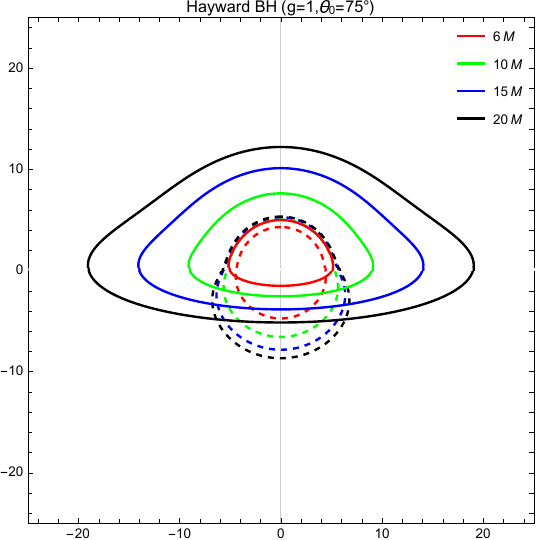}
  \includegraphics[width=5cm,height=5cm]{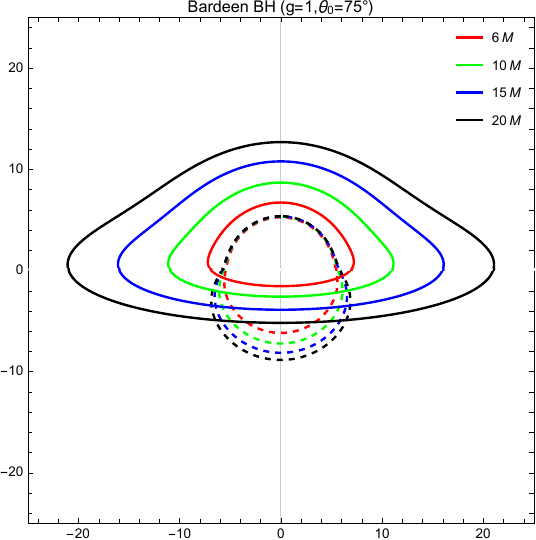}
\parbox[c]{15.0cm}{\footnotesize{\bf Fig~3.}  
The direct (solid line) and secondary (dashed line) images of three types of BHs accretion disks with different the observation angles. {\em Left Panel}: Schwarzschild BH. {\em Middle Panel}: Hayward BH. {\em Right Panel}: Bardeen BH. The BH mass is taken as $M = 3 M_{\odot}$ and magnetic charge is $g=1$.}
\label{fig3}
\end{center}

\section{Optical appearance of three types of BHs}
\label{sec:4}
\par
In this section, we examine the observable characteristics of three types of BHs and explore how observation inclination angles affect them. Our goal is to distinguish these characteristics from their optical appearance. Assuming an optically thick and geometrically thin accretion disk surrounding the BH, we can calculate the electromagnetic radiation flux ($\rm ergs^{-1}cm^{-2}str^{-1}Hz^{-1}$) emitted from a specific radial position $r$ on the accretion disk using the following formula \cite{Page}:
\begin{equation}
\label{4-1}
F = - \frac{\dot{M}}{4\pi \sqrt{\rm -I}} \frac{\Omega_{,\rm r}}{(E-\Omega L)^{2}} \int_{r_{\rm in}}^{r} (E- \Omega L)L_{,\rm r} {\rm d} r,
\end{equation}
where $\dot{M}$ refers to the mass accretion rate, $I$ is the determinant of the induced metric in the equatorial plane, and $r_{\rm in}$ denotes the inner edge of the accretion disk. The parameters $E$, $\Omega$, and $L$ represent the angular velocity, energy and angular momentum of the particles in a circular orbit, respectively, and they can be expressed as follows:
\begin{eqnarray}
\label{4-2}
&&E= - \frac{g_{\rm tt} + g_{\rm t \phi} \Omega}{\sqrt{-g_{\rm tt} + 2g_{\rm t \phi}\Omega - g_{\rm \phi \phi}\Omega^{2}}},\\
\label{4-3}
&&L= \frac{g_{\rm t \phi} + g_{\rm \phi \phi} \Omega}{\sqrt{-g_{\rm tt} + 2g_{\rm t \phi}\Omega - g_{\rm \phi \phi}\Omega^{2}}},\\
\label{4-4}
&&\Omega=\frac{{\rm d}\phi}{{\rm d}t}=\frac{-g'_{\rm t \phi} + \sqrt{(g'_{\rm t \phi})^{2} - g'_{\rm tt}g'_{\rm \phi \phi}}}{g'_{\rm \phi \phi}}.
\end{eqnarray}

\par
It is worth noting that the Novikov-Thorne model is specifically designed for Kerr metrics. However, if the metric is changed to that of a regular BH, the structure of the thin disk should also be adjusted. Jaroszy\'{n}ski $et~al.$ calculated the radiation spectrum observed by observers at various positions relative to the equatorial plane of the disk \cite{Jaroszynski}. They investigated the emission line profiles from self-gravitating toroids around BHs \cite{Usui}. Some literature on Kerr metric accretion disk images and ray-tracing can be found in references \cite{Harko,Harko-1,Bambi}. The pioneering work on the four-dimensional spherical symmetry metric originated from reference \cite{13}. It is important to note that the spherical symmetry metric component $g_{\rm t \phi}=0$ leads to changes in angular velocity, energy, and angular momentum forms. A discussion on spherically symmetric ray-tracing and accretion disk images can be found in references \cite{Muller,Atamurotov,Eiroa,Schee,Falco,Poutanen,Vincent,Vagnozzi}. In the case of spherical and axisymmetric scenarios, the received light intensity is inconsistent for a distant observer.

\subsection{Radiation flux}
\label{sec:4-1}
Utilizing the Eqs. (\ref{4-1})-(\ref{4-4}), the radiation flux over the disk for three BHs are obtained, we have\\
$\emph{Schwarzschild~BH}:$
\begin{equation}
\label{4-1-1}
F_{1}(r) = - \frac{3 \dot{M} \sqrt{\frac{M}{r^{3}}}}{\pi r(24M - 8r)} \int_{r_{\rm in}}^{r} \frac{(6M-r)\sqrt{\frac{M}{r^{3}}}r}{6M-2r} {\rm d}r,
\end{equation}
$\emph{Hayward~BH}:$
\begin{eqnarray}
\label{4-1-2}
F_{2}(r) &&=\frac{3 \dot{M} Mr(r^3-5g^3)}{8\pi \sqrt{\frac{M(r^3-2g^3)}{(g^3+r^3)^2}} (g^3+r^3) H} \times  \nonumber \\
&& \int_{r_{\rm in}}^{r} \frac{M r \Big(-8g^6 +11 g^3 r^3 + r^5(r-6M)\Big)}{2\sqrt{\frac{M(r^3-2g^3)}{(g^3+r^3)^2}} (g^3+r^3) H} {\rm d}r,
\end{eqnarray}
where $H \equiv g^6 + 2g^3 r^3 + r^5(r-3M)$.\\
$\emph{Bardeen~BH}:$
\begin{eqnarray}
\label{4-1-3}
F_{3}(r) &&= \frac{3 \dot{M} Mr(r^2 - 4g^2)}{8 \pi r \sqrt{\frac{M(r^2 - 2g^2)}{(g^2 + r^2)^\frac{5}{2}}} B \Big(-3M r^4 + (g^2 + r^2)^{\frac{5}{2}}\Big)} \nonumber \\
&&\times \int_{r_{\rm in}}^{r} -\frac{M r C}{2 \sqrt{\frac{M(r^2 - 2g^2)}{(g^2 + r^2)^\frac{5}{2}}} B^2 D} {\rm d}r,
\end{eqnarray}
where $B \equiv (g^2+r^2)$, $C \equiv (8g^8 + 8g^6 r^2 -9g^4 r^4 -10 g^2 r^6 -r^8 +6M r^6 \sqrt{g^2 + r^2})(-3Mr^4 + (g^2+r^2)^\frac{5}{2})$, $D \equiv (g^6 + 3g^4 r^2 + 3 g^2 r^4 +r^6 -3M r^4 \sqrt{g^2+r^2})^2$.

\par
Figure 4 illustrates the radiant intensity of the accretion disk in three different scenarios involving BHs. As the observation tilt angle increases, the direct and secondary images gradually separate. Notably, the luminosity in the region near the BH on the inner side of the accretion disk is higher than that on the outer side due to higher material density in closer proximity to the BH. This proximity makes it more susceptible to the BH's gravity, resulting in more intense heating and radiation. Furthermore, the appearance of the three types of BHs can be distinguished from one another. In comparison to the Schwarzschild BH image, the Hayward BH exhibits a feature of minimal distance between the innermost region of the direct image and the outermost region of the secondary image. The appearance of the Bardeen BH closely resembles that of the Schwarzschild BH, but there is a significant difference in radiance. Consequently, our results suggest that observing such an effect can be beneficial when considering an external metric consistent with the regular BH solution.
\begin{center}
  \includegraphics[width=5cm,height=4.4cm]{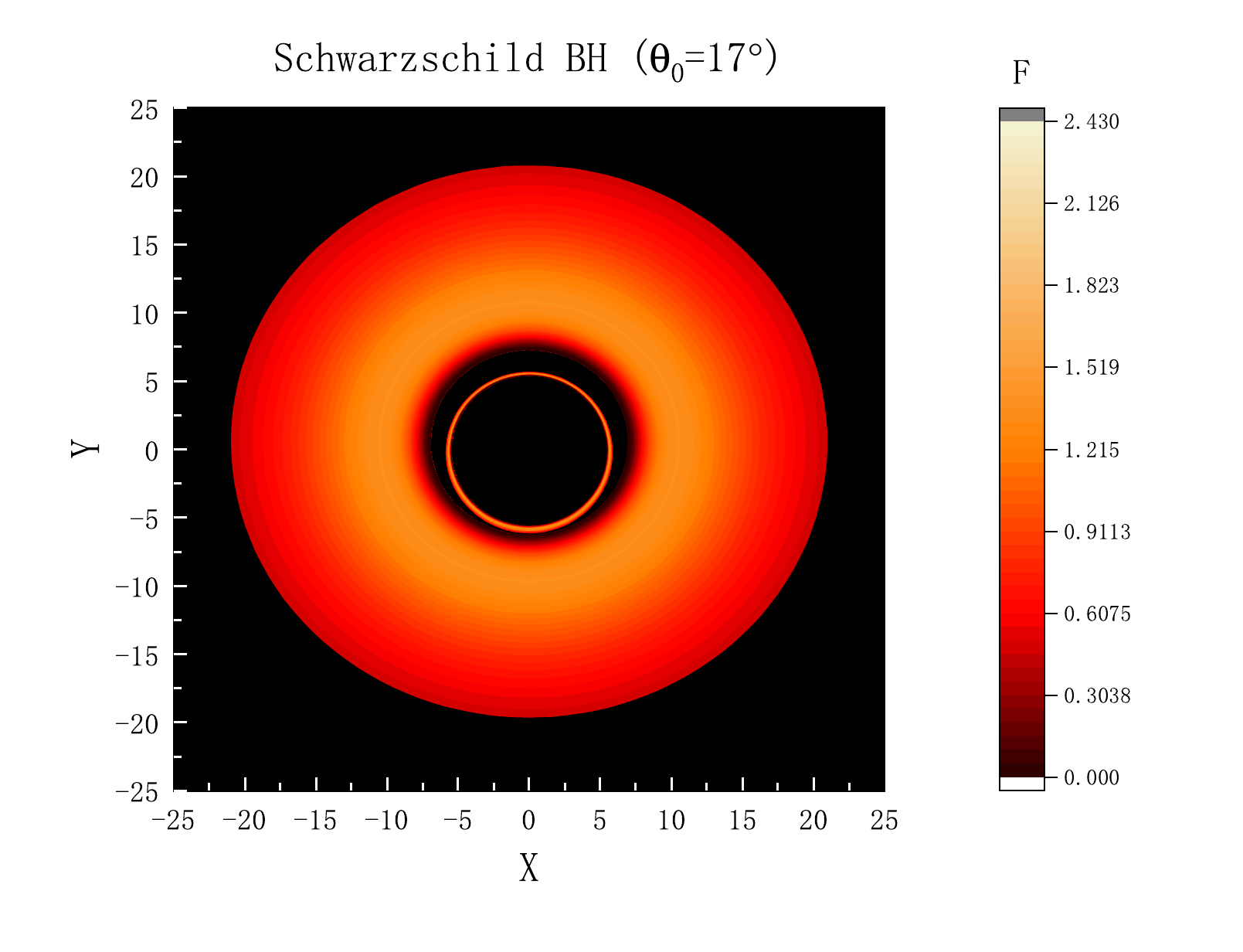}
  \includegraphics[width=5cm,height=4.4cm]{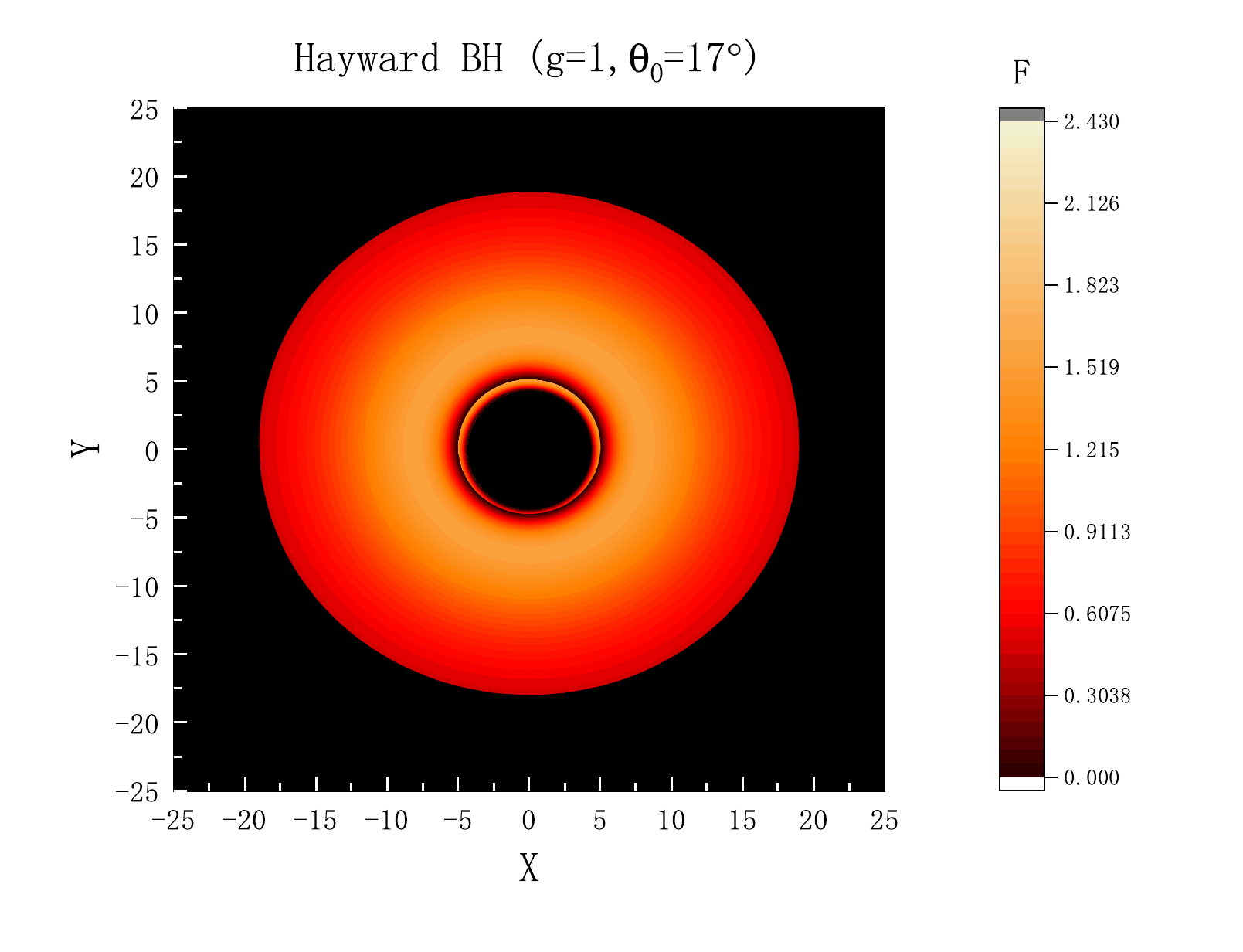}
  \includegraphics[width=5cm,height=4.4cm]{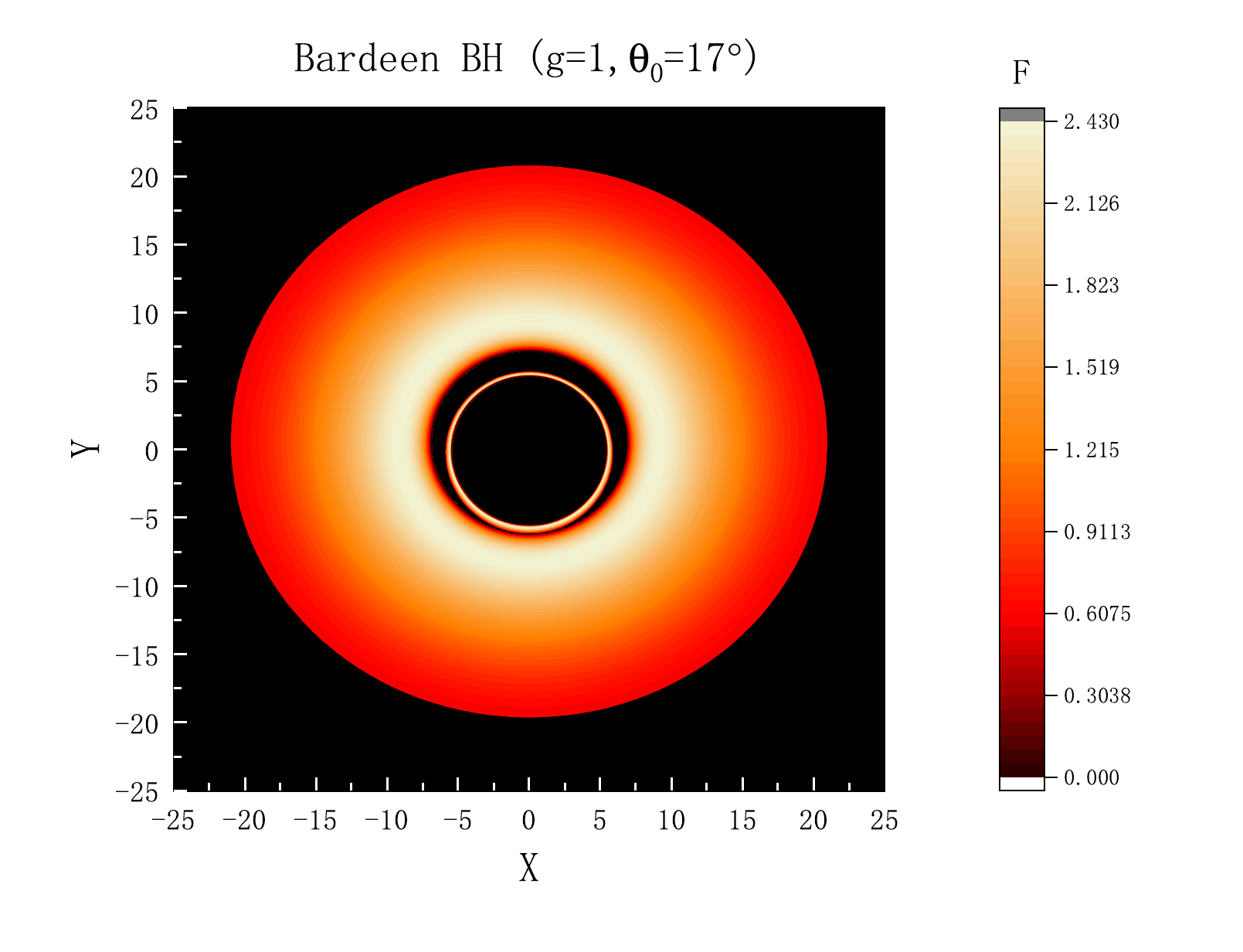}
  \includegraphics[width=5cm,height=4.4cm]{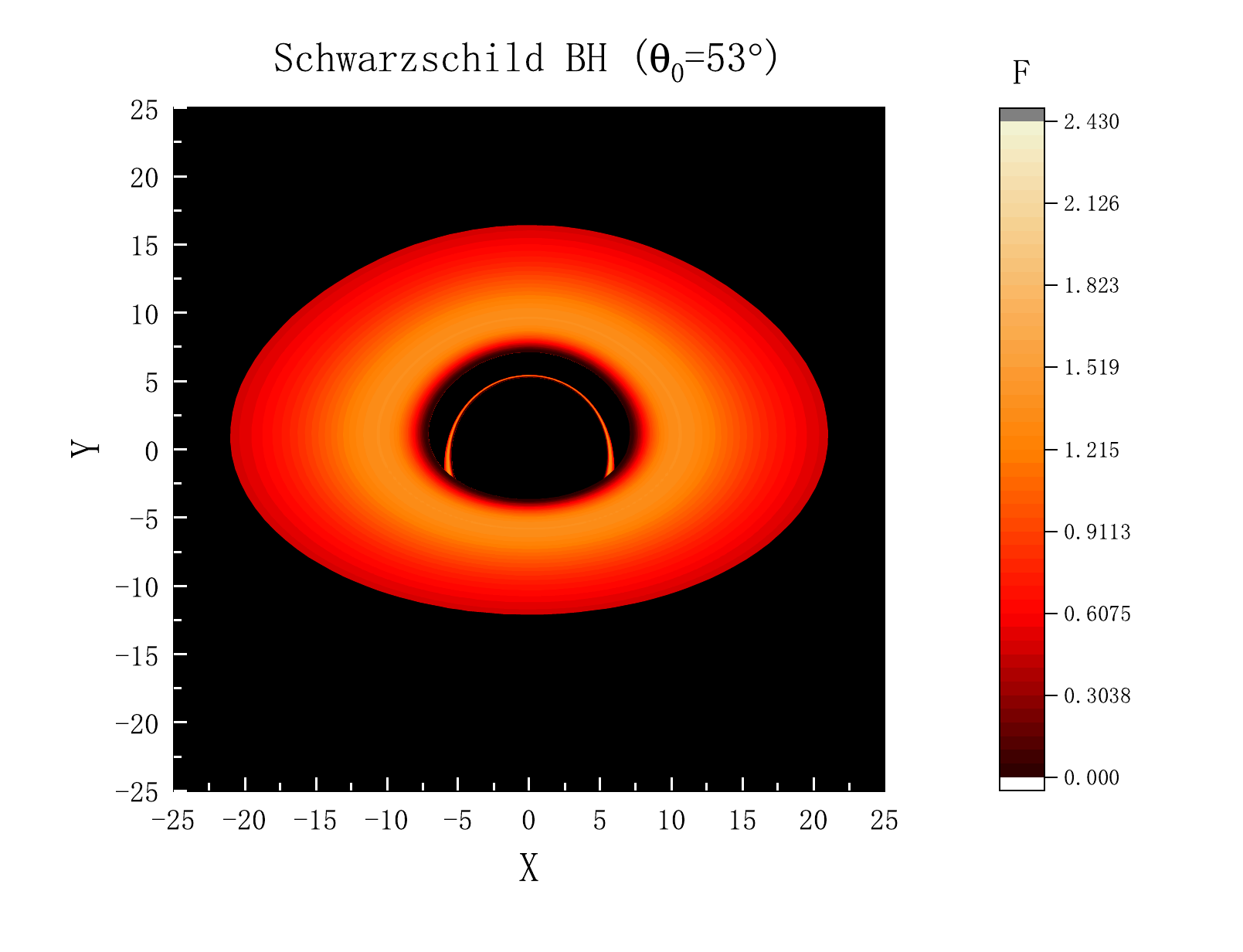}
  \includegraphics[width=5cm,height=4.4cm]{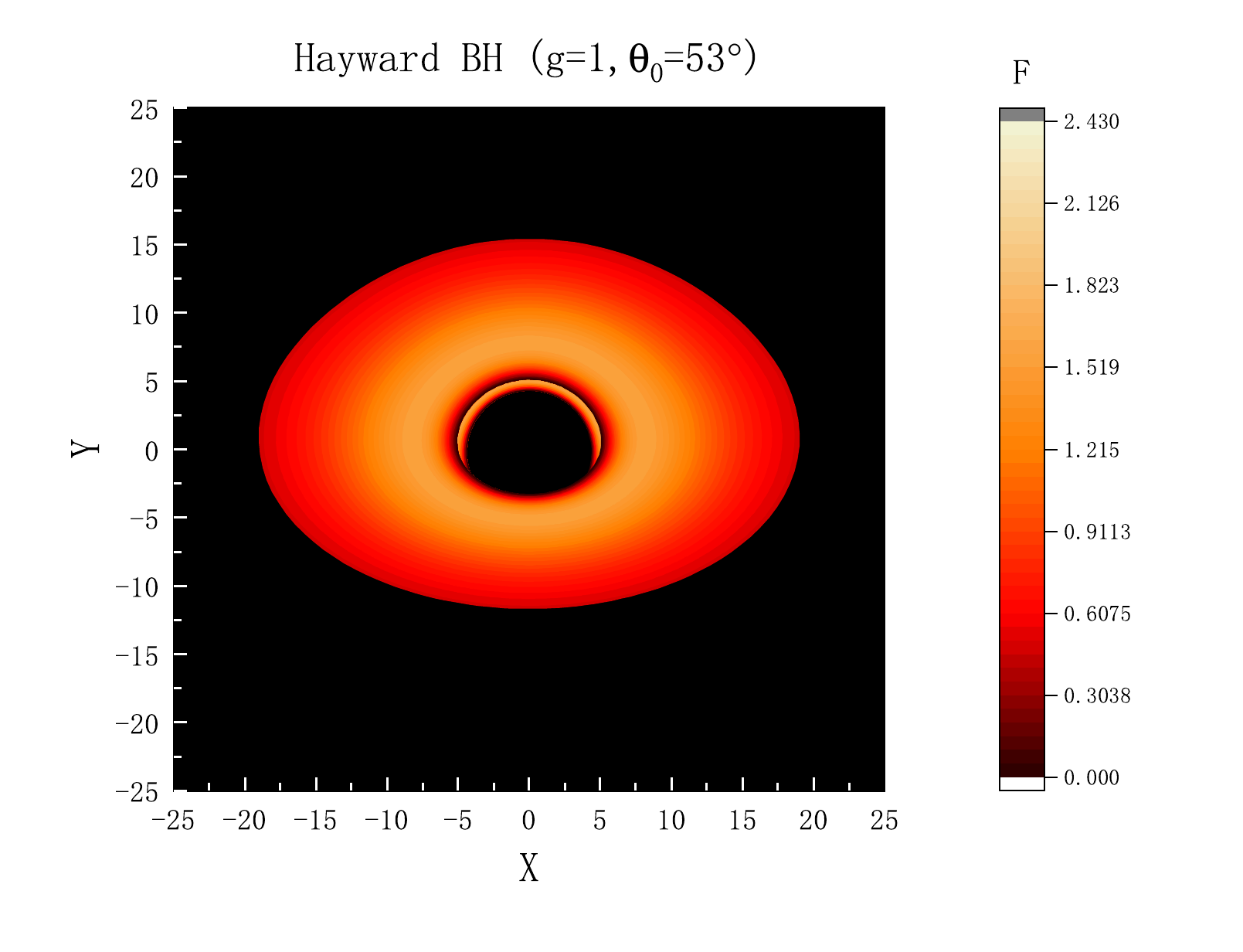}
  \includegraphics[width=5cm,height=4.4cm]{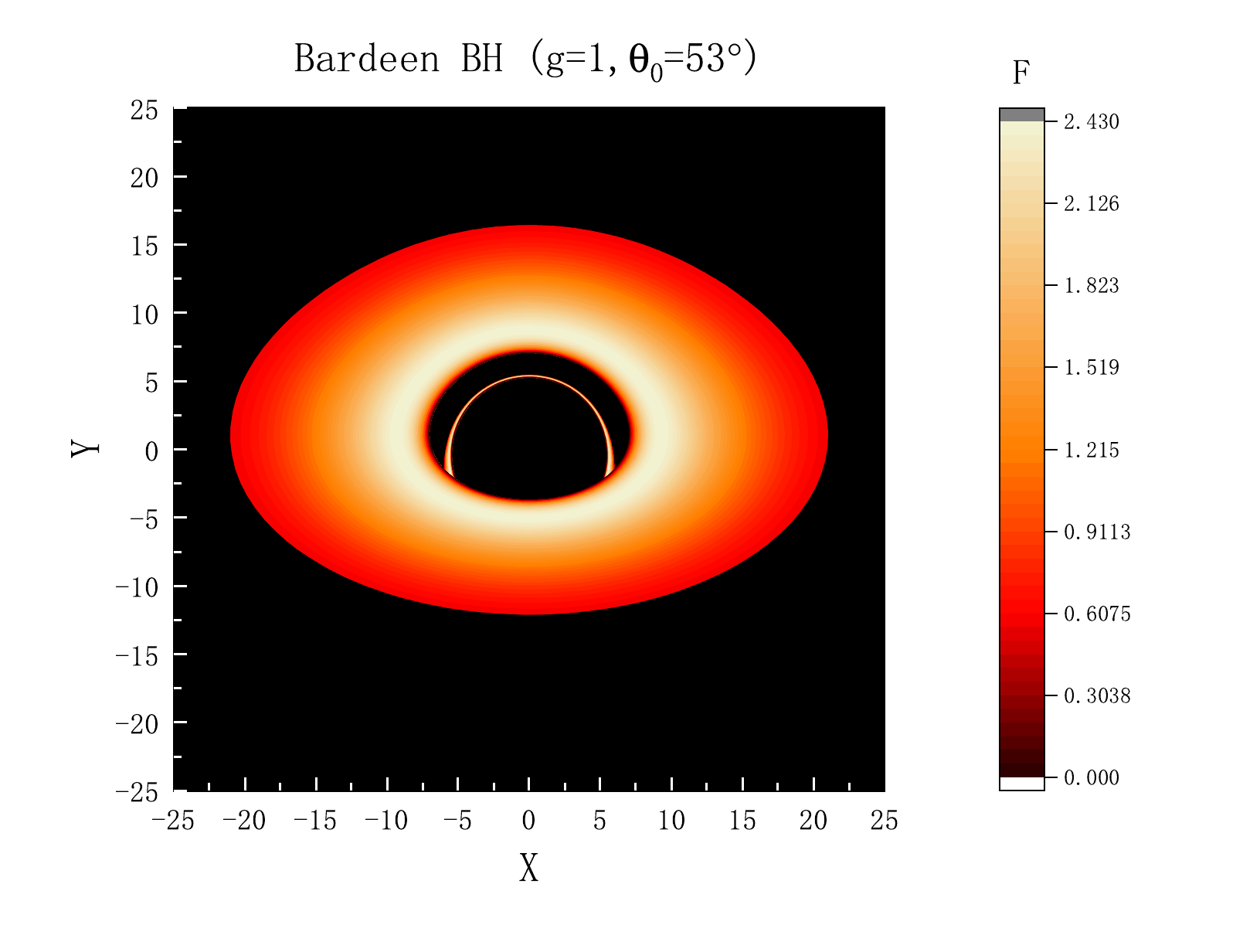}
  \includegraphics[width=5cm,height=4.4cm]{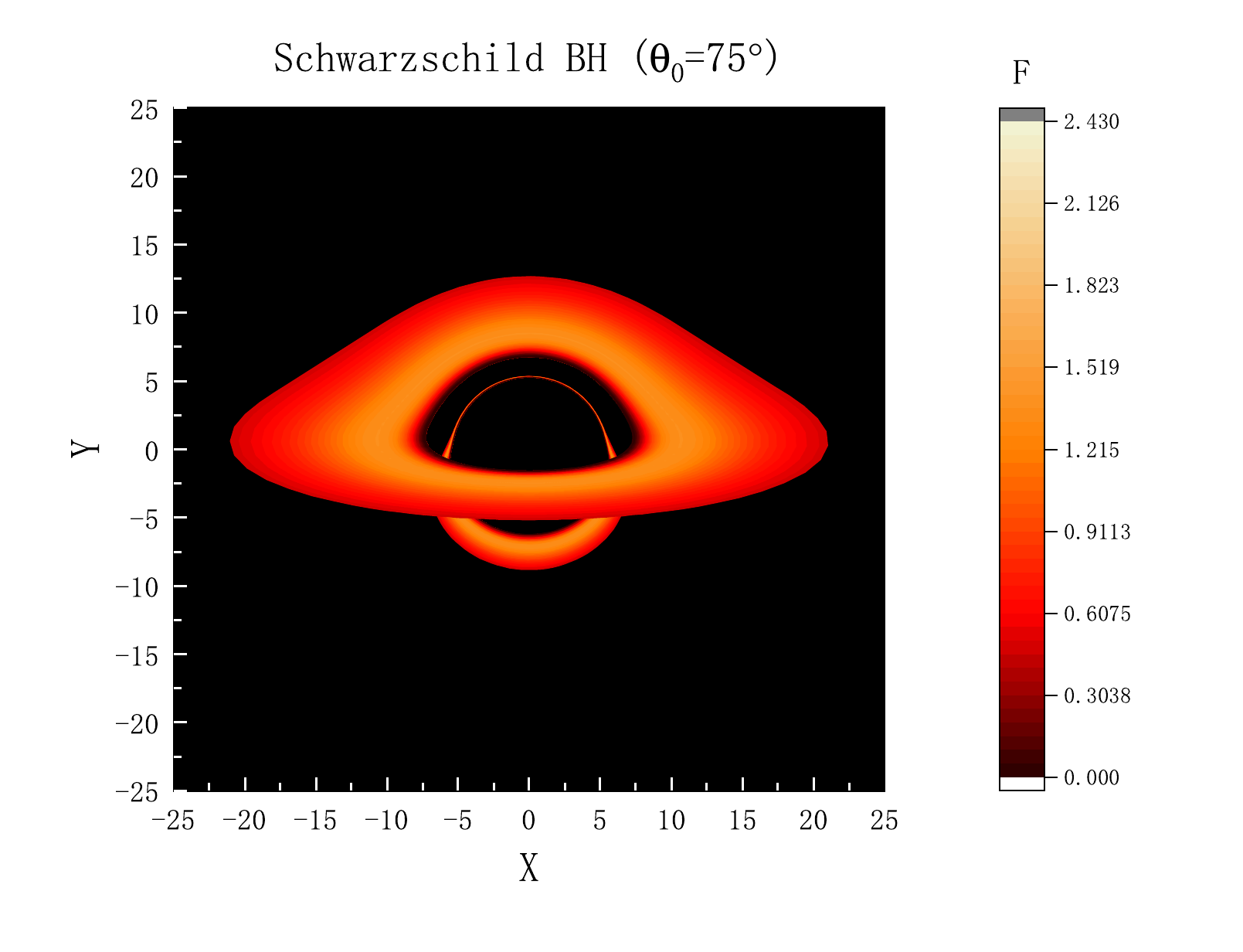}
  \includegraphics[width=5cm,height=4.4cm]{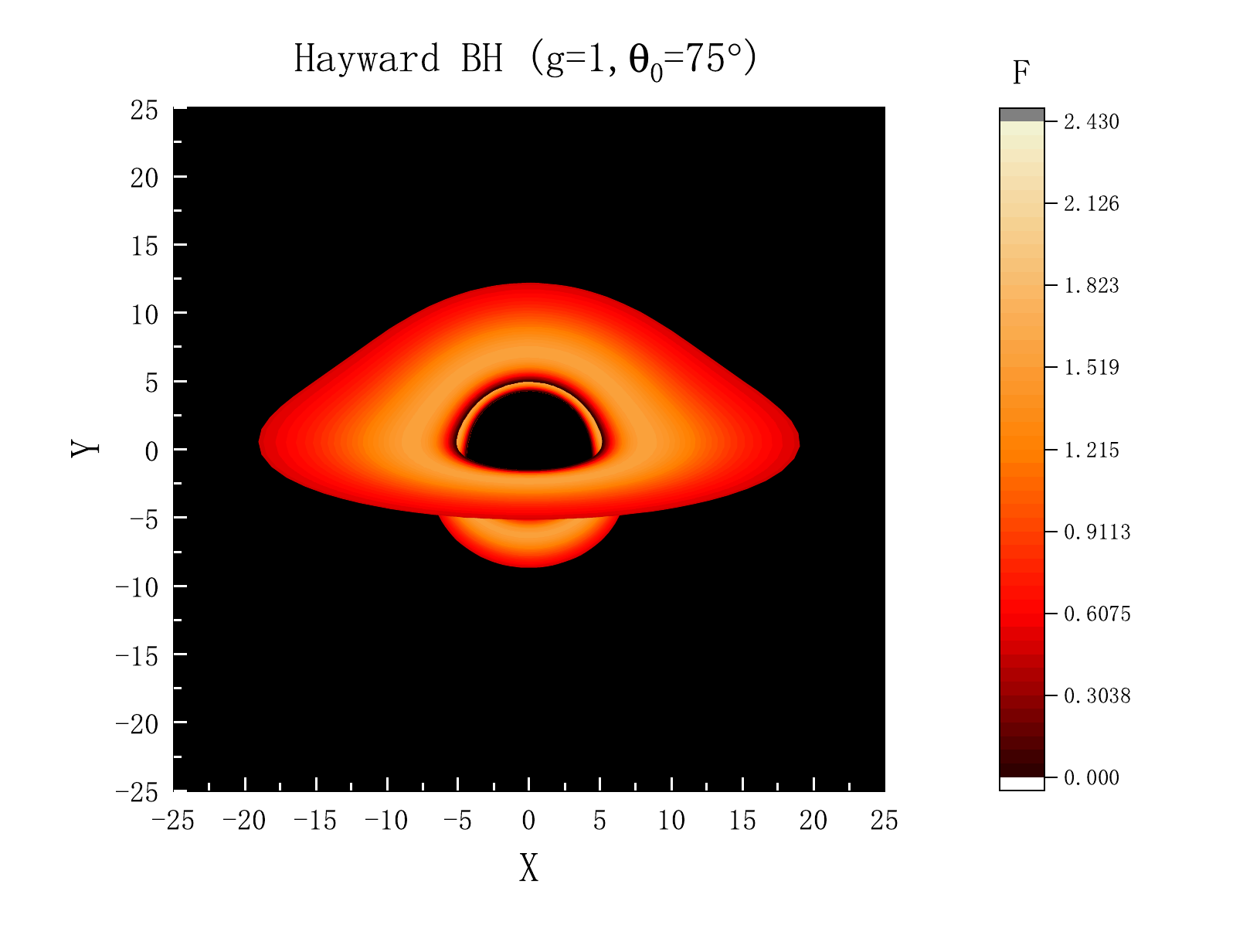}
  \includegraphics[width=5cm,height=4.4cm]{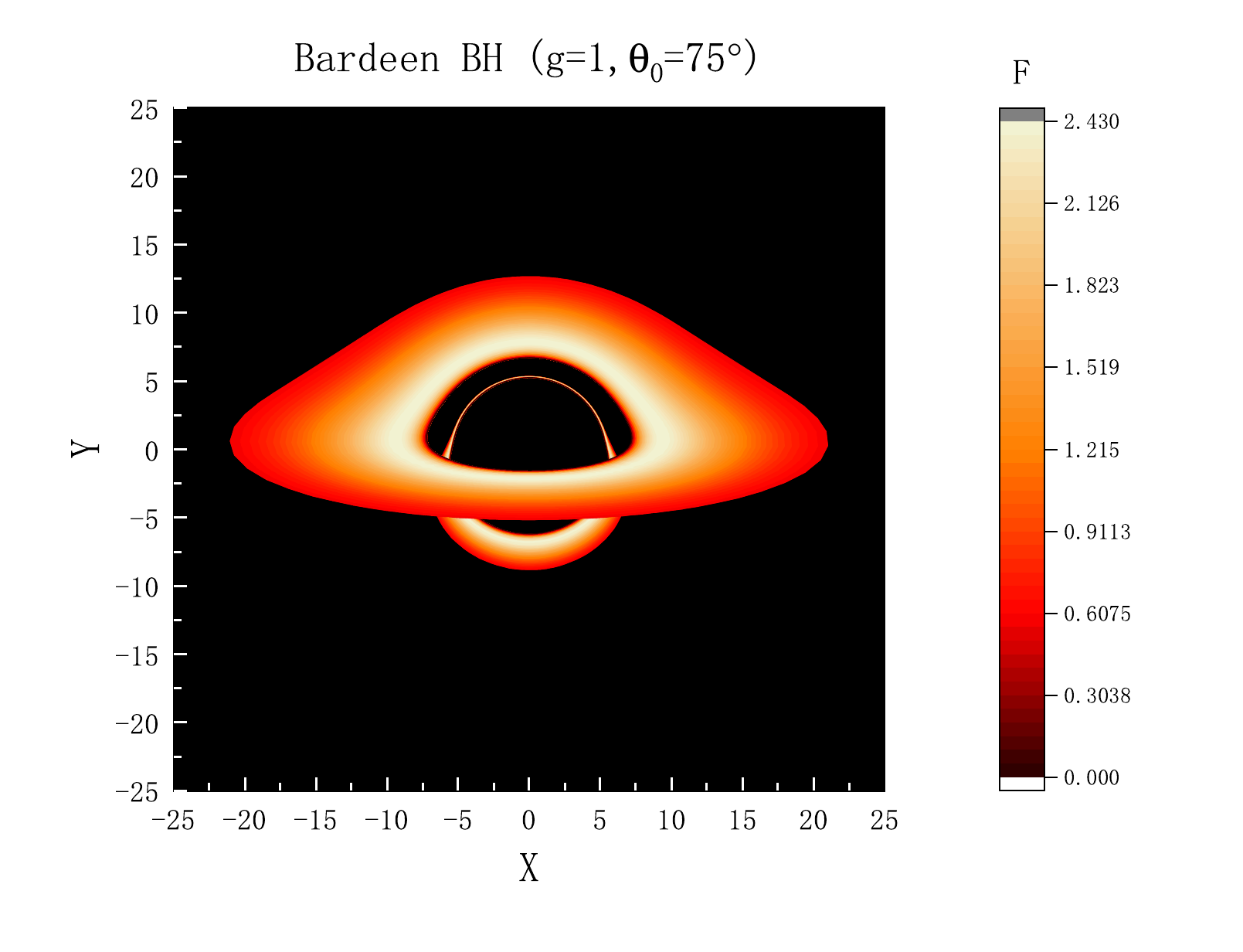}
\parbox[c]{15.0cm}{\footnotesize{\bf Fig~4.}  
The radiation flux images of three types of BHs accretion disks with different observation angles. {\em Left Panel}: Schwarzschild BH. {\em Middle Panel}: Hayward BH. {\em Right Panel}: Bardeen BH. The BH mass is taken as $M = 3 M_{\odot}$ and magnetic charge is $g=1$.}
\label{fig4}
\end{center}

\par
Figure 5 shows that the total observed intensity function, the observed intensity has an extremely sharp rise before the peak. Fig. 4 illustrates that the two-dimensional shadows cast on the celestial coordinates. Our result indicates that the size and position of the BH shadows remain unchanged in different accretion flows, implying that shadows are the spatiotemporal geometric characteristics of BHs. The primary distinction between regular BHs and Schwarzschild BHs is the presence of magnetic charges. To better illustrate the correlation between radiant flux and magnetic charge, we have employed three functional models: the Fourier Function (Case 1), the Gaussian Function (Case 2), and the Exponential Function (Case 3). Section \ref{sec:4-4} provides detailed information on the form and accuracy of these functions. Our objective is to represent the accurate value of $F$ as a function of $g$ in a more analytical manner. Fig. 5 displays the red dot, which represents the precise value of $F$ as a function of $g$, indicating an increase in $F$. With the Hayward BH, the slope of the fitting curve grows with an increase in $g$, whereas the fitting curve of the Bardeen BH exhibits a linear relationship with the magnetic charge. Additionally, we have observed that, under specific fixed parameters, the exponential function provides the best fit to the numerical results.
\begin{center}
\includegraphics[width=6.5cm,height=5cm]{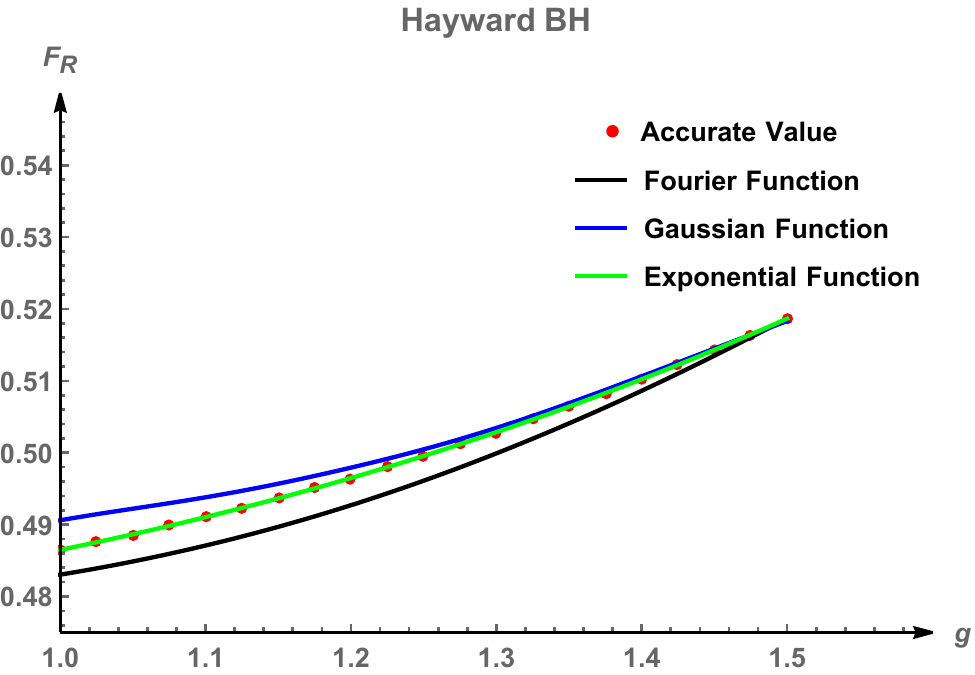}
\hspace{0.8cm}
\includegraphics[width=6.5cm,height=5cm]{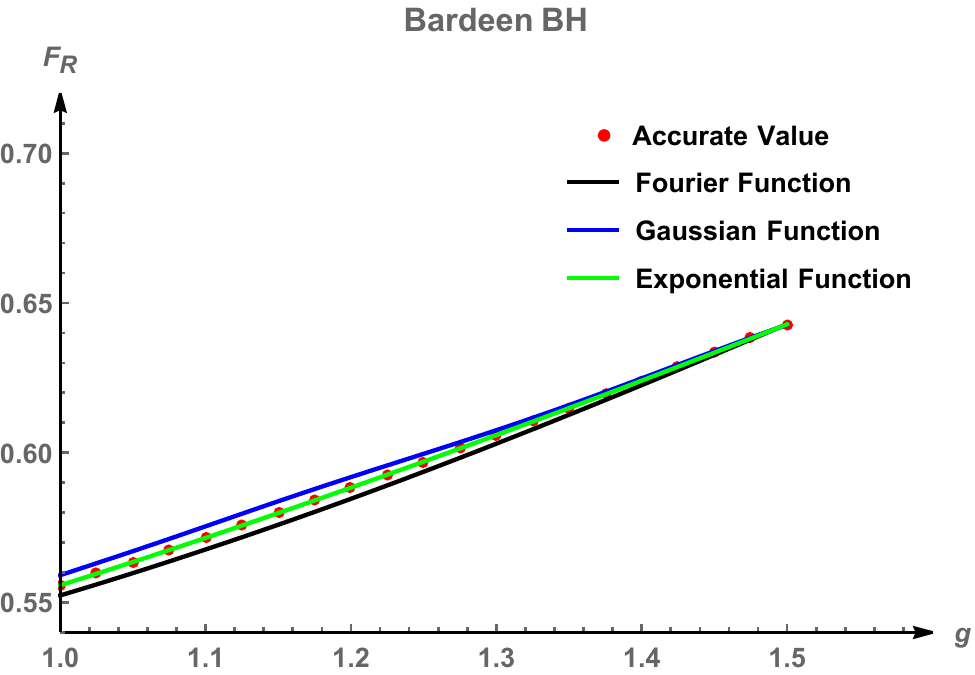}
\parbox[c]{15.0cm}{\footnotesize{\bf Fig~5.}  
Fitting of Fourier function (black solid line), Gaussian function (bule solid line), and Exponential function (green solid line) to the accurate value of the radiation flux $F$ and magnetic charge $g$ (red dots). The BH mass is taken as $M = M_{\odot}$.}
\label{fig5}
\end{center}

\subsection{Redshift factor}
\label{sec:4-2}
To obtain the observable radiant flux for a remote observer, it is necessary to consider the correction for the redshift effect. It is important to note that the redshift effect in this context encompasses both gravitational redshift and the Doppler effect. This correction involves taking into account the redshift factor, denoted as $d$, which is expressed as follows:
\begin{equation}
\label{4-2-1}
d=1+z= \frac{E_{\rm em}}{E_{\rm obs}},
\end{equation}
where $E_{\rm em}$ and $E_{\rm obs}$ respectively represent the projection of photon four-momentum $k_{\mu}$ on the four-velocities of source $p_{\rm source}^{\rm \mu}$ and observer $p_{\rm obs}^{\rm \mu}$ at a distance. The $E_{\rm em}$ is
\begin{equation}
\label{4-2-2}
E_{\rm em} = p_{\rm t} \mu^{\rm t} + p_{\rm \phi} \mu^{\rm \phi} = p_{\rm t} \mu^{\rm t} \Bigg(1 + \Omega \frac{p_{\rm \phi}}{p_{\rm t}}\Bigg),
\end{equation}
in which $p_{\rm t}$ and $p_{\rm \phi}$ represent the photon 4-momentum. For observers at a large distance, the ratio $p_{t}/p_{\phi}$ represents the impact parameter of the photons relative to the z-axis. This can be related to the trigonometric functions as follows
\begin{equation}
\label{4-2-3}
\sin \theta_{0} \cos \alpha = \cos \gamma \sin \beta,
\end{equation}
the $p_{t}/p_{\phi}$ can be rewritten as
\begin{equation}
\label{4-2-4}
\frac{p_{\rm t}}{p_{\rm \phi}}= b \sin \theta_{0} \sin \alpha.
\end{equation}
Therefore, the redshift factor $d$ is obtained, one can get
\begin{equation}
\label{4-2-5}
d = 1 + z = \frac{E_{\rm em}}{E_{\rm obs}} = \frac{1 + b \Omega \cos \beta}{\sqrt{-g_{\rm tt} - 2 g_{\rm t \phi} - g_{\rm \phi\phi}}}.
\end{equation}
For the three types of spherical symmetry metrics we are considering, their redshift factors are written as\\
$\emph{Schwarzschild~BH}:$
\begin{equation}
\label{4-2-6}
d_{1} = \frac{1+b \sqrt{\frac{M}{r^3}} \sin \alpha \sin \theta_{0}}{\sqrt{1-\frac{3M}{r}}},
\end{equation}
$\emph{Hayward~BH}:$
\begin{equation}
\label{4-2-7}
d_{2} = \frac{1+b\sqrt{\frac{M(r^3 - 2g^3)}{(g^3+r^3)^2}} \sin \alpha \sin \theta_{0}}{\sqrt{\frac{g^6 +2 g^3 r^3 +r^5(r-3M)}{(g^3 + r^3)^2}}},
\end{equation}
$\emph{Bardeen~BH}:$
\begin{equation}
\label{4-2-8}
d_{3} = \frac{1 + b \sqrt{\frac{M(r^2 - 2g^2)}{(g^2 + r^2)^{\frac{5}{2}}}} \sin \alpha \sin \theta_{0}}{\sqrt{1- \frac{3Mr^4}{(g^2 + r^2)^{\frac{5}{2}}}}}.
\end{equation}

\par
We assume that the inner and outer boundaries of the accretion disk consist of stable circular orbits at $r=6M$ and $r=20M$, respectively. As illustrated in Fig. 6, we have observed a significant asymmetry between the redshift area ($z>0$) and blueshift area ($z<0$). Furthermore, the redshift area gradually increases as the observer's inclination angle increases. In comparison to Schwarzschild BH, Hayward BHs exhibit a smaller redshift region, whereas Bardeen BHs display a larger redshift region.
\begin{center}
  \includegraphics[width=5cm,height=4.4cm]{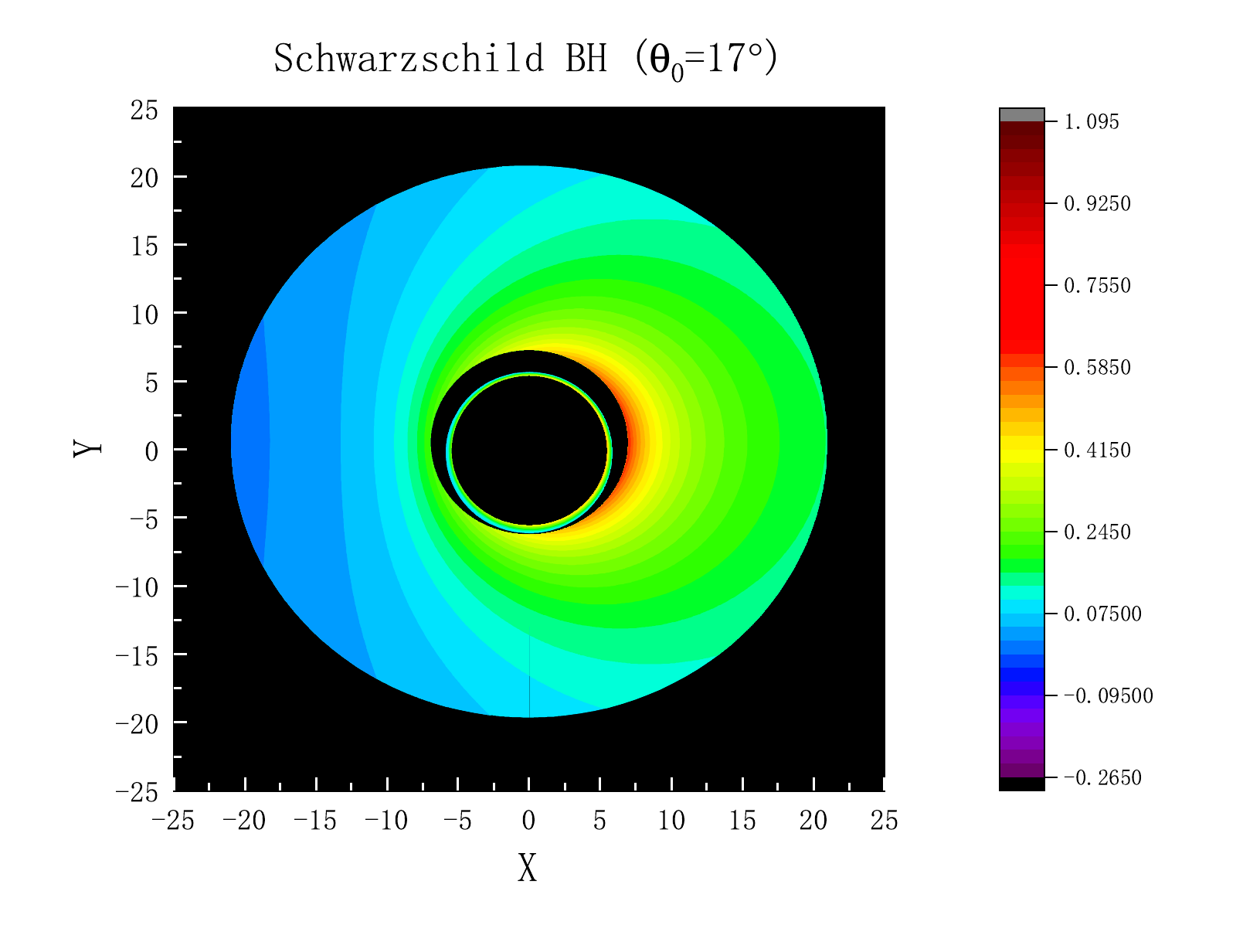}
  \includegraphics[width=5cm,height=4.4cm]{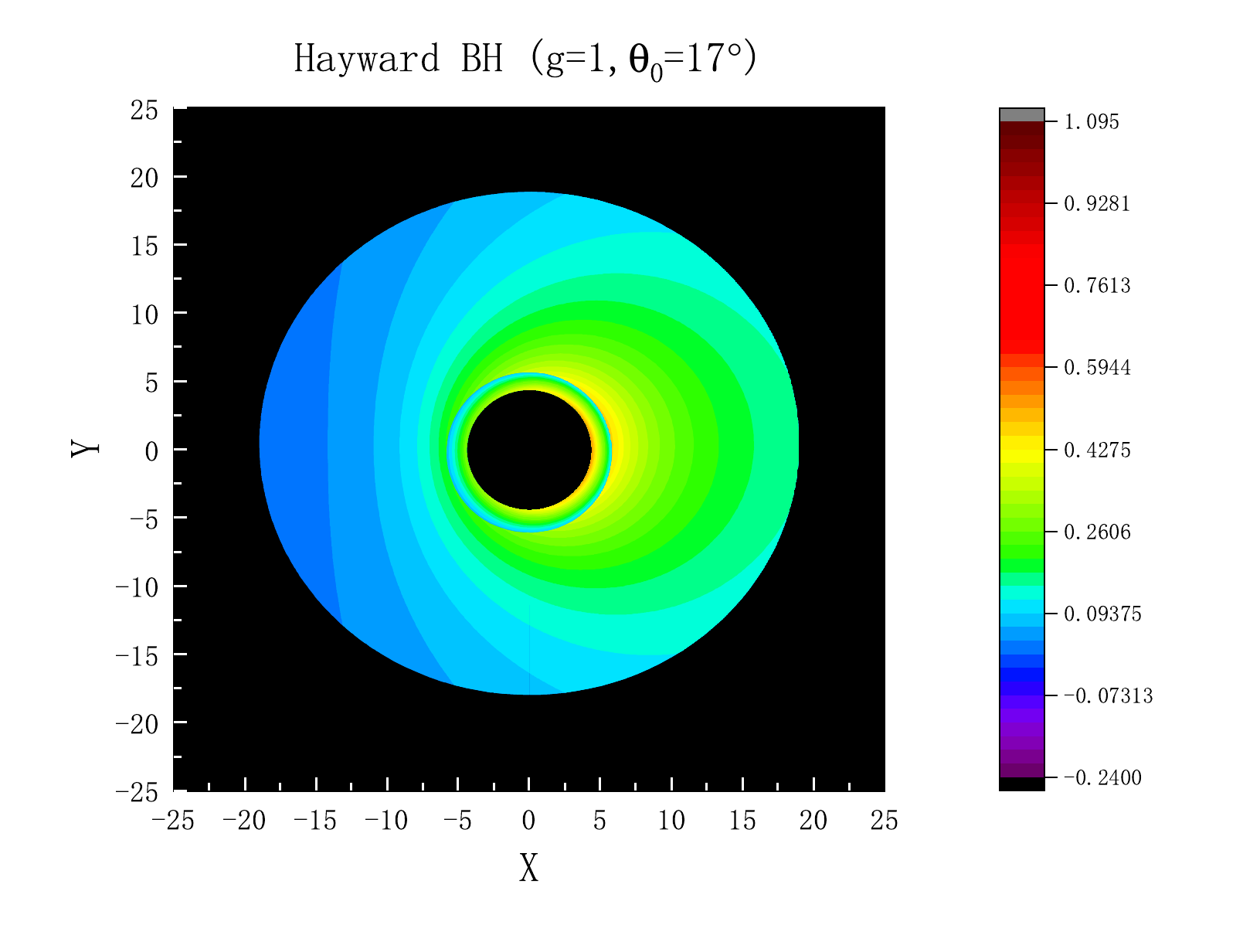}
  \includegraphics[width=5cm,height=4.4cm]{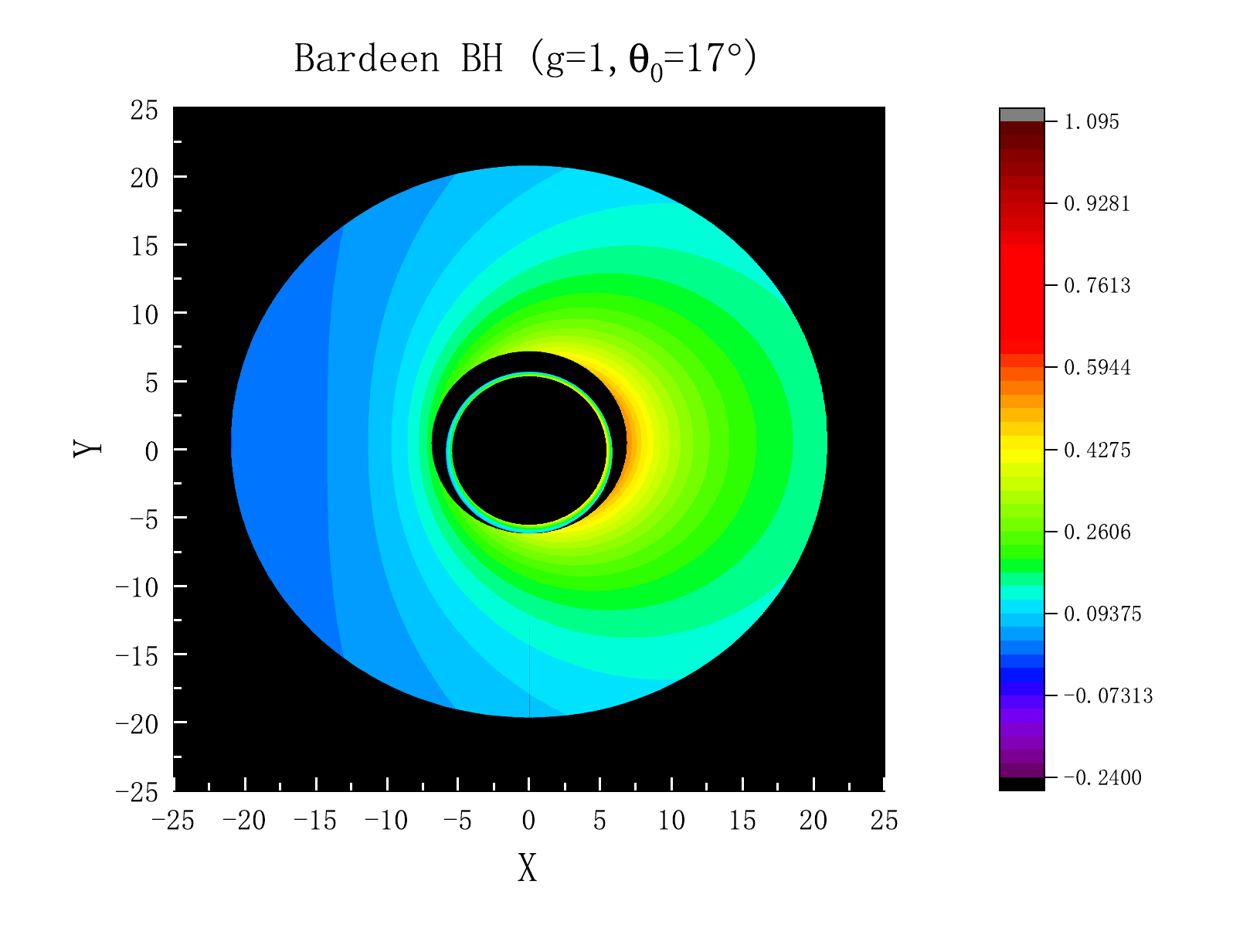}
  \includegraphics[width=5cm,height=4.4cm]{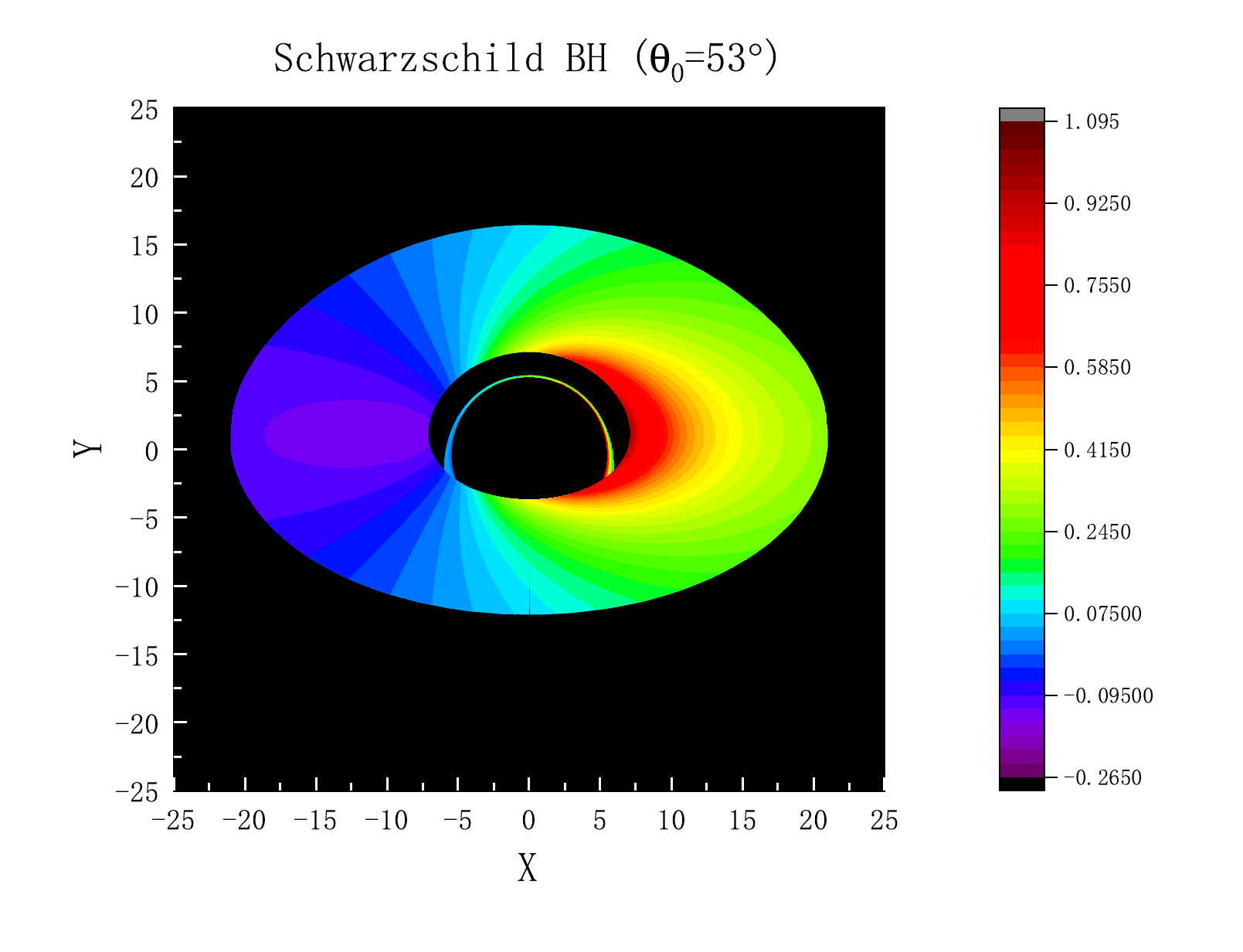}
  \includegraphics[width=5cm,height=4.4cm]{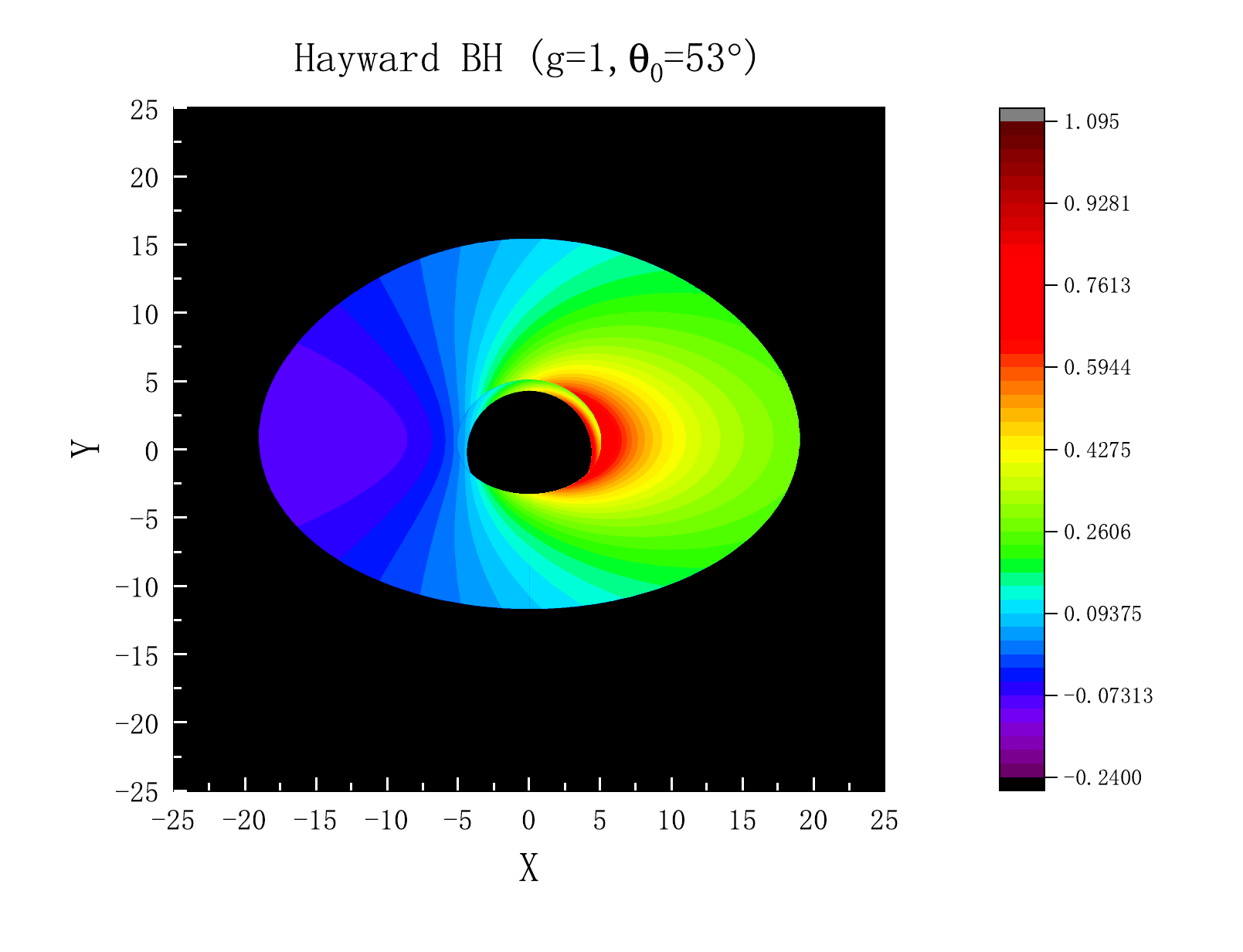}
  \includegraphics[width=5cm,height=4.4cm]{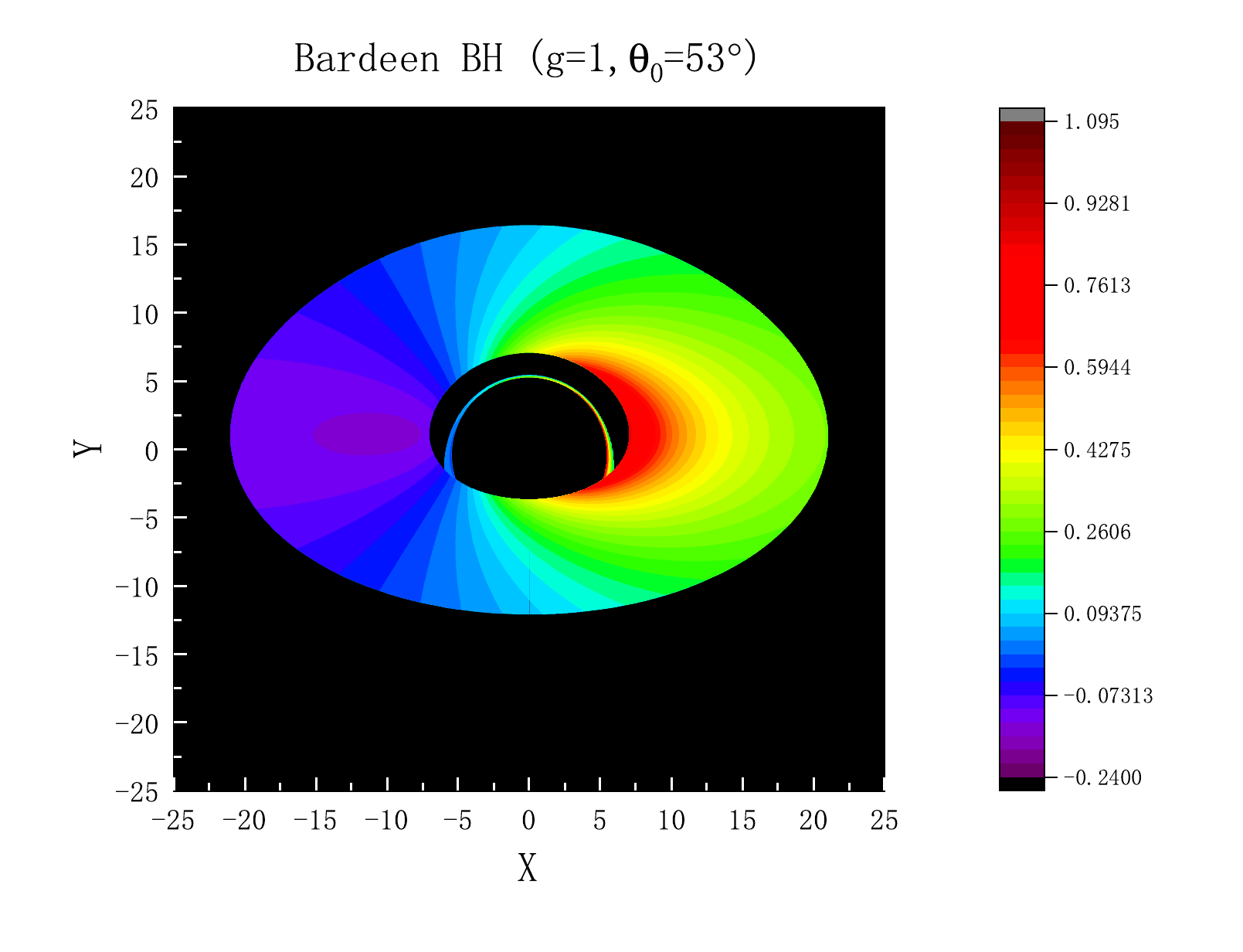}
  \includegraphics[width=5cm,height=4.4cm]{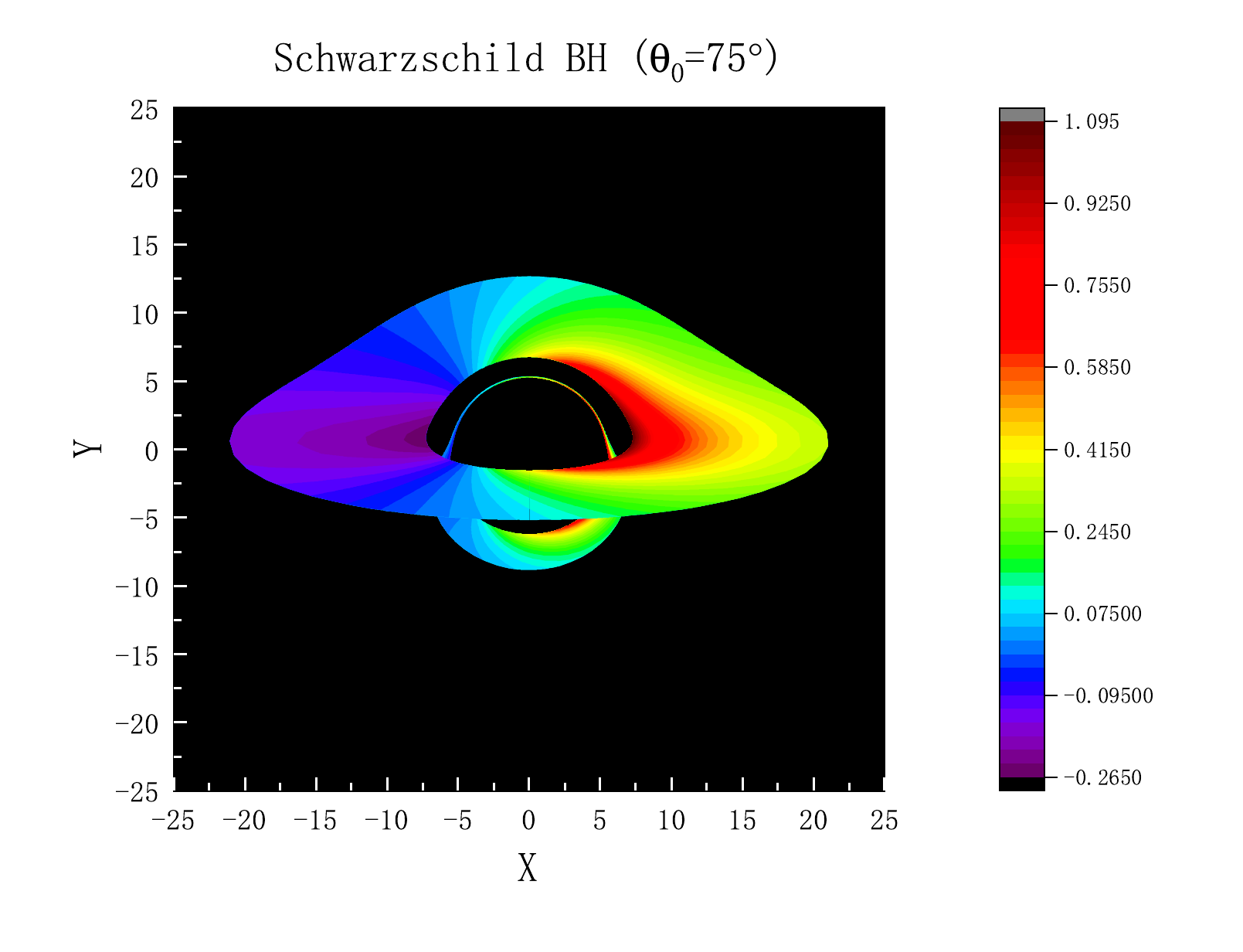}
  \includegraphics[width=5cm,height=4.4cm]{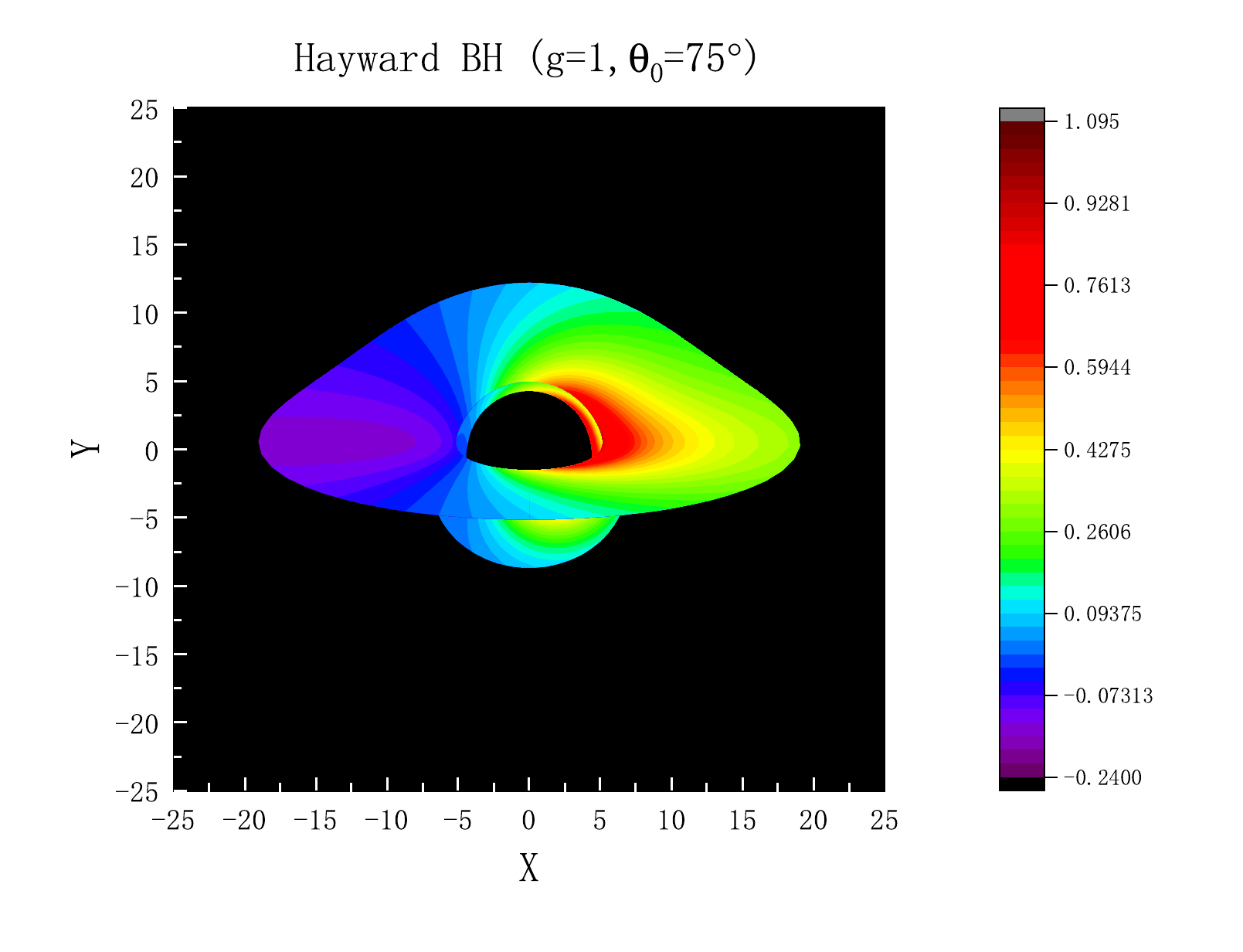}
  \includegraphics[width=5cm,height=4.4cm]{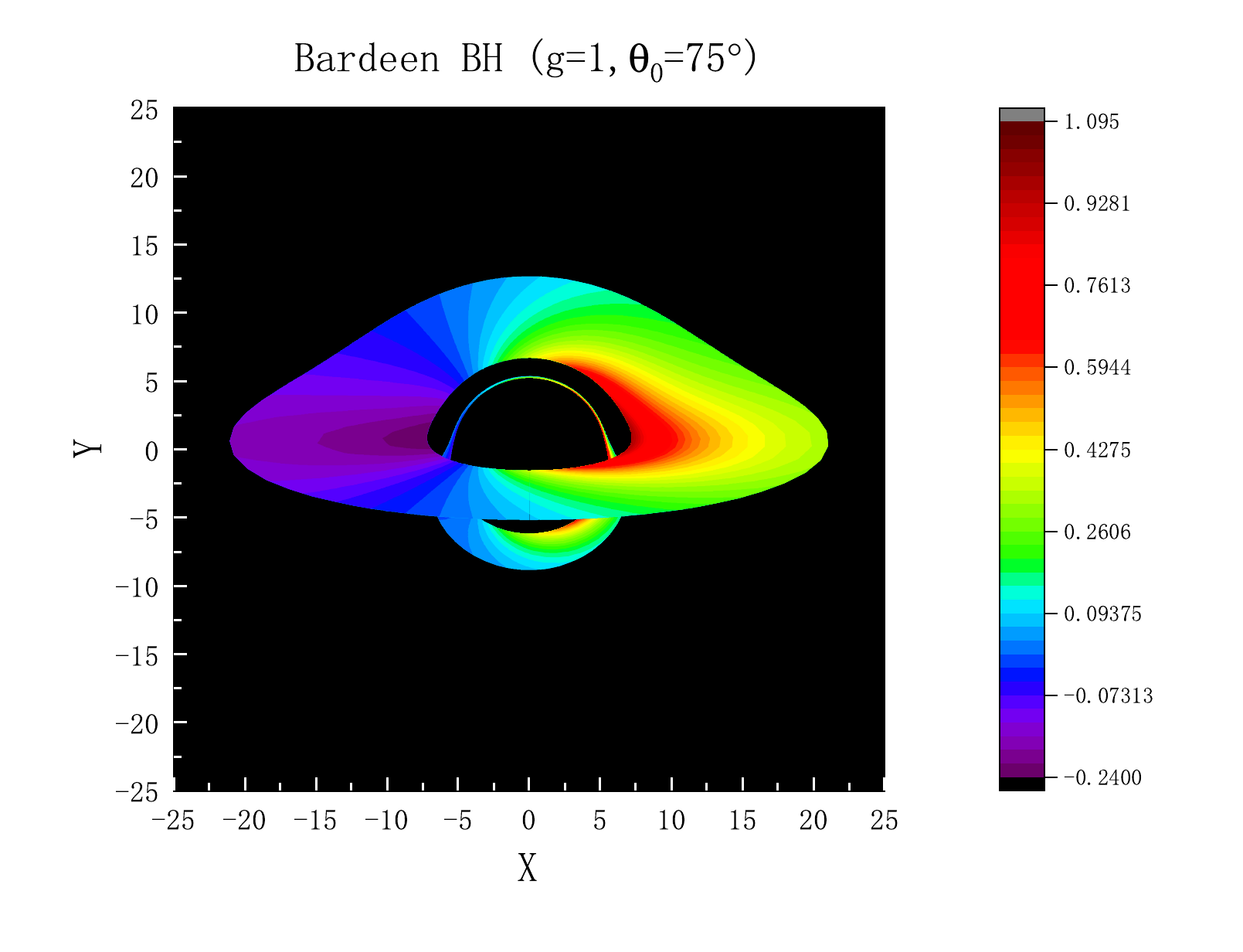}
\parbox[c]{15.0cm}{\footnotesize{\bf Fig~6.}  
Red shift distribution of direct and secondary images of accretion disk at different observation angles. {\em Left Panel}: Schwarzschild BH. {\em Middle Panel}: Hayward BH. {\em Right Panel}: Bardeen BH.}
\label{fig6}
\end{center}

\subsection{Observation images}
\label{sec:4-3}
\par
In this section, we will derive observational images of an accretion disk under three different BH spacetime backgrounds. Building on the discussions in the previous subsections, we can determine the observed flux of the accretion disk
\begin{equation}
\label{4-3-1}
F_{obs} = \frac{F}{d^{4}}.
\end{equation}

\par
Figure 7 presents the results obtained for the three different BH scenarios. It is evident that the inclination angle of observation has a significant impact on the final result. As the inclination angle increases, the observable image of the accretion disk takes on a hat-like shape. It is worth noting that according to the Novikov-Thorne model and Eq. (\ref{4-1}), we obtain the bolometric luminosity. Additionally, it is noteworthy that, regardless of whether the dip angle increases or decreases, there is a prominent accumulation of brightness on the left side of the accretion disk. This phenomenon may have two contributing factors. Firstly, as the light passes through the material surrounding the BH, it gets absorbed and re-emitted, resulting in the observed brightness distribution. The velocity and density distribution of matter as it moves towards the BH along the accretion disk can influence the probability of light being absorbed and re-emitted, leading to the observed brightness accumulation. On the other hand, the light may pass through the top and bottom of the disk, but due to the line-of-sight effect, we are more likely to observe the light emitted from the bottom of the BH, as it is closer to our line of sight. Therefore, the brightness accumulation often appears on the left side, which is a geometric effect dependent on the path and angle of light emission.

\par
Furthermore, we can observe that the Bardeen BH more closely resembles the Schwarzschild BH, whereas the Hayward BH exhibits a feature of minimal distance between the innermost region of the direct image and the outermost region of the secondary image, which can serve as a significant characteristic for distinguishing Hayward BHs from other types of BHs. Notably, the accretion disk surrounding the Bardeen BH is brighter than that around the Schwarzschild BH. This is due to the fact that the accretion disk around the Bardeen BH experiences stronger gravitational forces, leading to a closer accumulation of matter and the release of more energy.
\begin{center}
  \includegraphics[width=5cm,height=4.4cm]{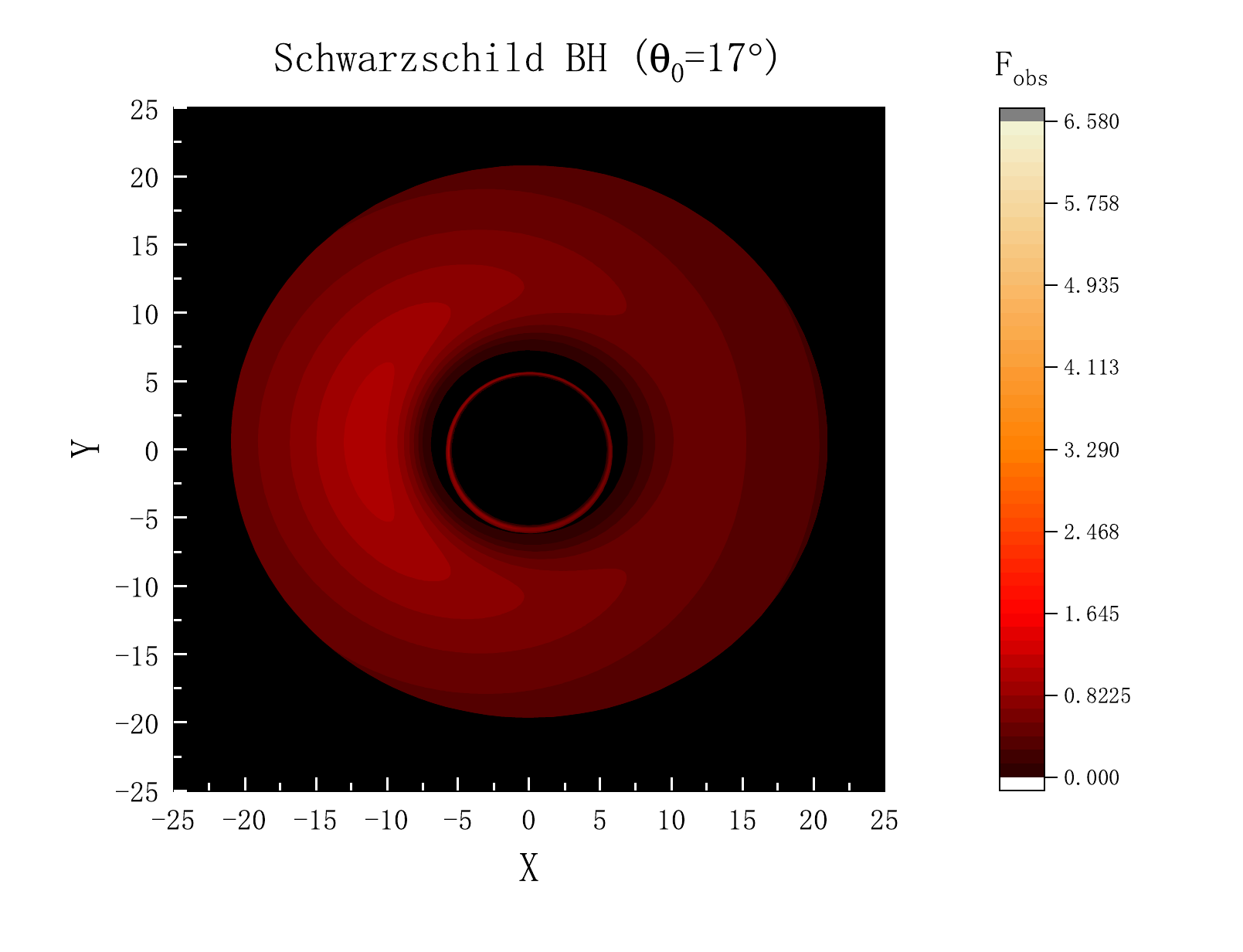}
  \includegraphics[width=5cm,height=4.4cm]{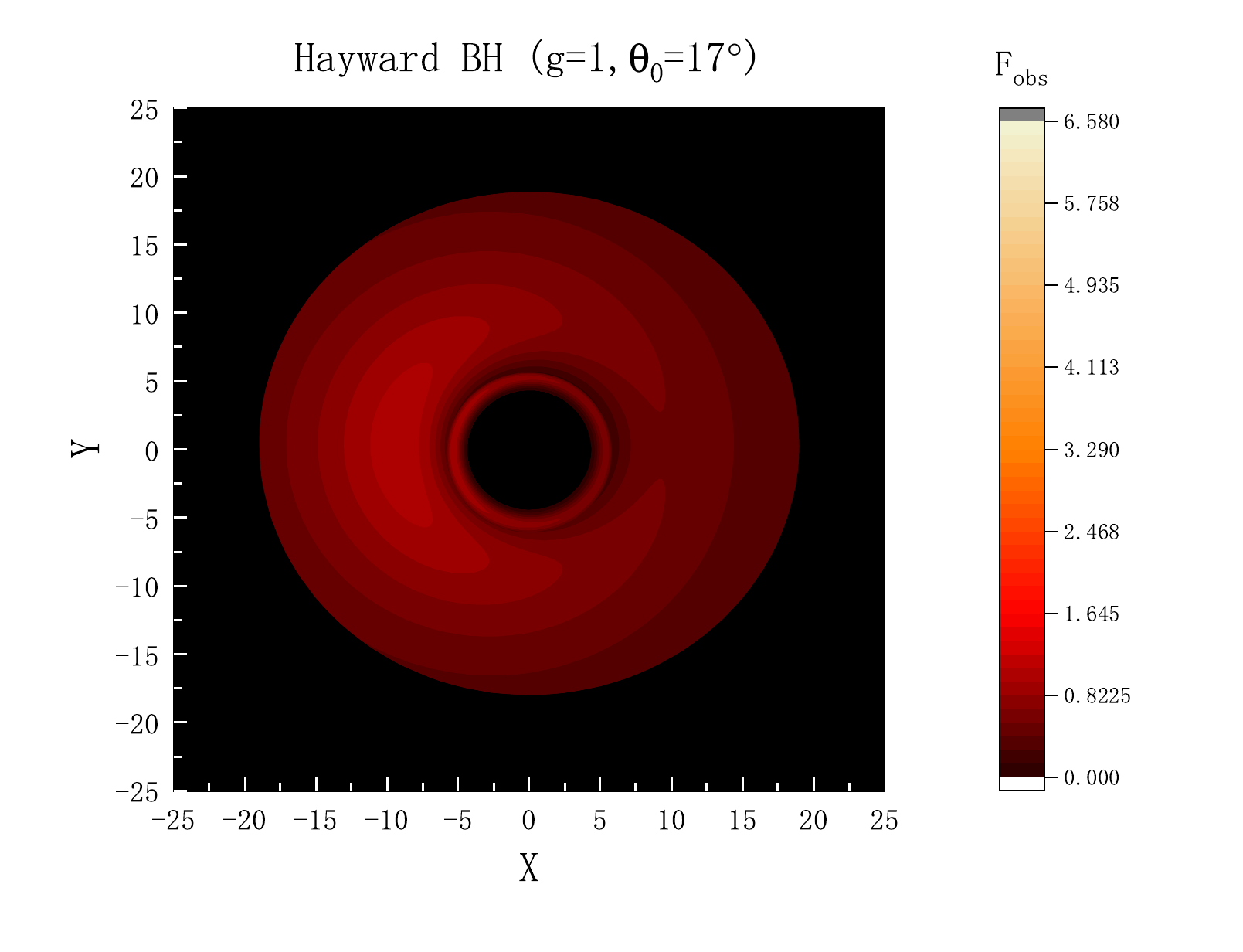}
  \includegraphics[width=5cm,height=4.4cm]{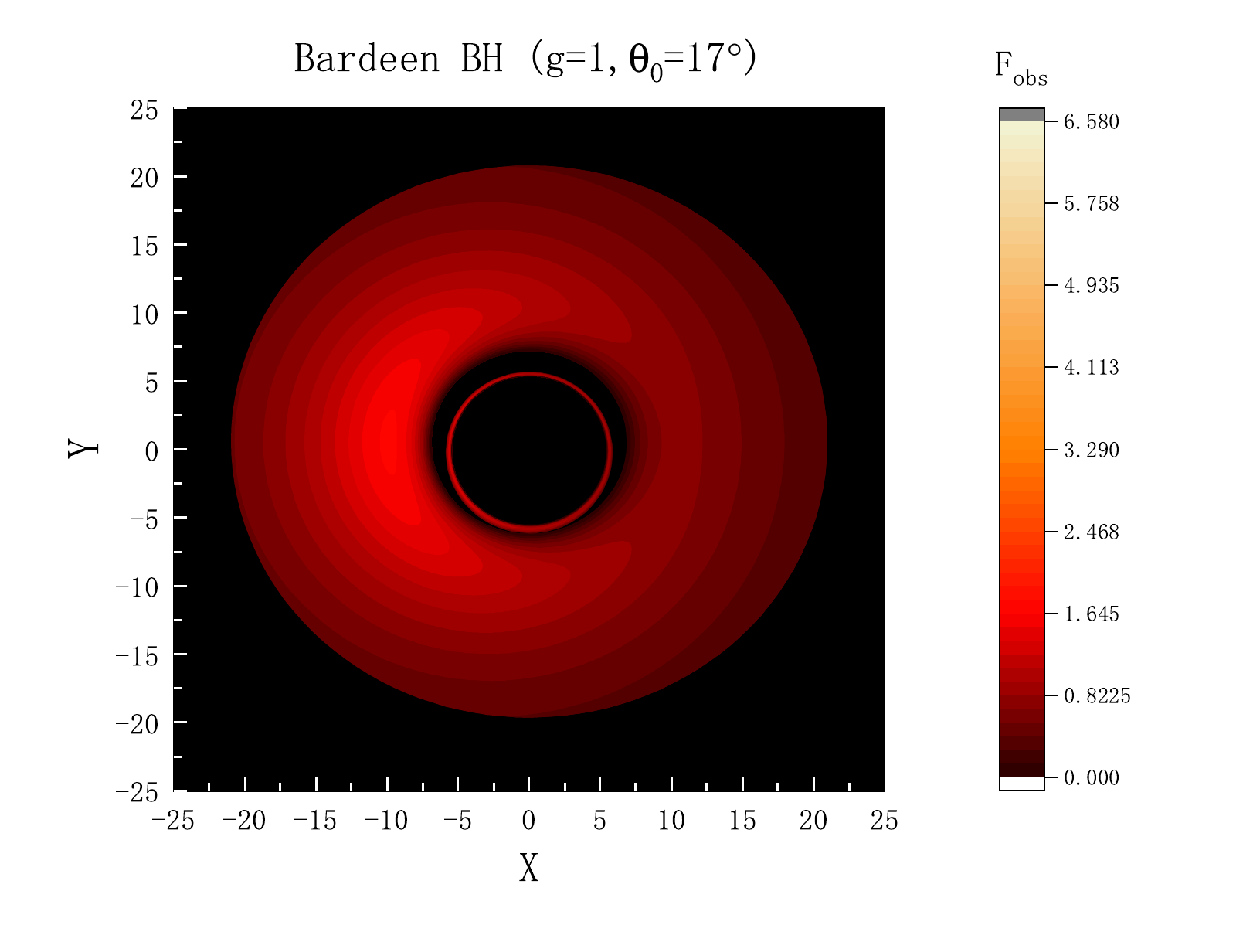}
  \includegraphics[width=5cm,height=4.4cm]{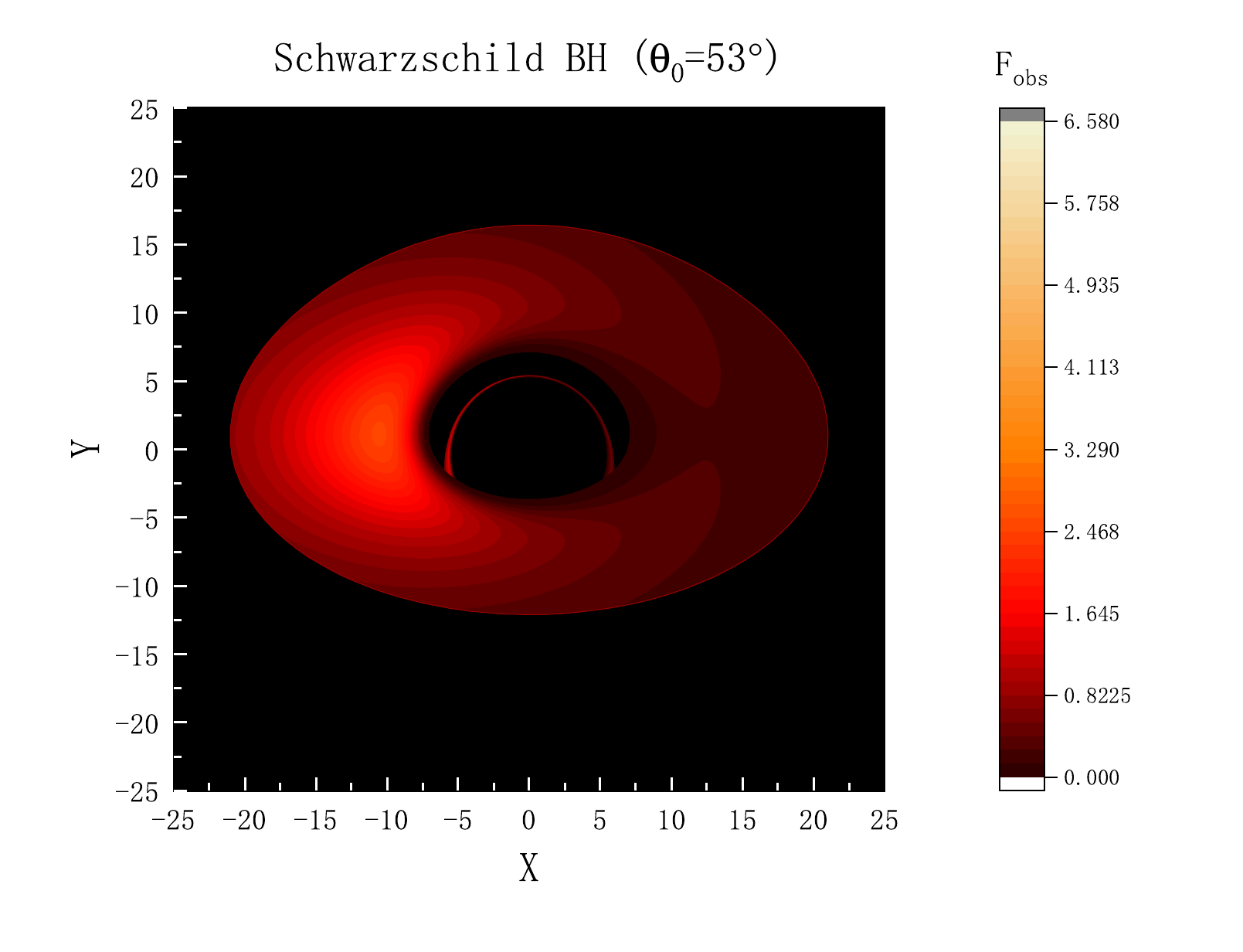}
  \includegraphics[width=5cm,height=4.4cm]{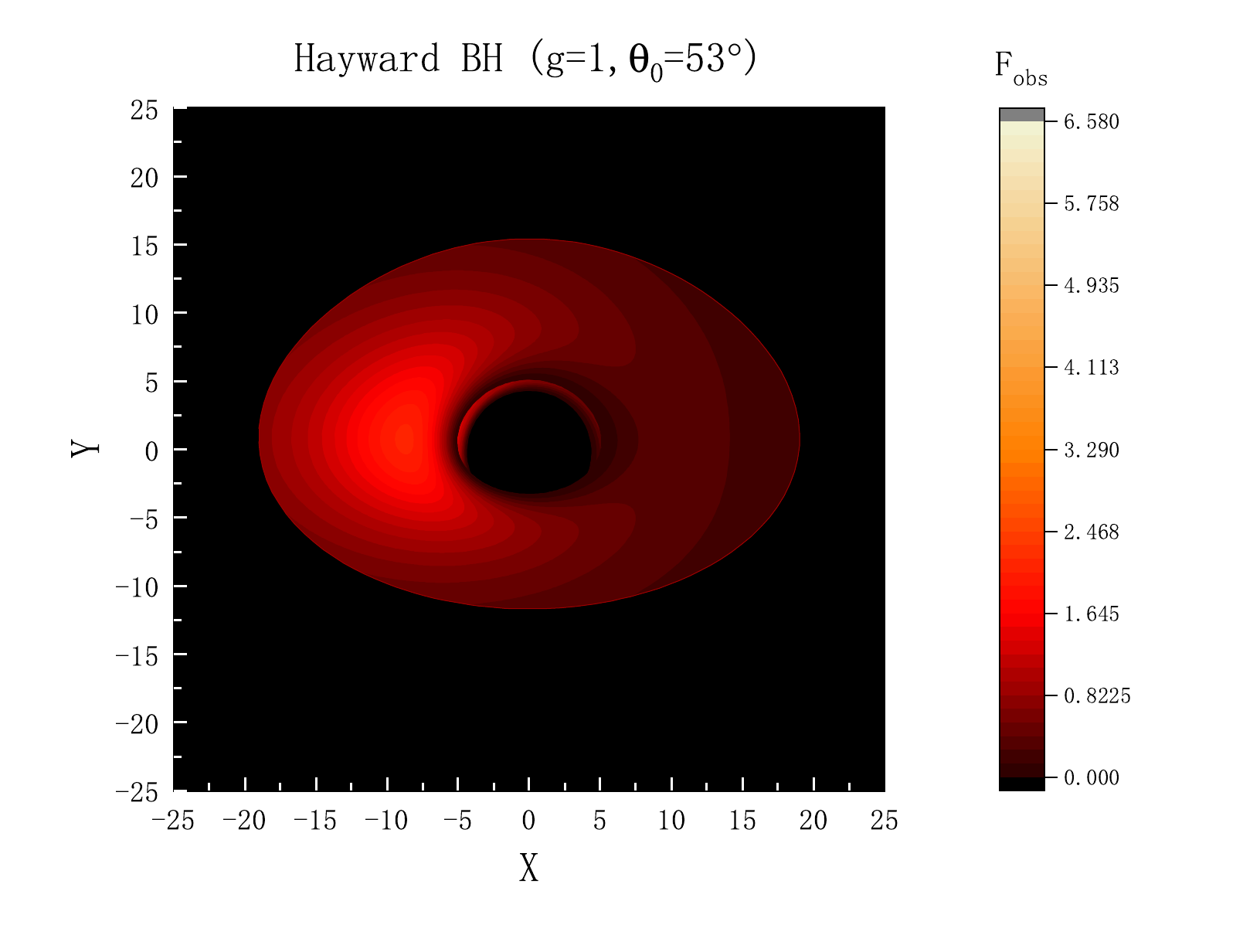}
  \includegraphics[width=5cm,height=4.4cm]{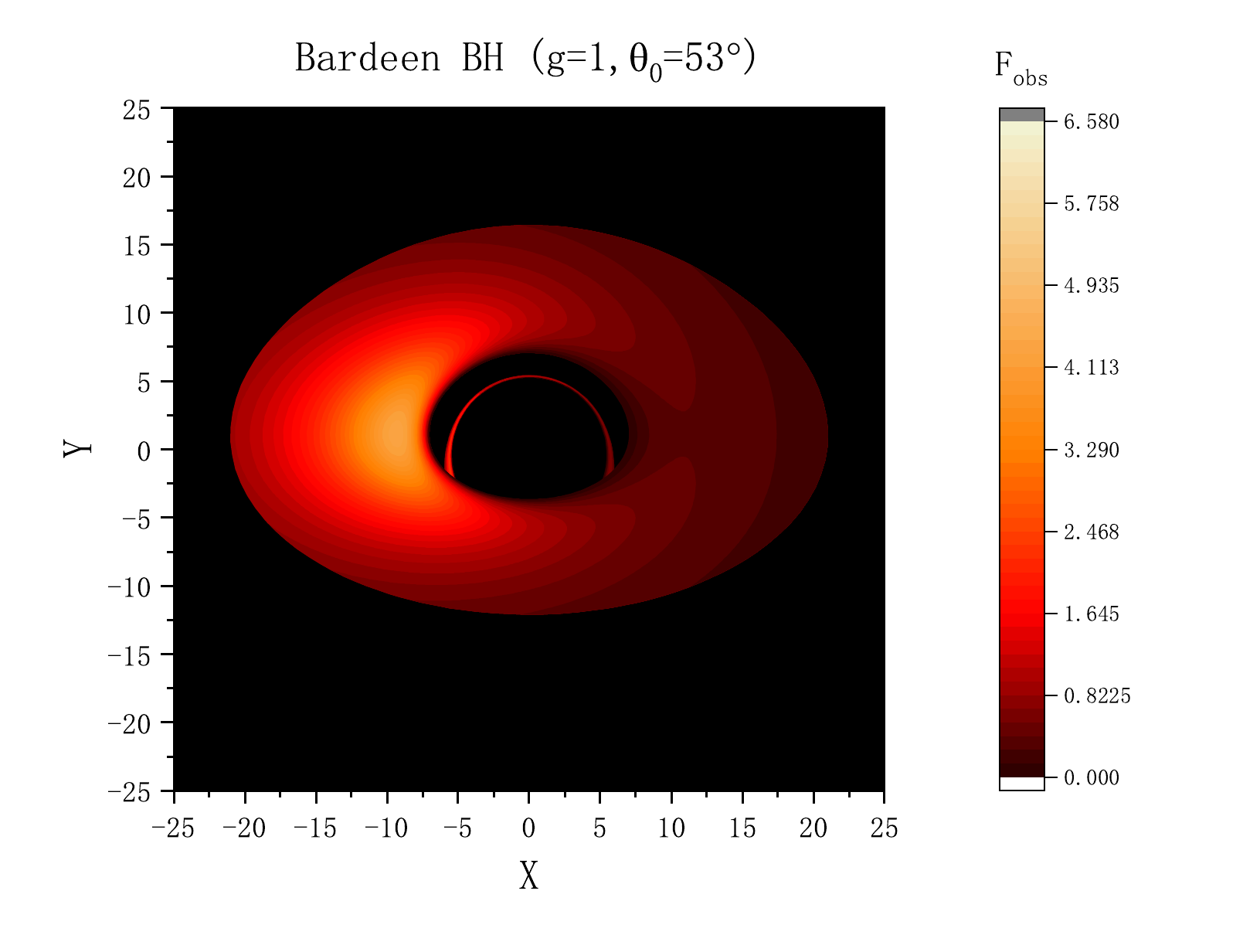}
  \includegraphics[width=5cm,height=4.4cm]{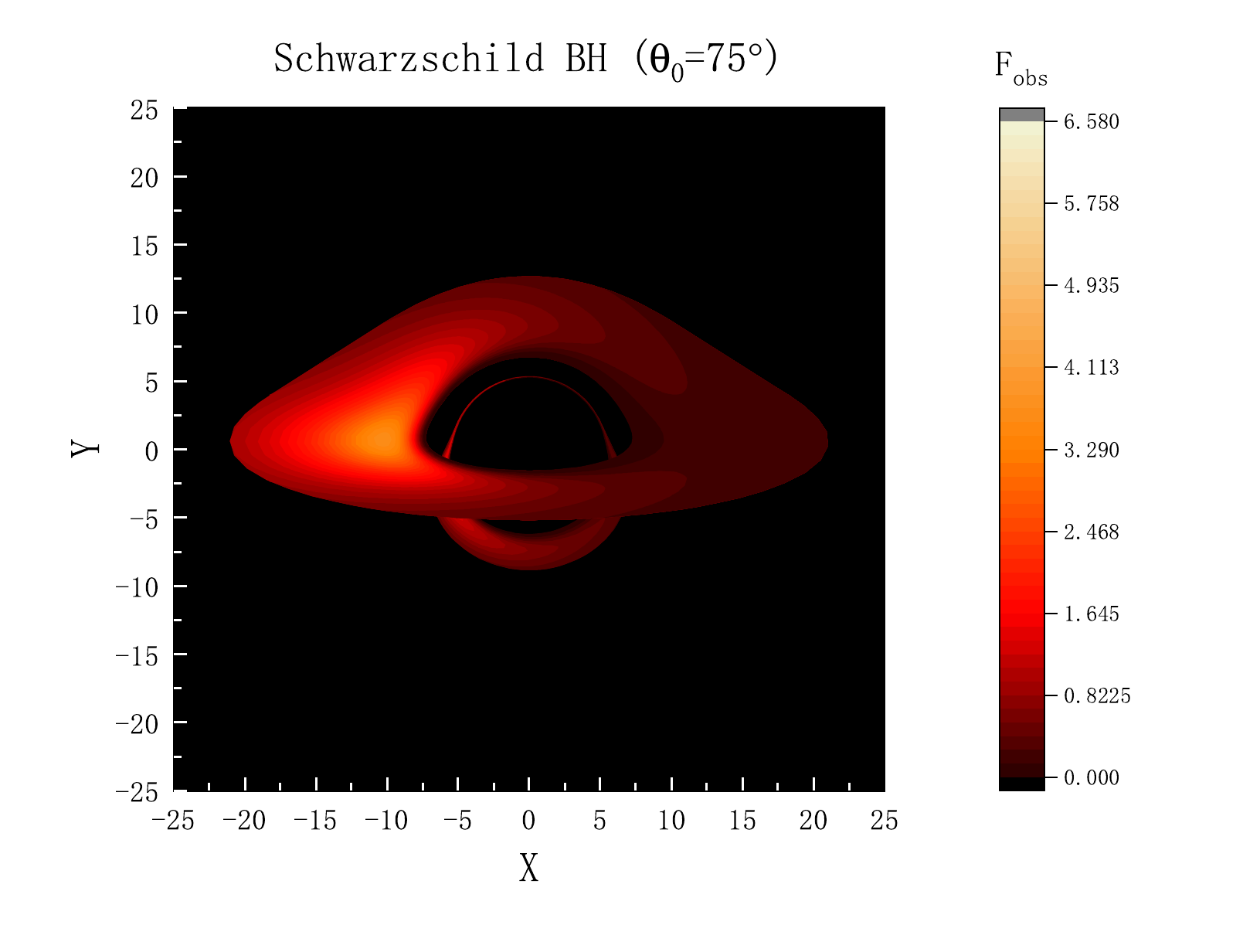}
  \includegraphics[width=5cm,height=4.4cm]{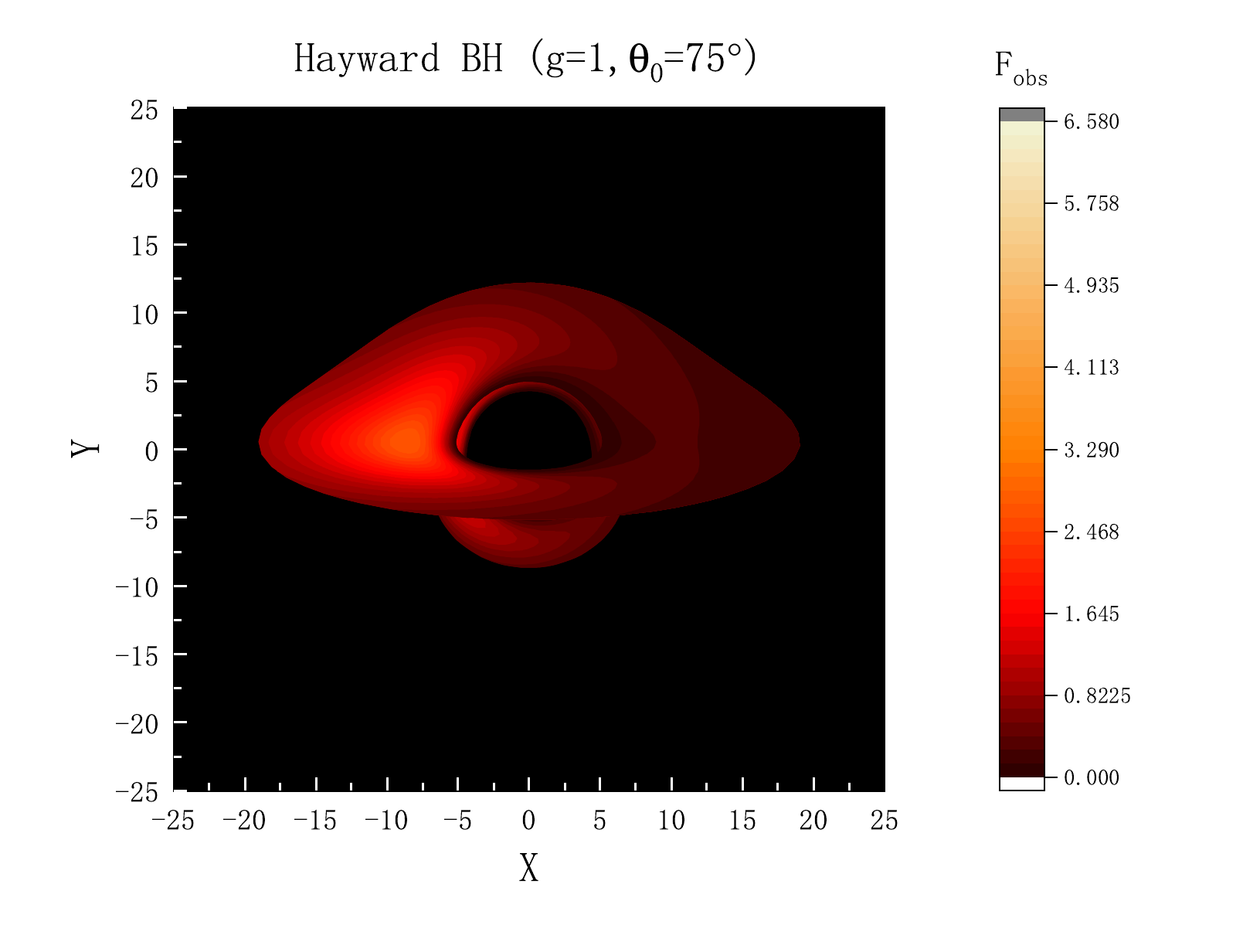}
  \includegraphics[width=5cm,height=4.4cm]{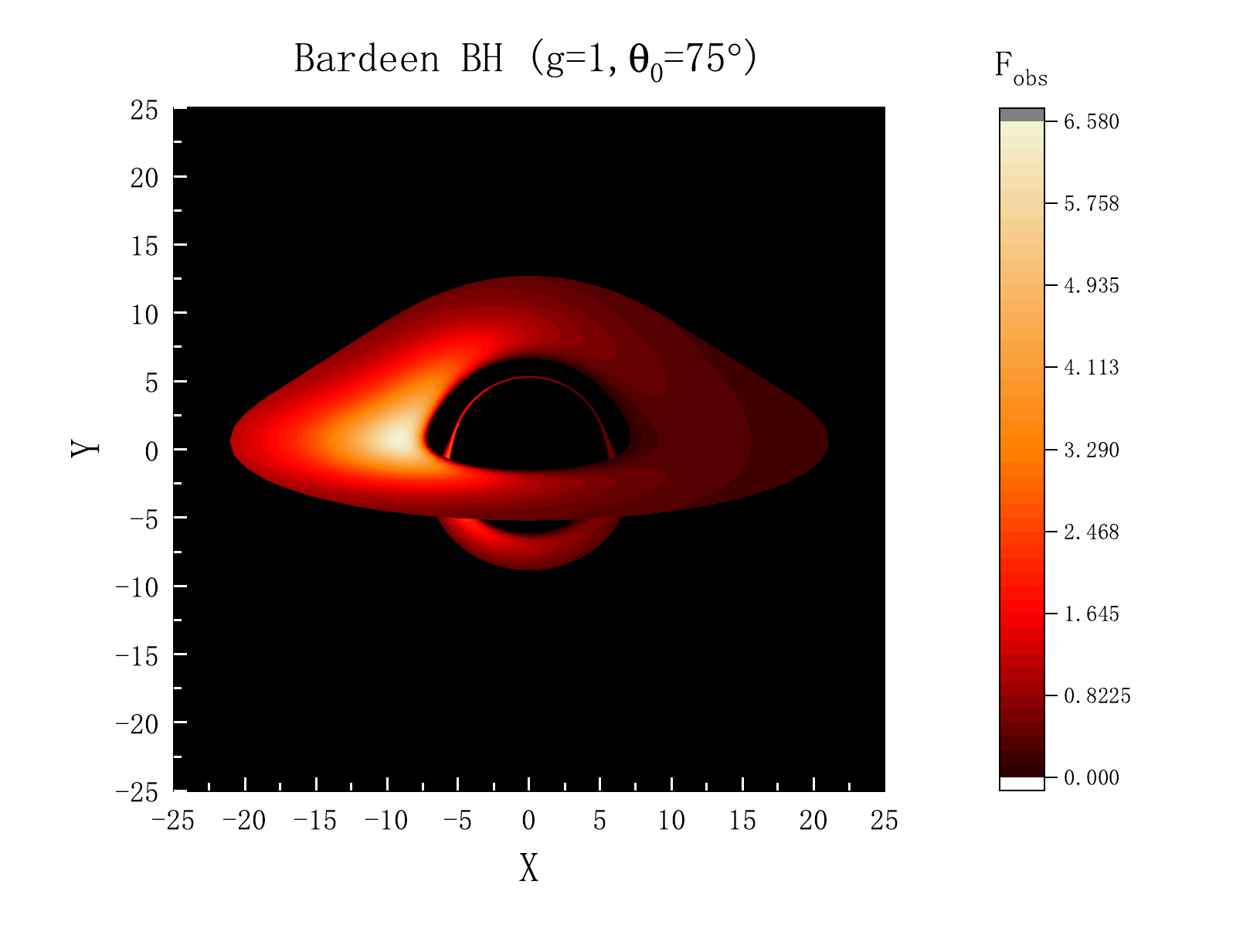}
\parbox[c]{15.0cm}{\footnotesize{\bf Fig~7.}  
The observation images of three types of BHs accretion disks with different the observation angles. {\em Left Panel}: Schwarzschild BH. {\em Middle Panel}: Hayward BH. {\em Right Panel}: Bardeen BH. The BH mass is taken as $M = 3 M_{\odot}$ and magnetic charge is $g=1$.}
\label{fig7}
\end{center}

\subsection{Data analysis}
\label{sec:4-4}
\par
In order to better illustrate the observational characteristics of regular BHs, we aim to establish a relationship between magnetic charges and observed flux. To achieve this, we propose three hypothetical functional models and fit them to our data\\
$\emph{Fourier~function}:$
\begin{eqnarray}
\label{4-2-6}
F_{1} (x) = && a_{1} + b_{1} \cos(hx) + c_{1} \sin(hx) + d_{1} \cos(2hx) \nonumber \\
&&+ e_{1} \sin(2hx) + f_{1} \cos(3hx) + g_{1} \sin(3hx),
\end{eqnarray}
$\emph{Gaussian~function}:$
\begin{eqnarray}
\label{4-2-7}
F_{2} = a_{2} e^{-\big(\frac{x-b_{2}}{c_{2}}\big)^{2}} + d_{2} e^{-\big(\frac{x-e_{2}}{f_{2}}\big)^{2}} + g_{2} e^{-\big(\frac{x-h_{2}}{i_{2}}\big)^{2}},
\end{eqnarray}
$\emph{Exponential~function}:$
\begin{eqnarray}
\label{4-2-8}
F_{3} = a_{3} e^{b_{3} x} + c_{3} e^{d_{3} x},
\end{eqnarray}
where $a-g$ with subscripts are the coefficient in the fitting function.

\par
Tables. 1 - 2 present the numerical relationship between the radiation flux $F_{\rm R}$, the observed flux $F_{\rm obs}$, and the magnetic charge $g$ at various inclination angle $\theta_{0}$. In Fig. 8 shows the results obtained from fitting the data using three hypothetical functions: the Fourier function represented by the black solid line, the Gaussian function represented by the blue solid line, and the exponential function represented by the green solid line. It is evident from the figure that Bardeen's observed flux $F_{\rm obs}$ exhibits a more linear relationship with magnetic charge, showing an increasing trend.
\begin{center}
\label{table:1}
{\footnotesize{\bf Table 1.} (Bardeen BH) numerical relationship between observed flux and the magnetic charge with $\alpha=\frac{\pi}{2}$ and $r=20M$.\\
\vspace{2mm}
\begin{tabular}{ccccccccccc}
\hline
  $g$ &${F_{\rm R}}$ &${F_{\rm obs}(\theta_{0}=17^{\circ})}$ &${F_{\rm obs}(\theta_{0}=53^{\circ})}$ &${F_{\rm obs}(\theta_{0}=75^{\circ})}$ \\
  \hline
  $1.000$ &$0.555780$   &$0.309124$   &$0.203257$   &$0.178584$   \\
  \hline
  $1.025$ &$0.559649$   &$0.311333$   &$0.204732$   &$0.179885$   \\
 \hline
  $1.050$ &$0.563580$   &$0.312868$   &$0.206232$   &$0.181209$   \\
  \hline
  $1.075$ &$0.567573$   &$0.315862$   &$0.207632$   &$0.182426$   \\
  \hline
  $1.100$ &$0.571625$   &$0.318181$   &$0.209180$   &$0.183793$   \\
  \hline
  $1.125$ &$0.575735$   &$0.320533$   &$0.210878$   &$0.185310$   \\
  \hline
  $1.150$ &$0.579901$   &$0.322919$   &$0.212474$   &$0.186718$   \\
  \hline
  $1.175$ &$0.584119$   &$0.325337$   &$0.214091$   &$0.188146$   \\
  \hline
  $1.200$ &$0.588389$   &$0.327786$   &$0.215730$   &$0.189593$   \\
  \hline
  $1.225$ &$0.592709$   &$0.330265$   &$0.217389$   &$0.191059$   \\
  \hline
  $1.250$ &$0.597076$   &$0.332773$   &$0.219069$   &$0.192543$   \\
  \hline
  $1.275$ &$0.601488$   &$0.335308$   &$0.220768$   &$0.194043$   \\
  \hline
  $1.300$ &$0.605943$   &$0.337871$   &$0.222485$   &$0.195561$   \\
  \hline
  $1.325$ &$0.610440$   &$0.340459$   &$0.224221$   &$0.197094$   \\
  \hline
  $1.350$ &$0.614976$   &$0.343072$   &$0.225974$   &$0.198643$   \\
  \hline
  $1.375$ &$0.619549$   &$0.345708$   &$0.227743$   &$0.200207$   \\
  \hline
  $1.400$ &$0.624157$   &$0.348367$   &$0.229528$   &$0.201785$   \\
  \hline
  $1.425$ &$0.628799$   &$0.351047$   &$0.231329$   &$0.203376$   \\
  \hline
  $1.450$ &$0.633472$   &$0.353748$   &$0.233144$   &$0.204981$   \\
  \hline
  $1.475$ &$0.638175$   &$0.356468$   &$0.234973$   &$0.206598$   \\
  \hline
  $1.500$ &$0.642905$   &$0.359206$   &$0.236815$   &$0.208228$   \\
  \hline
\end{tabular}}
\end{center}

\begin{center}
\label{table:2}
{\footnotesize{\bf Table 2.} (Hayward BH) numerical relationship between observed flux and the magnetic charge with $\alpha=\frac{\pi}{2}$ and $r=20M$.\\
\vspace{2mm}
\begin{tabular}{ccccccccccc}
\hline
  $g$ &${F}$ &${F_{obs}(\theta_{0}=17^{\circ})}$ &${F_{obs}(\theta_{0}=53^{\circ})}$ &${F_{obs}(\theta_{0}=75^{\circ})}$ \\
  \hline
  $1.000$ &$0.486465$   &$0.276265$   &$0.187880$   &$0.166693$   \\
  \hline
  $1.025$ &$0.487528$   &$0.276871$   &$0.188293$   &$0.167060$   \\
 \hline
  $1.050$ &$0.488642$   &$0.277507$   &$0.188727$   &$0.167445$   \\
  \hline
  $1.075$ &$0.489808$   &$0.278173$   &$0.189181$   &$0.167849$   \\
  \hline
  $1.100$ &$0.491029$   &$0.278870$   &$0.189656$   &$0.168271$   \\
  \hline
  $1.125$ &$0.492304$   &$0.279598$   &$0.190153$   &$0.168711$   \\
  \hline
  $1.150$ &$0.493635$   &$0.280357$   &$0.190671$   &$0.169172$   \\
  \hline
  $1.175$ &$0.495022$   &$0.281150$   &$0.191212$   &$0.169652$   \\
  \hline
  $1.200$ &$0.496467$   &$0.281975$   &$0.191775$   &$0.170151$   \\
  \hline
  $1.225$ &$0.497971$   &$0.282833$   &$0.192360$   &$0.170672$   \\
  \hline
  $1.250$ &$0.499535$   &$0.283726$   &$0.192969$   &$0.171212$   \\
  \hline
  $1.275$ &$0.501158$   &$0.284653$   &$0.193602$   &$0.171774$   \\
  \hline
  $1.300$ &$0.502844$   &$0.585616$   &$0.194259$   &$0.172357$   \\
  \hline
  $1.325$ &$0.504591$   &$0.286614$   &$0.194939$   &$0.172962$   \\
  \hline
  $1.350$ &$0.506402$   &$0.287648$   &$0.195645$   &$0.173589$   \\
  \hline
  $1.375$ &$0.508276$   &$0.288718$   &$0.196375$   &$0.174237$   \\
  \hline
  $1.400$ &$0.510215$   &$0.289826$   &$0.197131$   &$0.174908$   \\
  \hline
  $1.425$ &$0.512220$   &$0.290971$   &$0.197912$   &$0.175602$   \\
  \hline
  $1.450$ &$0.514290$   &$0.292154$   &$0.198719$   &$0.176319$   \\
  \hline
  $1.475$ &$0.516428$   &$0.293375$   &$0.199553$   &$0.177059$   \\
  \hline
  $1.500$ &$0.518633$   &$0.294634$   &$0.200412$   &$0.177823$   \\
  \hline
\end{tabular}}
\end{center}
\begin{center}
  \includegraphics[width=5cm,height=4.5cm]{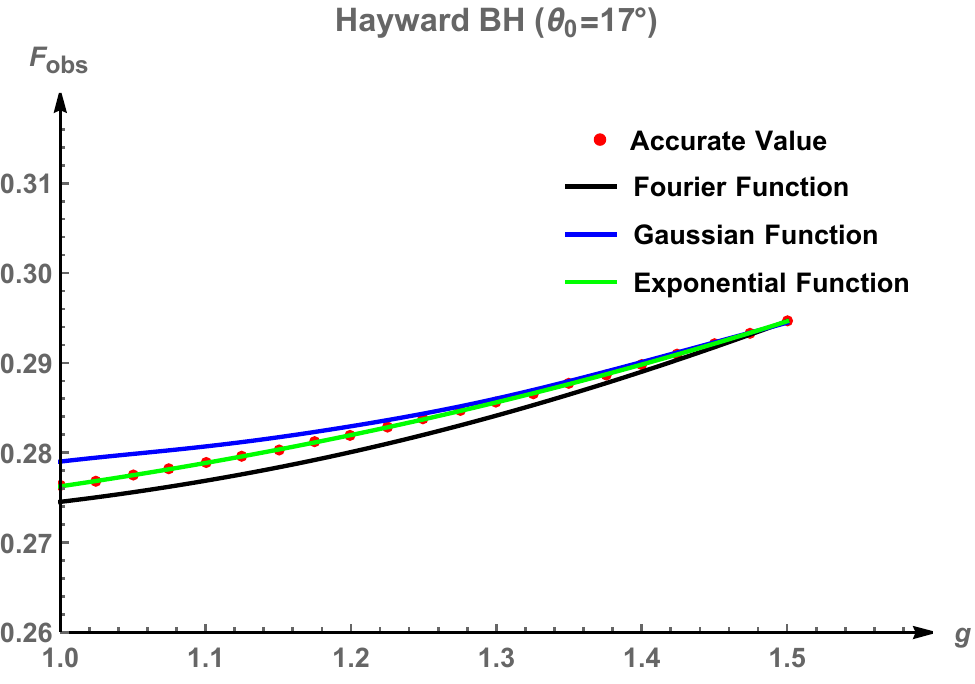}
  \includegraphics[width=5cm,height=4.5cm]{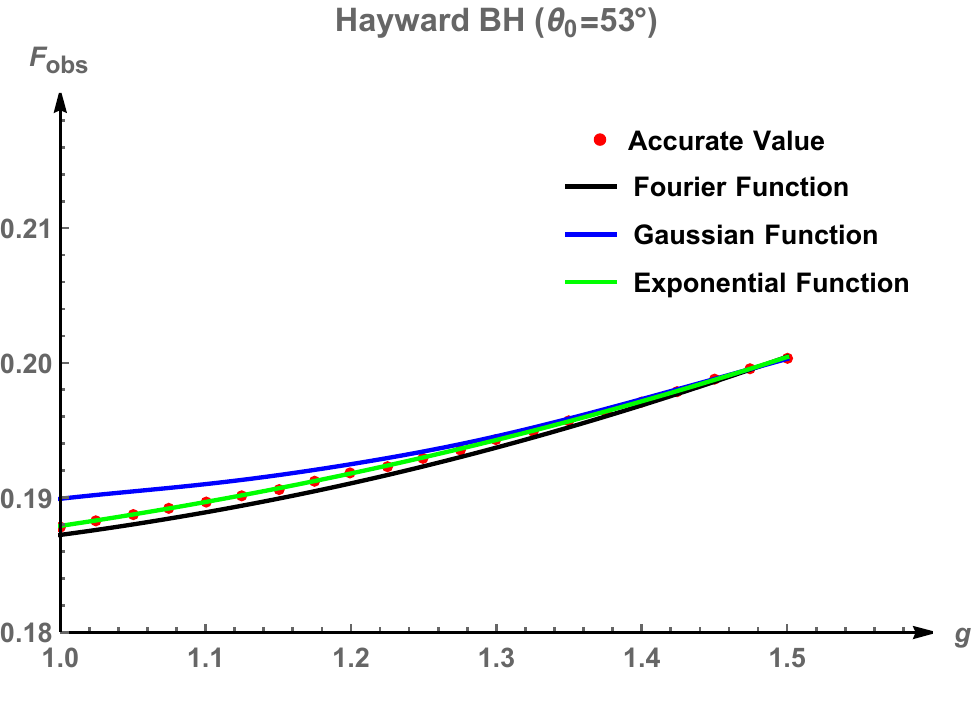}
  \includegraphics[width=5cm,height=4.5cm]{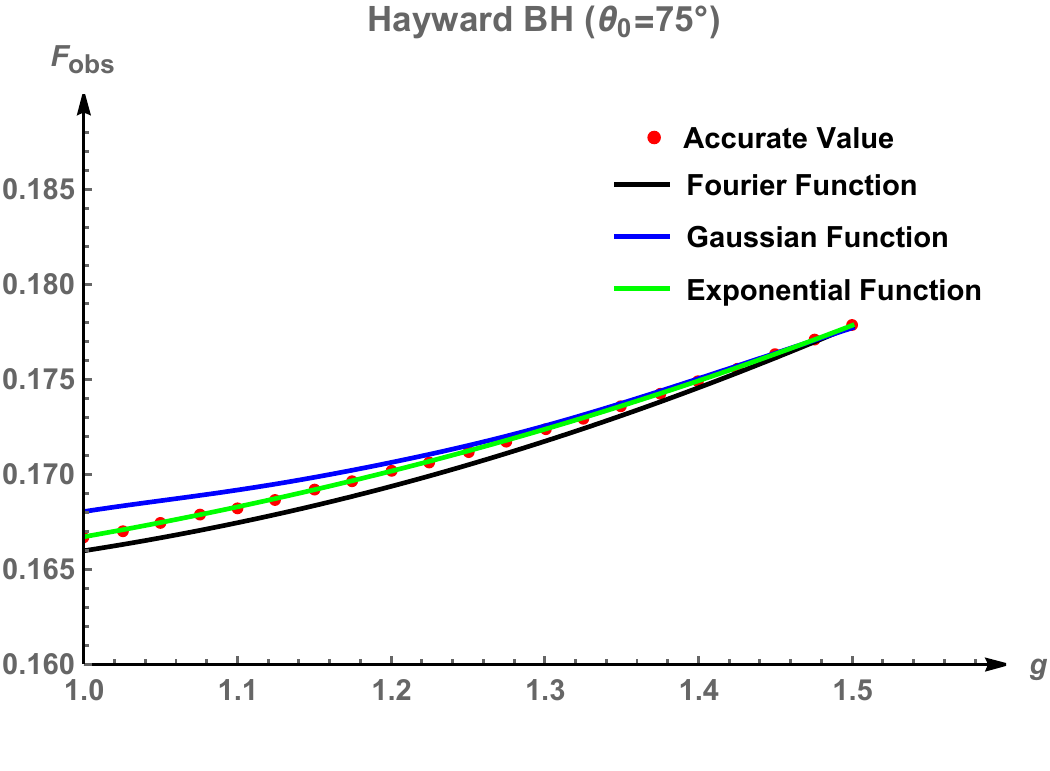}
  \includegraphics[width=5cm,height=4.5cm]{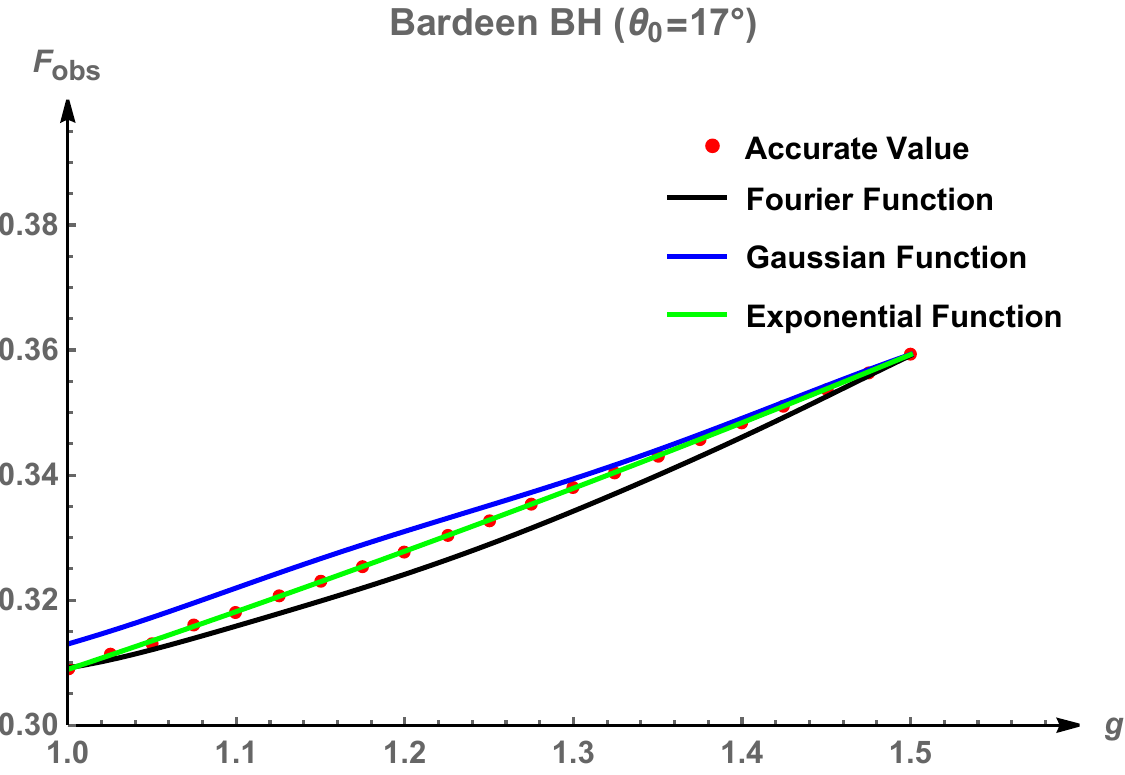}
  \includegraphics[width=5cm,height=4.5cm]{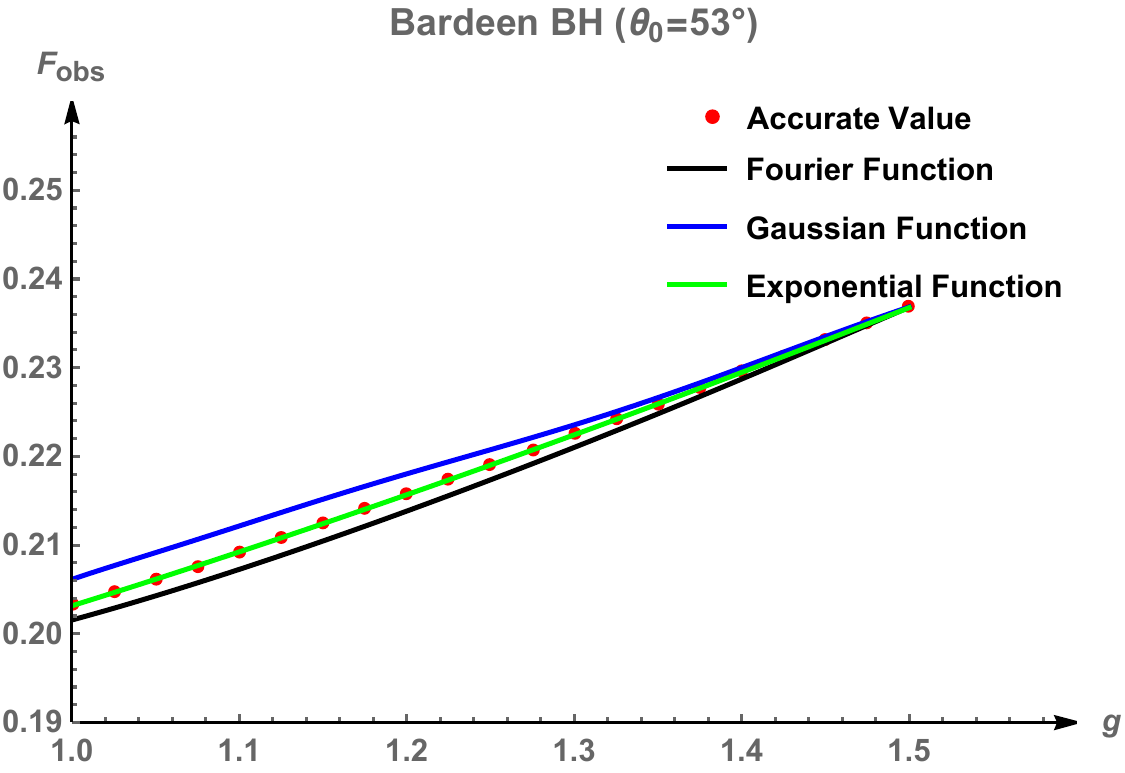}
  \includegraphics[width=5cm,height=4.5cm]{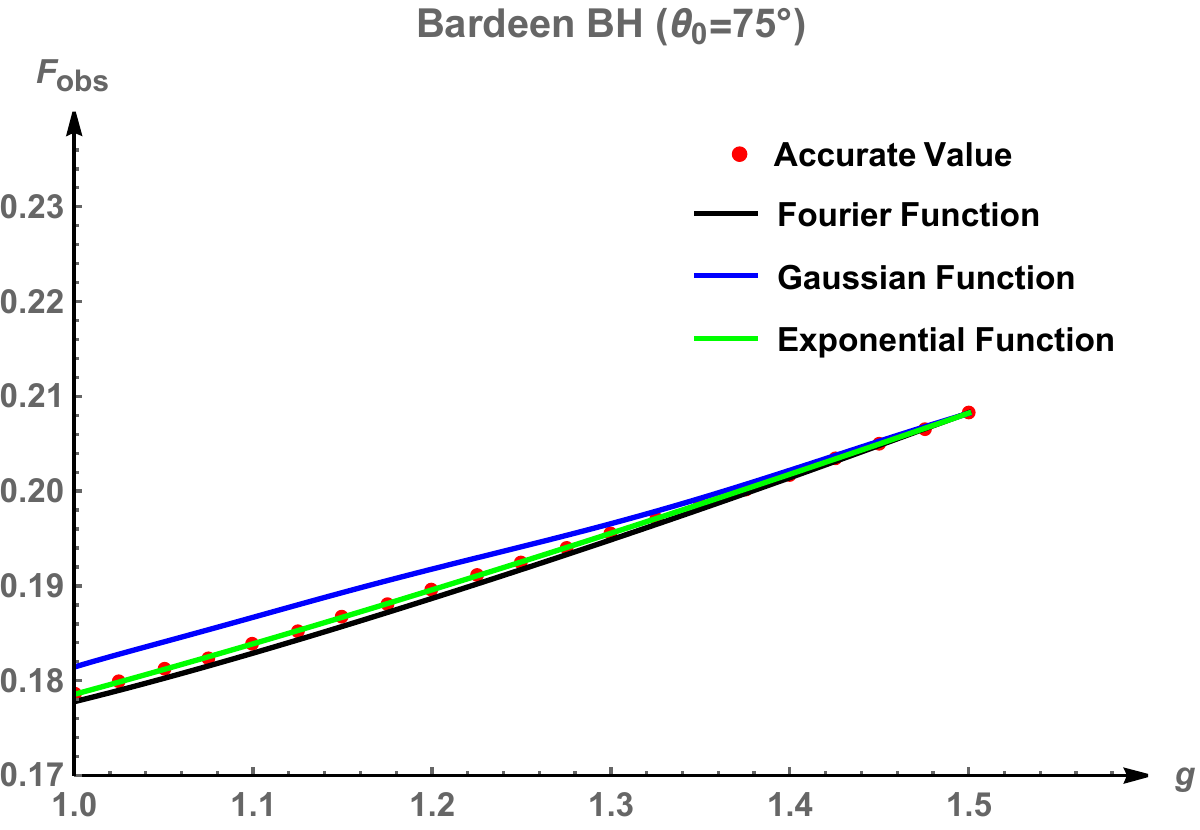}
\parbox[c]{15.0cm}{\footnotesize{\bf Fig~8.}  
Fitting of Fourier function (black solid line), Gaussian function (bule solid line), and Exponential function (green solid line) to the accurate value of the observed flux $F_{\rm obs}$ and magnetic charge $g$ (red dots). {\em Top Panel}: Hayward BH. {\em Bottom Panel}: Bardeen BH. The BH mass is taken as $M = M_{\odot}$.}
\label{fig8}
\end{center}

\par
Our current analysis relies on intuitive methods. However, for more accurate data fitting, we should employ the Mean Squared Error (MSE). This metric quantifies the difference between a model's predicted values and the actual values. A smaller MSE indicates a higher level of similarity between predicted and actual values, leading to a more accurate model. Conversely, a larger MSE indicates greater disparity and a less accurate model. To determine the best-fit function for our data, we computed the MSE for three hypothesis functions and compared them to the standard results.
\begin{center}
\label{table:3}
{\footnotesize{\bf Table 3.} (Bardeen BH) MSE of Fourier functions, exponential functions, and polynomial functions with $\alpha=\frac{\pi}{2}$ and $r=20M$.\\
\vspace{2mm}
\begin{tabular}{ccccccccccc}
\hline
  $Functions$ & ${F}$ &${F_{\rm obs}(\theta_{0}=17^{\circ})}$ & ${F_{\rm obs}(\theta_{0}=53^{\circ})}$ & ${F_{\rm obs}(\theta_{0}=75^{\circ})}$ \\
  \hline
  $fourier function$     &    $9.24229\times10^{-6}$   &  $7.07774\times10^{-6}$  &  $2.35592\times10^{-6}$  &  $5.37796\times10^{-7}$   \\
  \hline
  $exponential function$ &    $2.98562\times10^{-9}$   &  $2.20407\times10^{-8}$  &  $4.18114\times10^{-9}$  &  $1.14029\times10^{-9}$   \\
  \hline
  $gaussian function$    &    $6.98465\times10^{-6}$   &  $7.12659\times10^{-6}$  &  $3.90385\times10^{-6}$  &  $3.65842\times10^{-6}$  \\
  \hline
\end{tabular}}
\end{center}

\begin{center}
\label{table:4}
{\footnotesize{\bf Table 4.} (Hayward BH) MSE of Fourier functions, exponential functions, and polynomial functions with $\alpha=\frac{\pi}{2}$ and $r=20M$.\\
\vspace{2mm}
\begin{tabular}{ccccccccccc}
\hline
  $Functions$ & ${F}$ &${F_{\rm obs}(\theta_{0}=17^{\circ})}$ & ${F_{\rm obs}(\theta_{0}=53^{\circ})}$ & ${F_{\rm obs}(\theta_{0}=75^{\circ})}$ \\
  \hline
  $fourier function$ &$9.26748\times10^{-6}$     &$2.36647\times10^{-6}$   &$3.26572\times10^{-7}$    &$3.97203\times10^{-7}$   \\
  \hline
  $exponential function$ &$2.71226\times10^{-10}$       &$9.10981\times10^{-11}$       &$4.61284\times10^{-10}$     &$6.22655\times10^{-10}$   \\
  \hline
  $gaussian function$ &$3.92596\times10^{-6}$       &$1.71991\times10^{-6}$       &$9.32263\times10^{-7}$     &$4.02467\times10^{-7}$  \\
  \hline
\end{tabular}}
\end{center}

\par
Tables. 3 - 4 present the MSE values for different fitting functions. Notably, the MSE for the exponential function is significantly smaller than that of the Fourier and Gaussian functions, by approximately four orders of magnitude. This suggests that the exponential function provides the closest fit to the theoretical results. Consequently, we report the best fit achieved with the exponential function in Tables. 5 to 6. It is well known that the magnetic field of a BH accretion disk arises from the plasma within it. The size of the magnetic field depends on the magnetic charge density and the motion speed of the plasma. In this study, we establish a functional relationship between a BH's magnetic charge and the observed flux, which may guide further exploration of the effects of a BH's magnetic charge on optical appearance.

\begin{center}
\label{table:5}
{\footnotesize{\bf Table 5.} (Bardeen BH) coefficient value of exponential fitting function under different conditions for $\alpha=\frac{\pi}{2}$ and $r=20M$.\\
\vspace{2mm}
\begin{tabular}{ccccccccccc}
\hline
  $-$ &${a_{3}}$ &${b_{3}}$ &${c_{3}}$ &${d_{3}}$ \\
  \hline
  ${F}$  &  $2.0391$  &  $-7.0162$  &  $0.4113$  &  $0.2977$  \\
  \hline
  ${F_{\rm obs}(\theta_{0}=17^{\circ})}$  &  $0.0207$  &  $-1.0981$  &  $0.2182$  &  $0.3253$   \\
  \hline
  ${F_{\rm obs}(\theta_{0}=53^{\circ})}$  &  $0.1193$  &  $-4.6926$  &  $0.1473$  &  $0.3161$  \\
  \hline
  ${F_{\rm obs}(\theta_{0}=75^{\circ})}$  &  $0.3207$  &  $-5.9462$  &  $0.1295$  &  $0.3165$ \\
  \hline
\end{tabular}}
\end{center}

\begin{center}
\label{table:6}
{\footnotesize{\bf Table 6.} (Hayward BH) coefficient value of exponential fitting function under different conditions for $\alpha=\frac{\pi}{2}$ and $r=20M$.\\
\vspace{2mm}
\begin{tabular}{ccccccccccc}
\hline
  $-$ &${a_{3}}$ &${b_{3}}$ &${c_{3}}$ &${d_{3}}$ \\
  \hline
  ${F}$  &  $0.4074$  &  $-0.1822$  &  $0.0725$  &  $0.7019$  \\
  \hline
  ${F_{\rm obs}(\theta_{0}=17^{\circ})}$  &  $0.2317$  &  $-0.1793$  &  $0.0408$  &  $0.7062$   \\
  \hline
  ${F_{\rm obs}(\theta_{0}=53^{\circ})}$  &  $0.1578$  &  $-0.1784$  &  $0.0275$  &  $0.7095$  \\
  \hline
  ${F_{\rm obs}(\theta_{0}=75^{\circ})}$  &  $0.1412$  &  $-0.1787$  &  $0.0244$  &  $0.7096$ \\
  \hline
\end{tabular}}
\end{center}

\subsection{Magnetic charge constraint}
\label{sec:4-5}
\par
Note that the regular BH shadow radius depends on the magnetic charge parameter $g$. However, from an observational perspective, the diameter of the BH shadow, denoted as $d_{\rm sh}$, can be measured using data from the EHT. It is well-known that the angular size of the shadow of M87$^{*}$ is $\delta=(42 {\pm} 3)$ ${\rm {\mu as}}$, and its distance is $D=16.8_{-0.7}^{+0.8}$ ${\rm Mpc}$. The BH mass is estimated to be $M=(6.5 {\pm} 0.9) \times 10^{9}M_{\odot}$. Using these observational data, we can calculate the diameter of the shadow as $d_{\rm M87^{*}}=\frac{D\delta}M\simeq 11.0\pm 1.5$ \cite{42,43}.

\par
The EHT has not only measured the angular size of the emission ring of the Sgr A$^{*}$ BH, which is $\delta_{\rm d}=(51.8 {\pm} 2.3)$ ${\rm {\mu as}}$, but also estimated the angular size of its shadow to be $\delta=(48.7 {\pm} 7)$ ${\rm {\mu as}}$. Additionally, the distance and mass of Sgr A$^{*}$ are separately given as $D=(8.15 {\pm} 0.15)$ ${\rm kpc}$ and $M=(4.0_{-0.6}^{+1.1}) \times 10^{6}M_{\odot}$ \cite{7}. Using this observational data, we can calculate the diameter of the shadow, denoted as $d_{\rm Sgr A^{*}}$. By applying the constraints obtained from EHT observations, we can further investigate the magnetic charge parameter $g$.

\par
As depicted in Fig. 9, the result obtained for Hayward and Bardeen BHs are consistent with those derived from EHT observations within the observational uncertainty. Using the $1\sigma$ and $2\sigma$ confidence intervals of $d_{\rm M87^{*}}$, the magnetic charge parameter of the Hayward BH can be constrained as $g \leq 1.72$ at the $1\sigma$ level and $g \leq 2.11$ at the $2\sigma$ level. The magnetic charge parameter of the Bardeen BH can be constrained as $g \leq 0.92$ within $1\sigma$ and $g \leq 1.33$ within $2\sigma$. When applying the confidence intervals of $d_{\rm sgrA^{*}}$, the magnetic charge parameter of the Hayward BH can be constrained as $g \leq 1.89$ within $1\sigma$ and $g \leq 2.02$ within $2\sigma$. For the Bardeen BH, the corresponding constraints are $g \leq 1.09$ at $1\sigma$ and $g \leq 1.24$ at $2\sigma$.
\begin{center}
\includegraphics[width=7.5cm,height=5.5cm]{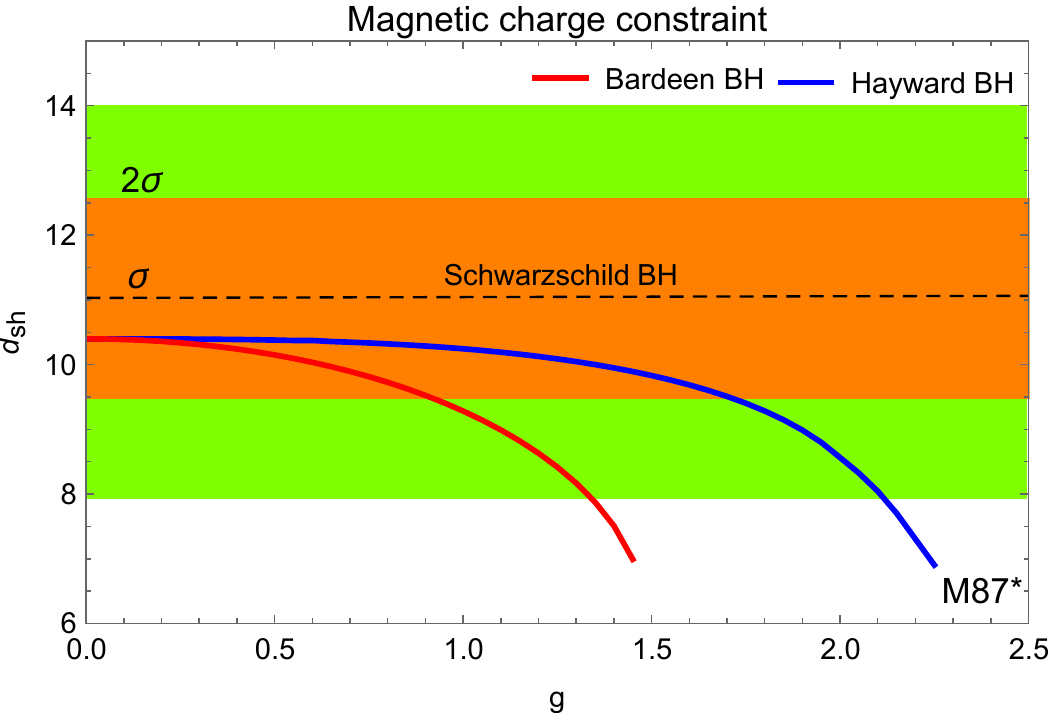}
\includegraphics[width=7.5cm,height=5.5cm]{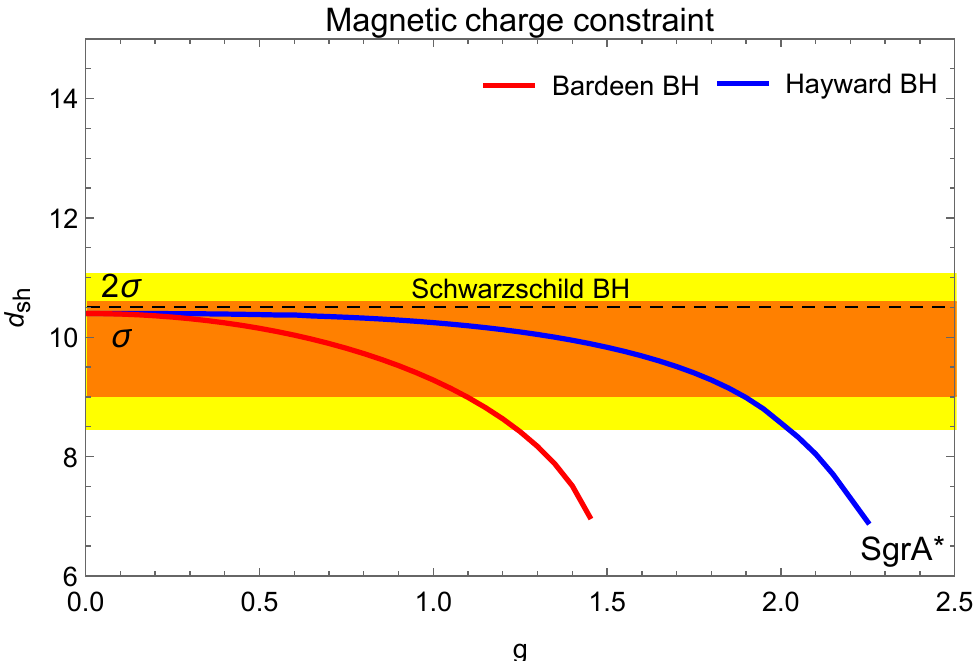}
\parbox[c]{15.0cm}{\footnotesize{\bf Fig~9.}  
Shadow diameter of the three types BHs as a function of the magnetic charge. The green and orange shaded regions represent the regions of $1\sigma$ and $2\sigma$ confidence intervals, respectively.}
\label{fig9}
\end{center}

\par
It is important to note that the accretion flow surrounding both Sgr A$^{*}$ and M87$^{*}$ is known to be optically thin and geometrically thick, while the models analyzed in this analysis assume optically thick and geometrically thin disks. However, our constraints on the magnetic charge are based only on the diameter of the central brightness depression, which can be reasonably expected to remain unchanged in both regimes. Hence, our results remain consistent across both the optically thin and geometrically thick model and the optically thick and geometrically thin model.

\subsection{Anti degeneracy of mass and magnetic charge}
\label{sec:4-6}
\par
Upon reviewing our previous discussion, it becomes evident that distinguishing between regular BHs and Schwarzschild BHs based on direct images presents a challenge. Moreover, a circular structure at the center of the image contributed by the secondary image can resemble the features of BH photon rings but with smaller observation angles. We previously discussed that the distance between the direct image and the secondary image of a BH can help distinguish BH types. Another distinguishing factor is the impact of luminosity. However, we have yet to delve into the topics of BH mass and magnetic charge. Previous research has indicated that the mass of BHs is directly proportional to the shadow radius, while the effect of magnetic charges is the opposite. It is pertinent to explore whether the mass and charge parameters of BHs can be adjusted to strike a balance between these two factors. If such an adjustment is possible, it might render it challenging to differentiate between the three types of BHs solely based on these parameters. In conclusion, differentiating between regular BHs and Schwarzschild BHs based solely on direct images is a daunting task, but a feature of minimal distance between the innermost region of the direct image and the outermost region of the secondary image can serve as a distinguishing feature. Luminosity also plays a significant role. Further research into the impact of BH mass and magnetic charge parameters is essential to determine their utility in distinguishing between the three types of BHs.

\par
We simulated the apparent images of three types of BHs: the Schwarzschild BH with a mass of $1.732 \times 10^{6} M \odot$, the Hayward BH with a mass of $1.068 \times 10^{6} M \odot$, and the Bardeen BH with the same mass as the Hayward BH. In all simulations, the magnetic charge was set to $g=1.5$. The celestial coordinates, $\delta(\xi)$, were obtained from $X(Y)$ \cite{bb}
\begin{equation}
\label{4-6-1}
\frac{\delta}{{\rm {\mu as}}} = \Big(\frac{6.191165 \times 10^{-8}}{2\pi} \frac{\sigma}{D / {\rm Mpc}}\Big)\Big(\frac{X}{M}\Big),
\end{equation}
where $\delta$ represents the mass ratio of the BH to the Sun, while $D$ represents the fixed distance of 5$\rm kpc$. Figure 10 displays images of three types of BHs corresponding to specific mass and magnetic charge parameters, with an observed inclination angle of $\theta_{0}=17^{\circ}$. It is apparent that the structures of the Schwarzschild and Bardeen BHs are essentially the same, featuring similar circle diameters of approximately 40${\rm {\mu as}}$. Therefore, we can conclude that the observed Bardeen BH image can be considered a degenerate projection of the Schwarzschild BH image. The anti-degeneracy effect of mass and magnetic charge eliminating the gap between them.
\begin{center}
\includegraphics[width=4.5cm,height=4.5cm]{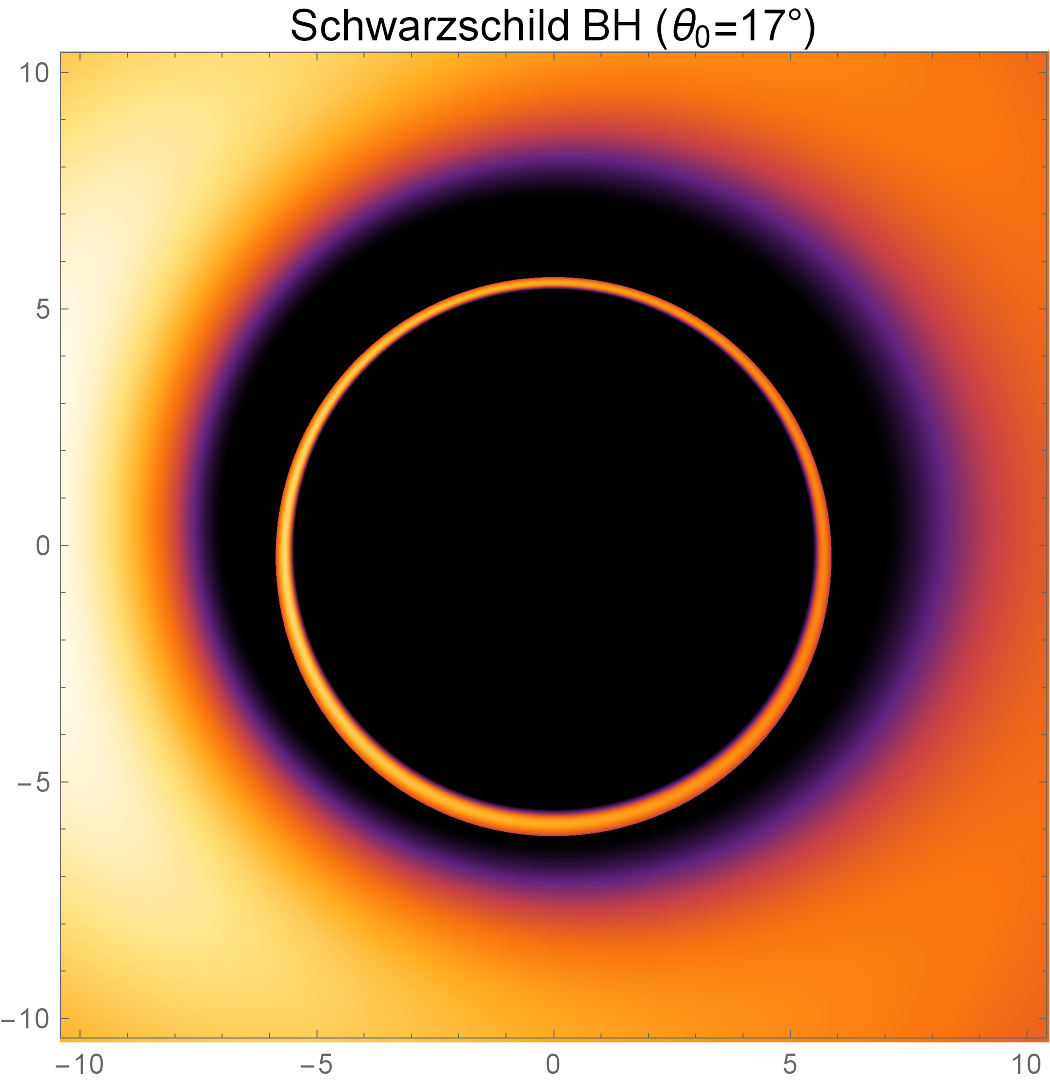}
\hspace{0.2cm}
\includegraphics[width=4.5cm,height=4.5cm]{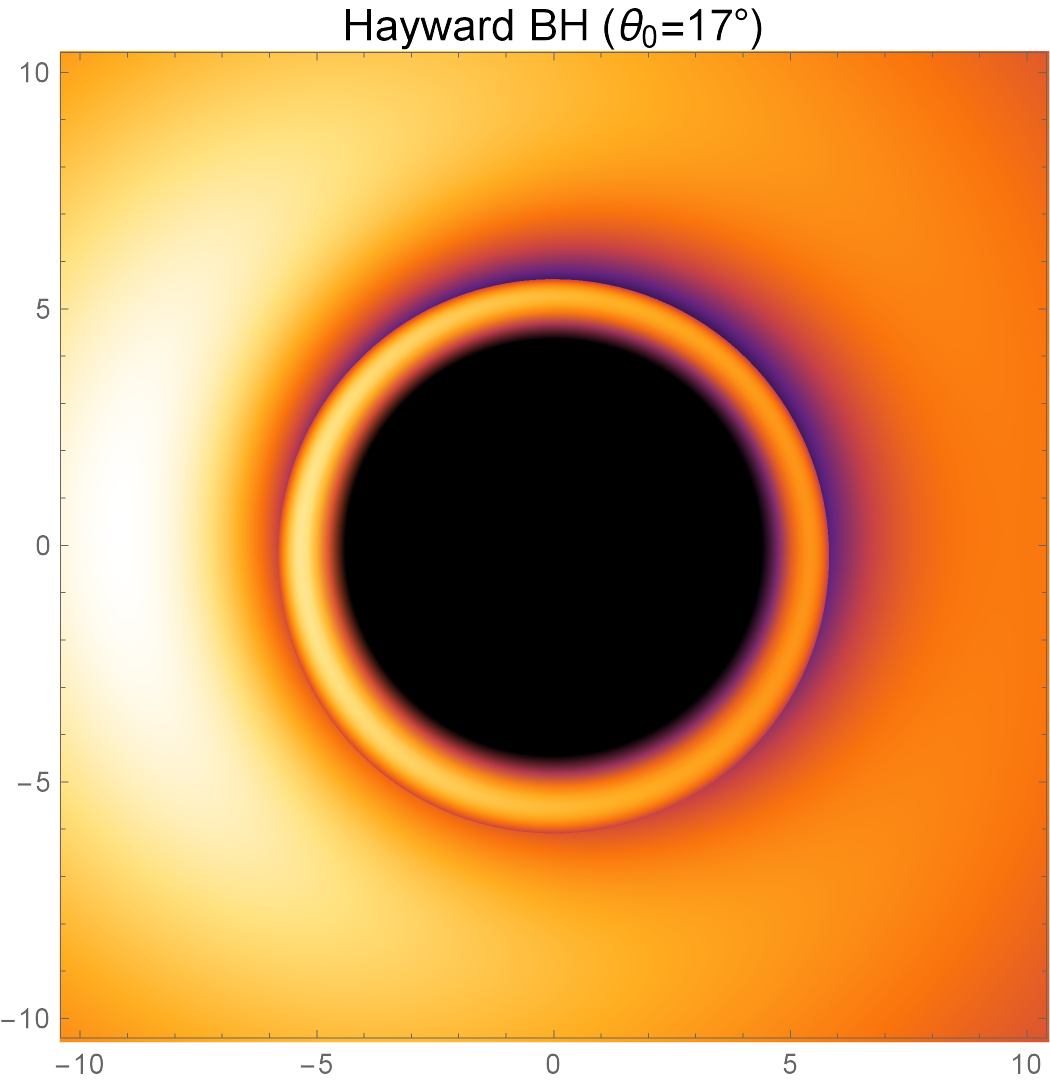}
\hspace{0.2cm}
\includegraphics[width=4.5cm,height=4.5cm]{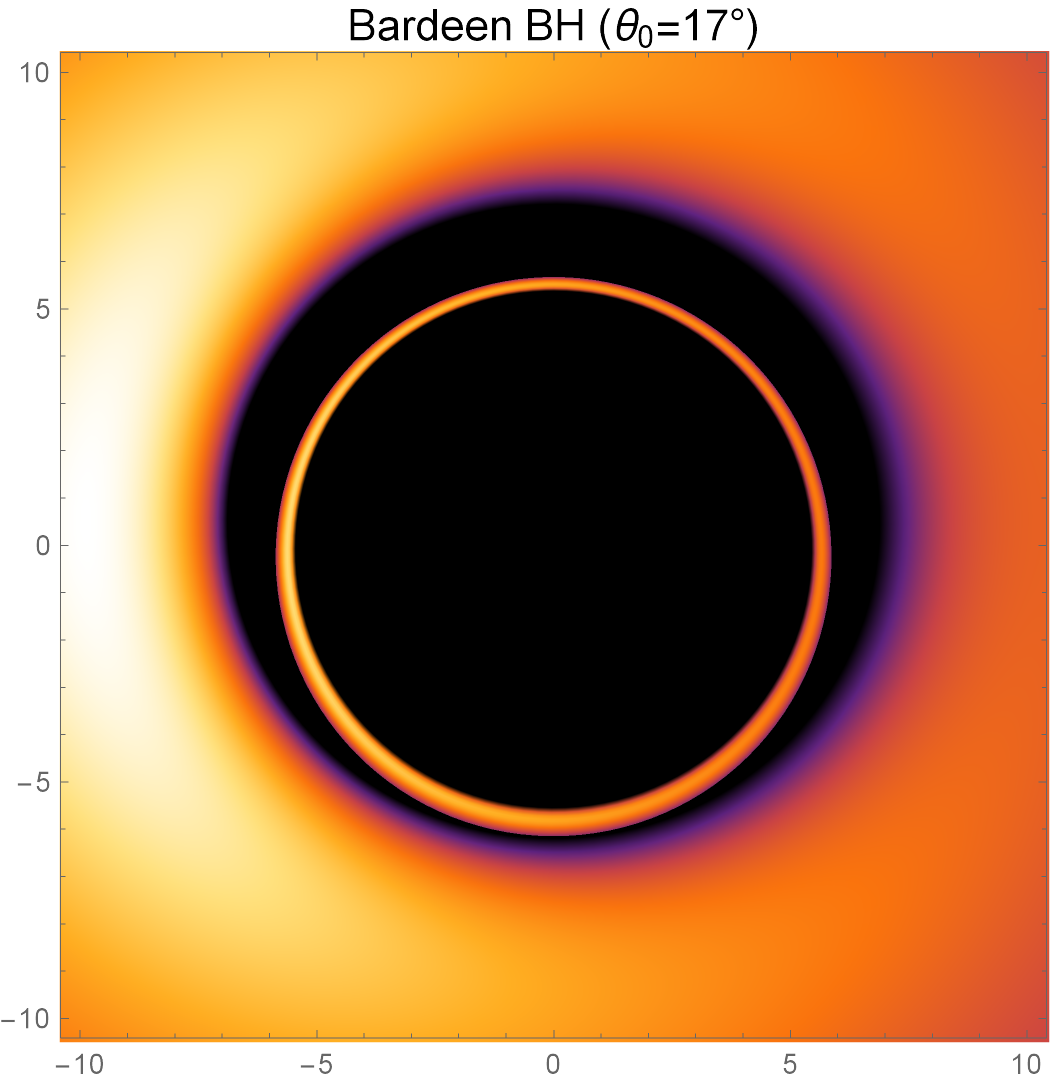}
\parbox[c]{15.0cm}{\footnotesize{\bf Fig~10.}  
The complete apparent images of the Schwarzschild BH with mass $1.732 \times 10^{6} M_\odot $ the Hayward and Bardeen BHs with mass $1.068 \times 10^{6} M \odot$, and the magnetic charge is $g=1.5$. the observation alangle and BH distance are $\theta_{0}=17^{\circ}$ and 5$\rm kpc$.}
\label{fig10}
\end{center}

\par
The spin of a rotating BH has a significant impact on the narrowing of the gap between the photon ring and the innermost stable circular orbit. This is due to changes in the BH's geometry and gravitational field. Importantly, this effect can potentially introduce degeneracies with magnetic charge. In certain situations, spin and magnetic charge can have similar effects, while our study primarily focuses on non-rotating BHs. In practical astronomical observations, the presence of rotating BHs needs to be taken into account. Understanding how spin affects the constraints on magnetic charge becomes crucial for accurately measuring BH properties. While our study does not delve into a detailed analysis of rotating black holes, further investigation into the influence of spin on this matter may prove beneficial for achieving a more comprehensive understanding of the problem and its astronomical implications.

\section{Conclusions and Discussion}
\label{sec:5}
\par
In this analysis, we have investigated the optical appearance of three distinct types of BHs surrounded by an optically thick accretion disk. Our differentiation among these BHs was based on their observable features and calculated their respective effective potential. Our analysis revealed that for a Schwarzschild BH, the peak of the effective potential occurs at approximately $\simeq 3r_{\rm g}$. The effective potential for a Bardeen BH is larger, and the radius corresponding to the peak of the effective potential is smaller than that of a Hayward BH. Furthermore, we determined the deflection angle for each type of BH using both numerical integration and semi-analytical methods. Our results demonstrated that although there were differences between the two methods, both successfully depicted the deflection of light around the BH. Furthermore, we employed a ray-tracing code to provide photon trajectories around these three types of BHs, providing further insights into their behavior in an optically thick accretion disk environment.

\par
We employed the Novikov-Thorne thin disk model to investigate BHs surrounded by an accretion disk, utilizing both numerical integration algorithms and semi-analytical methods. Our findings indicate that for smaller observation angles, the secondary image becomes enclosed within the direct image, giving rise to a structure resembling a ``photon ring''. Additionally, we observed that the degree of separation between the direct and secondary images increases with the inclination angle of observation. Upon comparing three types of BHs, we found that the Hayward BH has a smaller orbital size than the Bardeen BH under similar parameters.

\par
Based on our investigation into the optical appearance of the accretion disk and the corresponding observation images at various observation angles, we have discovered that the luminosity in the region near the BH on the inner side of the accretion disk is higher than that on the outer side. This phenomenon can be attributed to the higher material density in closer proximity to the BH, leading to a greater susceptibility to the BH's gravitational influence. Consequently, this results in more intense heating and radiation. Our findings also indicate that the Bardeen BH and Schwarzschild BH share a similar appearance. However, there is a significant difference in radiance between them. This discrepancy suggests that the spatiotemporal structure of a regular BH enables the surrounding accretion disk to emit more photons. This effect is driven by the more pronounced spatiotemporal curvature generated by magnetic charges. Furthermore, our analysis of gravitational redshift reveals that the redshift area gradually expands with increasing observation inclination angle. Specifically, our results demonstrate that Hayward BHs exhibit a smaller redshift region, while Bardeen BHs exhibit a larger redshift region.

\par
Based on the results presented in Fig. 7, we can conclude that the inclination angle of observation has a significant impact on the observed outcome. Notably, regardless of whether the dip angle increases or decreases, a distinct accumulation of brightness occurs on the left side of the accretion disk. This phenomenon can be attributed to the influence of matter movement along the accretion disk towards the BH. The velocity and density distribution of the matter affect the probability of light absorption and re-emitted, ultimately resulting in the observed brightness accumulation. Additionally, the brightness accumulation is a geometric effect that depends on the angle and path of light emission. Our research reveals that the relationship between the Bardeen BH and the Schwarzschild BH is closer, while the gap between the secondary image of the Hayward BH and the innermost stable circular orbit is very small, which is a characteristic different from other types of BHs. When fitting the functional model to the observed flux at various observation inclinations, we find that the exponential function effectively describes the relationship between the observed flux and the magnetic charge parameter.

\par
After conducting our study, we have concluded that distinguishing between regular BHs and Schwarzschild BHs based solely on direct images poses a significant challenging. However, we have identified that the small spacing feature between direct and secondary images can be utilized to differentiate Hayward BHs. Furthermore, we have observed that Bardeen BH images can function as a degenerate projection of Schwarzschild BHs, with the anti-degeneracy effect of mass and magnetic charge closing the gap between them.

\section*{Acknowledgments}
This work is supported by the National Natural Science Foundation of China (Grant No. 12133003, 42230207) and the Fundamental Research Funds for the Central Universities, China University of Geosciences (Wuhan) (Grant No. G1323523064), and the Sichuan Youth Science and Technology Innovation Research Team (Grant No. 21CXTD0038).

\clearpage

\end{CJK}

\section{References}
\addcontentsline{toc}{chapter}{References}
\begin{thebibliography}{99}\footnotesize
\itemsep=-3pt plus.2pt minus.2pt   

\bibitem{1}
R. Abbott et al, [LIGO Scientific Collaboration, Virgo Collaboration, and KAGRA Collaboration], Observation of Gravitational Waves from a Binary Black Hole Merger, Phys. Rev. Lett. \textbf{116}: 061102, (2016).

\bibitem{2}
R. Abbott et al, [LIGO Scientific Collaboration, Virgo Collaboration, and KAGRA Collaboration], GW170104: Observation of a 50-Solar-Mass Binary Black Hole Coalescence at Redshift 0.2, Phys. Rev. Lett. \textbf{118}: 221101, (2017).

\bibitem{3}
R. Abbott et al, [LIGO Scientific Collaboration, Virgo Collaboration, and KAGRA Collaboration], GW170608: Observation of a 19-Solar-Mass Binary Black Hole Coalescence, Phys. Rev. Lett. \textbf{851}: L35, (2017).

\bibitem{4}
R. Abbott et al, [LIGO Scientific Collaboration, Virgo Collaboration, and KAGRA Collaboration], GW150914: The Advanced LIGO Detectors in the Era of First Discoveries, Phys. Rev. Lett. \textbf{116}: 131103, (2018).

\bibitem{5}
R. Abbott et al, [LIGO Scientific Collaboration, Virgo Collaboration, and KAGRA Collaboration], GW190521: A Binary Black Hole Merger with a Total Mass of 150 Solar Masses, Phys. Rev. Lett. \textbf{125}: 101102, (2020).

\bibitem{6}
K. Akiyama et al. [Event Horizon Telescope Collaboration], First $M87$ Event Horizon Telescope Results. I. The Shadow of the Supermassive Black Hole, Astrophys. J. \textbf{L1}: 875 (2019); [Event Horizon Telescope Collaboration], First $M87$ Event Horizon Telescope Results. II. Array and Instrumentation, Astrophys. J. \textbf{L2}: 875, (2019); [Event Horizon Telescope Collaboration], First $M87$ Event Horizon Telescope Results. III. Data Processing and Calibration, Astrophys. J. \textbf{L3}: 875, (2019); [Event Horizon Telescope Collaboration], First $M87$ Event Horizon Telescope Results. IV. Imaging the Central Supermassive Black Hole, Astrophys. J. \textbf{L4}: 875, (2019); [Event Horizon Telescope Collaboration], First $M87$ Event Horizon Telescope Results. V. Physical Origin of the Asymmetric Ring, Astrophys. J. \textbf{L5}: 875, (2019); [Event Horizon Telescope Collaboration], First $M87$ Event Horizon Telescope Results. VI. The Shadow and Mass of the Central Black Hole, Astrophys. J. \textbf{L6}: 875, (2019).

\bibitem{7}
K. Akiyama et al, [Event Horizon Telescope Collaboration], First Sagittarius A* Event Horizon Telescope Results. I. The Shadow of the Supermassive Black Hole in the Center of the Milky Way, Astrophys. J. Lett. \textbf{L12}: 930, (2022); First Sagittarius A* Event Horizon Telescope Results. II. EHT and Multiwavelength Observations, Data Processing, and Calibration, Astrophys. J. Lett. \textbf{L13}: 930, (2022); First Sagittarius A* Event Horizon Telescope Results. III. Imaging of the Galactic Center Supermassive Black Hole, Astrophys. J. Lett. \textbf{L14}: 930, (2022); First Sagittarius A* Event Horizon Telescope Results. IV. Variability, Morphology, and Black Hole Mass, Astrophys. J. Lett. \textbf{L15}: 930, (2022); First Sagittarius A* Event Horizon Telescope Results. V. Testing Astrophysical Models of the Galactic Center Black Hole, Astrophys. J. Lett. \textbf{L16}: 930, (2022); First Sagittarius A* Event Horizon Telescope Results. VI. Testing the Black Hole Metric, Astrophys. J. Lett. \textbf{L17}: 930, (2022).

\bibitem{8}
K. Akiyama et al, [Event Horizon Telescope Collaboration], First $M87$ Event Horizon Telescope Results. VII. Polarization of the Ring, Astrophys. J. \textbf{L12}: 910, (2021); First $M87$ Event Horizon Telescope Results. VIII. Magnetic Field Structure near The Event Horizon, Astrophys. J. Lett. \textbf{L13}: 910, (2021).

\bibitem{9}
J. C. McKinney, A. Tchekhovskoy, R. D. Blandford, General Relativistic Magnetohydrodynamic Simulations of Magnetically Choked Accretion Flows around Black Holes, Mon. Not. R. Astron. Soc. \textbf{423}: 3083 (2012).

\bibitem{10}
T. Alexander, P. Natarajan, Rapid growth of seed black holes in the early universe by supra-exponential accretion, Science. \textbf{345}: 1330 (2014).

\bibitem{11}
S. E. Gralla, A. Lupsasca, M. J. Rodriguez, Electromagnetic Jets from Stars and Black Holes, Phys. Rev. D. \textbf{93}: 044038 (2016).

\bibitem{Porth}
O. Porth et al, The Event Horizon General Relativistic Magnetohydrodynamic Code Comparison Project, Astrophys. J. Suppl. \textbf{243}: 26 (2019).

\bibitem{12}
S. C. Noble, J. H. Krolik, J. F. Hawley, Direct calculation of the radiative efficiency of an accretion disk around a black hole, Astrophys. J \textbf{692}: 411-421 (2009).

\bibitem{13}
J. P. Luminet, Image of a spherical black hole with thin accretion disk, Astron. Astrophys. \textbf{75}: 228 (1979).

\bibitem{14}
C. Bambi, A code to compute the emission of thin accretion disks in non-Kerr space-times and test the nature of black hole candidates, Astrophys. J \textbf{761}: 174 (2012).

\bibitem{15}
A. Grenzebach, V. Perlick, C. L?mmerzahl, Photon Regions and Shadows of Kerr-Newman-NUT Black Holes with a Cosmological Constant, Phys. Rev. D. \textbf{89}: 124004 (2014).

\bibitem{Gold}
R. Gold et al, [Event Horizon Telescope Collaboration], Verification of Radiative Transfer Schemes for the EHT, Astrophys. J. \textbf{897}: 148 (2020).

\bibitem{Prather}
B. S. Prather et al, [Event Horizon Telescope Collaboration], Comparison of Polarized Radiative Transfer Codes Used by the EHT Collaboration, Astrophys. J. \textbf{950}: 35 (2023).

\bibitem{16}
S. E. Gralla, A. Lupsasca, A. Strominger, Observational Signature of High Spin at the Event Horizon Telescope, Mon. Not. R. Astron. Soc. \textbf{475}: 3829 (2018).

\bibitem{17}
P. V. P. Cunha, N. A. Eir\'{o}, C. A. R. Herdeiro, J. P. S. Lemos, Lensing and shadow of a black hole surrounded by a heavy accretion disk, JCAP, \textbf{03}: 035 (2020).

\bibitem{18}
R. Kumar, S. G. Ghosh, Rotating black holes in 4D Einstein-Gauss-Bonnet gravity and its shadow, JCAP, \textbf{07}: 053 (2020).

\bibitem{19}
R. Kumar, S. G. Ghosh, Gravitational lensing by charged black hole in regularized 4D Einstein-Gauss-Bonnet gravity, Eur. Phys. J. C, \textbf{80}: 1128 (2020).

\bibitem{20}
M. Okyay, A. \"{O}vg\"{u}n, Nonlinear electrodynamics effects on the black hole shadow, deflection angle, quasinormal modes and greybody factors, JCAP, \textbf{01}: 009 (2020).

\bibitem{21}
R. C. Pantig, A. \"{O}vg\"{u}n, Dark matter effect on the weak deflection angle by black holes at the center of Milky Way and M87 galaxies, Eur. Phys. J. C, \textbf{82}: 391 (2022).

\bibitem{22}
A. Chael, M. D. Johnson, A. Lupsasca, Observing the Inner Shadow of a Black Hole: A Direct View of the Event Horizon, Astrophys. J. \textbf{918}: 6 (2021).

\bibitem{23}
S. E. Gralla, D. E. Holz and R. M. Wald, Black hole shadows, photon rings, and lensing rings, Phys. Rev. D. \textbf{100}: 024018 (2019).

\bibitem{24}
X. X. Zeng, H. Q. Zhang, Influence of quintessence dark energy on the shadow of black hole, Eur. Phys. J. C. \textbf{80}: 1058 (2020).

\bibitem{25}
S. Guo, G. R. Li, E. W. Liang, Influence of accretion flow and magnetic charge on the observed shadows and rings of the Hayward black hole, Phys. Rev. D. \textbf{105}: 023024 (2022).

\bibitem{26}
S. Guo, K. J. He, G. R. Li and G. P. Li, The shadow and photon sphere of the charged black hole in Rastall gravity, Class. Quant. Grav. \textbf{38}: 165013 (2021).

\bibitem{27}
S. Guo, G. R. Li, E. W. Liang, Observable characteristics of the charged black hole surrounded by thin disk accretion in Rastall gravity, Class. Quant. Grav. \textbf{39}: 135004 (2020).

\bibitem{28}
X. X. Zeng, K. J. He, G. P. Li, E. W. Liang and S. Guo, QED and accretion flow models effect on optical appearance of Euler-Heisenberg black holes, Eur. Phys. J. C. \textbf{82}: 764 (2022).

\bibitem{29}
G. P. Li and K. J. He, Observational appearances of a f(R) global monopole black hole illuminated by various accretions, Eur. Phys. J. C. \textbf{81}: 1018 (2021).

\bibitem{30}
X. X. Zeng, K. J. He, G. P. Li, E. W. Liang and S. Guo, QED and accretion flow models effect on optical appearance of Euler-Heisenberg black holes, Eur. Phys. J. C. \textbf{82}: 764 (2022).

\bibitem{Wang}
X. Y. Wang, Y. H. Hou, M. Y. Guo, How different are shadows of compact objects with and without horizons? JCAP \textbf{05}: 036 (2023).

\bibitem{Zhang}
Z. Y. Zhang, H. P. Yan, M. Y. Guo and B. Chen, Shadows of Kerr black holes with a Gaussian-distributed plasma in the polar direction, Phys. Rev. D, \textbf{107}: 024027 (2023).

\bibitem{Hou}
Y. H. Hou, P. Liu, M. Y. Guo, H. P. Yan and B. Chen, Multi-level images around Kerr-Newman black holes, Class. Quant. Grav. \textbf{39}: 194001 (2022).

\bibitem{Yan}
H. P. Yan, Z. Z. Hu, M. Y. Guo and B. Chen, Photon emissions from near-horizon extremal and near-extremal Kerr equatorial emitters
, Phys. Rev. D, \textbf{104}: 124005 (2021).

\bibitem{Meng}
K. Meng, X. L. Fan, S. Li, W. B. Han, and H. S. Zhang, Dynamics of null particles and shadow for general rotating black hole, arXiv:2307.08953 [gr-qc].

\bibitem{Meng-1}
Y. Meng, X. M. Kuang, X. J. Wang, B. Wang, and J. P. Wu, Images from disk and spherical accretions of hairy Schwarzschild black holes, Phys. Rev. D, \textbf{108}: 064013 (2023).

\bibitem{31}
G. Gyulchev, P. Nedkova, T. Vetsov, S. Yazadjiev, Image of the thin accretion disk around compact objects in the Einstein-Gauss-Bonnet gravity, Eur. Phys. J. C. \textbf{81}: 885 (2021).

\bibitem{32}
G. Gyulchev, J. Kunz, P. Nedkova, T. Vetsov, S. Yazadjiev, Observational signatures of strongly naked singularities: image of the thin accretion disk, Eur. Phys. J. C. \textbf{80}: 1017 (2020).

\bibitem{33}
C. Liu, S. Yang, Q. Wu, T. Zhu, Thin accretion disk onto slowly rotating black holes in Einstein-${\AE}$ther theory, JCAP \textbf{02}: 034 (2022).

\bibitem{34}
R. Shaikh, S. Paul, P. Banerjee, T. Sarkar, Shadows and thin accretion disk images of the $\gamma$-metric, Eur. Phys. J. C. \textbf{82}: 696 (2022).

\bibitem{35}
S. Paul, R. Shaikh, P. Banerjee, T. Sarkar, Observational signatures of wormholes with thin accretion disks, JCAP \textbf{03}: 055 (2020).

\bibitem{36}
C. Q. Liu, L. Tang, J. L. Jing, Image of the Schwarzschild black hole pierced by a cosmic string with a thin accretion disk, Int. J. Mod. Phys. D, \textbf{31}: 2250041 (2022).

\bibitem{aa}
S. Guo, Y. X, Huang, G. P. Li, Optical appearance of the Schwarzschild black hole in the string cloud context, Chin. Phys. C, \textbf{37}: 6 (2023).

\bibitem{bb}
S. Y. Hu, C. Deng, S. Guo, X. Wu, E. W. Liang, Observational signatures of Schwarzschild-MOG black holes in scalar-tensor-vector gravity: images of the accretion disk, Eur. Phys. J. C. \textbf{83}: 264 (2023).

\bibitem{37}
J. M. Bardeen, Proc. of GR5, Tiflis, Georgia: U.S.S.R. (1968).

\bibitem{38}
S. A. Hayward. Formation and evaporation of nonsingular black holes, Phys. Rev. Lett. \textbf{96}: 031103 (2006).

\bibitem{39}
S. Guo, G. R. Li, E. W. Liang, Optical appearance of a thin-shell wormhole with a Hayward profile, Eur. J. Phys. \textbf{83}: 663 (2023).

\bibitem{40}
K. J. He, S Guo, S. C. Tan, G. P. Li, Shadow images and observed luminosity of the Bardeen black hole surrounded by different accretions, Chin. Phys. C, \textbf{46}: 085106 (2022).

\bibitem{Falcke}
H. Falcke, F. Melia, and E. Agol, Viewing the shadow of the black hole at the galactic center, Astrophys. J. \textbf{528}: L13 (1999).

\bibitem{Lemos}
J. P. S. Lemos, V. T. Zanchin, Regular black holes: Guilfoyle's electrically charged solutions with a perfect fluid phantom core, Phys. Rev. D. \textbf{93}: 124012 (2016).

\bibitem{Frolov}
V. P. Frolov, Notes on nonsingular models of black holes, Phys. Rev. D. \textbf{94}: 104056 (2016).

\bibitem{Leon}
J. P. d. Leon, Regular Reissner-Nordstr\"{o}m black hole solutions from linear electrodynamics, Phys. Rev. D. \textbf{95}: 124015 (2017).

\bibitem{Masa}
A. D. D. Masa, E. S. d. Oliveira, V. T. Zanchin, Stability of regular black holes and other compact objects with a charged de Sitter core and a surface matter layer, Phys. Rev. D. \textbf{103}: 104051 (2021).

\bibitem{Novikov}
I. D. Novikov and K. S. Thorne, in ``Black Holes'', ed. C. DeWitt and B. DeWitt, New York: Gordon and Breach (1973).

\bibitem{Page}
D. N. Page and K. S. Thorne, Disk-accretion onto a black hole. Time-averaged structure of accretion disk, Astrophys. J. \textbf{191}: 499 (1974).

\bibitem{Jaroszynski}
M. Jaroszynski and A. Kurpiewski, Optics near kerr black holes: spectra of advection dominated accretion flows, Astron. Astrophys \textbf{326}: 419 (1997).

\bibitem{Usui}
F. Usui, S. Nishida and Y. Eriguchi, Emission line profiles from self-gravitating toroids around black holes, Mon. Not. Roy. Astron. Soc \textbf{301}: 721 (1998).

\bibitem{Harko}
T. Harko, Z. Kovacs, F. S. N. Lobo, Testing Ho\v{i}ava-Lifshitz gravity using thin accretion disk properties, Phys. Rev. D. \textbf{80}: 044021 (2009).

\bibitem{Harko-1}
T. Harko, Z. Kovacs, F. S. N. Lobo, Can accretion disk properties distinguish gravastars from black holes? Class. Quant. Grav. \textbf{26}: 215006 (2009).

\bibitem{Bambi}
C. Bambi, A code to compute the emission of thin accretion disks in non-Kerr space-times and test the nature of black hole candidates. Astrophys. J. \textbf{761}: 174 (2012).

\bibitem{Muller}
T. Muller, J. Frauendiener, Interactive visualization of a thin disc around a Schwarzschild black hole. Eur. J. Phys. \textbf{33}: 955 (2012).

\bibitem{Atamurotov}
F. Atamurotov, A. Abdujabbarov and B. Ahmedov, Shadow of rotating non-Kerr black hole. Phys. Rev. D. \textbf{88}: 064004 (2013).

\bibitem{Eiroa}
E. F. Eiroa and C. M. Sendra, Strong deflection lensing by charged black holes in scalar-tensor gravity. Eur. Phys. J. C. \textbf{74}: 11 (2014).

\bibitem{Schee}
J. Schee and Z. Stuchlik, Gravitational lensing and ghost images in the regular Bardeen no-horizon spacetimes. JCAP. \textbf{06}: 048 (2015).

\bibitem{Falco}
V. D. Falco, M. Falanga, L. Stella, Approximate analytical calculations of photon geodesics in the Schwarzschild metric. Astron. Astrophys. \textbf{595}: A38 (2016).

\bibitem{Poutanen}
J. Poutanen, Accurate analytic formula for light bending in Schwarzschild metric. Astron. Astrophys. \textbf{640}: A24 (2020).

\bibitem{Vincent}
F. H. Vincent et al, Geometric modeling of M87* as a Kerr black hole or a non-Kerr compact object. Astron. Astrophys. \textbf{646}: A37 (2020).

\bibitem{Vagnozzi}
S. Vagnozzi et al, Horizon-scale tests of gravity theories and fundamental physics from the Event Horizon Telescope image of Sagittarius A. Class. Quant. Grav. \textbf{40}: 165007 (2023).

\bibitem{42}
C. Bambi, K. Freese, S. V agnozzi, L. Visinelli, Testing the rotational nature of the supermassive object M87$^{*}$ from the circularity and size of its first image. Phys. Rev. D. \textbf{100}: 044057 (2019).

\bibitem{43}
A. Allahyari, M. Khodadi, S. V agnozzi, D.F. Mota, Magnetically charged black holes from non-linear electrodynamics and the Event Horizon Telescope. JCAP. \textbf{02}: 003 (2020).



\end{thebibliography}
\end{document}